\documentclass[a4paper,12pt]{article}
\pdfoutput=1
\usepackage{epsfig,graphicx,xcolor,amsbsy,amssymb,latexsym,amsfonts,amsmath,,tcolorbox,setspace}
\usepackage{pstricks}
\usepackage{color}
\usepackage{soul}
\usepackage[numbers,square,comma, compress]{natbib}
\usepackage{placeins}
\usepackage{wrapfig}

\usepackage{eurosym}
\usepackage[a4paper]{geometry}
\usepackage{graphicx}
\usepackage{tikz}
\usepackage{xcolor}
\usepackage{amsmath,amssymb,amsfonts,pstricks,setspace}
\usepackage[enableskew]{youngtab}

\numberwithin{equation}{section}

\newcommand{\bea}{\begin{eqnarray}\displaystyle}
\newcommand{\eea}{\end{eqnarray}}

\newcommand{\figref}[1]{Fig.~\protect\ref{#1}}

\newcommand{\wprim}{\mathcal{T}}
\newcommand{\tder}{\mathfrak{t}}

\newcommand{\nome}{Q_\rho}
\newcommand{\nom}{\rho}

\newcommand{\subg}{\mathfrak{k}}

\title{
\begin{flushright}{\vspace{-2.5cm}\small LYCEN 2019-02\\}\end{flushright}
\vspace{2.3cm}
{\bf Symmetries in A-Type Little String Theories, Part II}\\[40pt]
{\Large Eisenstein Series and Generating Functions of Multiple Divisor Sums}\\[45pt]}

\author{\large \textsc{Brice Bastian\footnote{\tt b.bastian@uu.nl}}\,\,\, and\, \textsc{Stefan~Hohenegger\footnote{\tt s.hohenegger@ipnl.in2p3.fr}}}
\date{}

\begin{document}

\maketitle

\begin{center}
\renewcommand{\thefootnote}{\fnsymbol{footnote}}\vspace{-0.5cm}
${}^{\footnotemark[1]\footnotemark[2]}$ Univ Lyon, Univ Claude Bernard Lyon 1, CNRS/IN2P3, IP2I Lyon,\\ UMR 5822, F-69622, Villeurbanne, France\\[0.5cm]
\renewcommand{\thefootnote}{\fnsymbol{footnote}}
${}^{\footnotemark[1]}$ Institute~for~Theoretical~Physics\\ Utrecht University, Princetonplein 5, 3584 CE Utrecht, The Netherlands\\[2.5cm]
\end{center}

\begin{abstract}
We continue our study of symmetries of a class of little string theories of A-type, which are engineered by $N$ parallel M5-branes probing a flat transverse space. Extending the analysis of the companion paper \cite{Companion1}, we discuss the part of the free energy that is sensitive to the details of the $\mathfrak{a}_{N-1}$ gauge structure, by computing explicit series expansions for the cases $N=2,3,4$. Based on these examples, we find a class of functions that we conjecture to resum whole sectors in the instanton expansion of the free energy and which combine in a natural manner its modular properties as well as the gauge symmetry. These functions have previously been introduced in the literature as the generating functions of multi-divisor sums and in the case $N=2$ can also be cast into the form of a generalised Eisenstein series. We use these resummed contributions to the free energy to perform a number of non-trivial consistency checks for our results. 
\end{abstract}

\newpage

\tableofcontents

\onehalfspacing

\vskip1cm

%%%%%%%%%%%%%%%%%%%%%%%%%%%%%%%%%%%%%%
%%%%%%%%%%%%%%%%%%%%%%%%%%%%%%%%%%%%%%%%%%%%%%%%
\section{Introduction}
This paper constitutes the second part in our study of symmetries in a class of supersymmetric quantum field theories, that are constructed from $N$ parallel M5-branes spaced out on a circle and probing a transverse $\mathbb{Z}_M$ orbifold background \cite{Haghighat:2013gba,Haghighat:2013tka,Hohenegger:2013ala,Hohenegger:2015cba,Hohenegger:2015btj}. Continuing the work started in the companion paper \cite{Companion1} as well as \cite{Bastian:2018jlf}, we focus on $M=1$ and use recent insights into dualities among these configurations (which have been described in detail in \cite{Hohenegger:2016yuv,Bastian:2017ing,Bastian:2017ary,Bastian:2018dfu}) to gain a better picture of the symmetries realised within a single such theory. 

The theories we are interested in, are in fact little string theories (LSTs) \cite{Seiberg:1997zk,Intriligator:1997dh} of A-type: while their low-energy descriptions are six-dimensional supersymmetric gauge theories, with gauge groups of $A$-type (\emph{i.e.} $U(N)$ in the current case) and with different matter contents, their UV completion contains not only point like degrees of freedom, but also string-like excitations. They thus correspond to a simplified version of string theory, in which notably gravity is decoupled. The study of such theories has attracted a lot of attention recently, with a focus on classifying them \cite{Bhardwaj:2015oru,Bhardwaj:2019hhd} and exploring their duality structures \cite{Bastian:2017ing,Bastian:2017ary,Bastian:2018dfu}. The study of the A-type LSTs has been particularly fruitful\footnote{For explicit computations in theories with other gauge groups, see \emph{e.g.} \cite{Haouzi:2017vec,Haghighat:2018dwe,Haghighat:2018gqf}.} due to the fact that if one considers them on $\mathbb{R}^4 \times T^2$, the (refined) BPS partition function $\mathcal{Z}_{N,M}(\omega,\epsilon_{1,2})$ can be computed and analysed in a very explicit and direct fashion. Here $\omega$ denotes a set of gauge couplings, mass- and Coulomb branch parameters that are intrinsic to the theory, while $\epsilon_{1,2}$ are regularisation parameters needed to render $\mathcal{Z}_{N,M}$ well-defined and which can be identified with the parameters of the $\Omega$-background \cite{Nekrasov:2002qd}. To obtain $\mathcal{Z}_{N,M}(\omega,\epsilon_{1,2})$, we can use the fact that apart from the M-brane setup mentioned above\footnote{See \cite{Companion1} for more details on this viewpoint.}, these theories allow for an alternative geometric description in terms of F-theory compactified on a toric, non-compact Calabi-Yau threefold $X_{N,M}$. It was argued in \cite{Haghighat:2013gba,Haghighat:2013tka,Hohenegger:2013ala,Hohenegger:2015btj} that $\mathcal{Z}_{N,M}(\omega,\epsilon_{1,2})$ is captured by the topological string partition function of $X_{N,M}$. The latter in turn can be computed efficiently with the help of the refined topological vertex formalism \cite{Aganagic:2003db,Iqbal:2007ii} and $\omega$ corresponds to a suitable basis of K\"ahler parameters of $X_{N,M}$. 

In \cite{Hohenegger:2016yuv,Bastian:2017ary,Bastian:2018dfu} it was argued (and explicitly demonstrated for a number of examples) that there are dualities among the Calabi-Yau manifolds of the type $X_{N,M}\sim X_{N',M'}$ if $NM=N'M'$ and $\text{gcd}(N,M)=\text{gcd}(N',M')$, leading to $\mathcal{Z}_{N,M}(\omega,\epsilon_{1,2})=\mathcal{Z}_{N,M}(\omega',\epsilon_{1,2})$, where $\omega$ and $\omega'$ are related by a suitable duality transformation. This was shown explicitly for $M=1$ in \cite{Bastian:2018dfu} and for generic $(N,M)$ (but in the limit of vanishing $\epsilon_{1,2}$) in \cite{Haghighat:2018gqf}, using a more geometric approach. Focusing on $M=1$, it was furthermore argued in \cite{Bastian:2018jlf} that this duality also leads to additional symmetries for any single such theory. More concretely, it was demonstrated (and verified explicitly in a number of examples) that $\mathcal{Z}_{N,1}(\omega,\epsilon_{1,2})$ is invariant under the symmetry group $\widetilde{\mathbb{G}}(N)\cong\mathbb{G}(N)\times \mathcal{S}_N$ where $\mathcal{S}_N\cong \text{Dih}_{N}\subset S_N$ acts on the gauge parameters, while
\begin{align}
\mathbb{G}(N)\cong\left\{\begin{array}{lcl}\text{Dih}_3 & \text{if} & N=1\,, \\ \text{Dih}_N & \text{if} & N=2,3\,,\\\text{Dih}_\infty & \text{if} & N\geq 4\,.\end{array}\right.\label{FirstIntroDihedral}
\end{align}
Here $\text{Dih}_\infty$ can also be characterised as the group that is freely generated by two elements of order 2, with no additional braid relations. The group $\widetilde{\mathbb{G}}(N)$ has a natural action as a matrix group on the Fourier coefficients appearing in the expansion of the free energy $F_{N,1}(\omega,\epsilon_{1,2})$ after a suitable choice of basis. 

In \cite{Companion1}, we have further analysed consequences of $\widetilde{\mathbb{G}}(N)$ on the so-called reduced free energy, which captures a subsector of the BPS spectrum that is uncharged with respect to the Cartan subalgebra of the maximal $\mathfrak{a}_{N-1}$ gauge algebra.\footnote{Another way of phrasing this is as follows: from the LST perspective, the parameters $\omega$ can be grouped into a coupling constant (referred to as $R$), a mass parameter (called $S$), $N-1$ simple roots of the $\mathfrak{a}_{N-1}$ gauge algebra (called $\widehat{a}_{1,\ldots,{N-1}}$) and a single affine root (called $\rho$) that enhances the latter to affine $\widehat{\mathfrak{a}}_{N-1}$ and which is inversely proportional to the radius of the six-dimensional compactification circle. The reduced free energy studied in \cite{Companion1} is the sector that is independent of $\widehat{a}_{1,\ldots,N-1}$.} In previous works \cite{Ahmed:2017hfr} it has already been argued, that these BPS states in fact constitute a symmetric orbifold CFT. In \cite{Companion1}, we furthermore argued that $\widetilde{\mathbb{G}}(N)$ acting on this sector, along with two $SL(2,\mathbb{Z})$ symmetries, is in fact promoted to a paramodular group $\Sigma_N$, which in the Nekrasov-Shatashvili-limit \cite{Nekrasov:2009rc,Mironov:2009uv} (corresponding to $\epsilon_2\to 0$), is further extended to $\Sigma_N^*\subset Sp(4,\mathbb{R})$. The latter is indeed the hallmark of a symmetric orbifold CFT, thus corroborating the observation of \cite{Ahmed:2017hfr}.

In the current paper we are focusing on the complement of this 'orbifold sector' within the spectrum of the LSTs, \emph{i.e.} all BPS states that are not captured by the above mentioned reduced free energy. Our strategy is to analyse explicitly the cases $N=2$ and $N=3$ (as well as $N=4$, albeit to a lesser degree of detail), from where we shall observe several emerging patterns (for which we also provide non-trivial checks). The latter allow us to reformulate the free energy in terms of a class of generating functions of multiple divisor sums, first introduced in \cite{Bachmann:2013wba}, as well as, for $N=2$ in terms of a generalised Eisenstein series introduced by G.~Eisenstein and reviewed by A.~Weil in \cite{Weil}. The former presentation, in turn can generically be cast into a form which combines modular properties with the transformations under an $\mathfrak{a}_{N-1}$ gauge algebra. In the simplest case $N=2$, the basic building block can be written akin to an Eisenstein series where the usual divisor sigma is replaced by a sum over the root lattice of $\mathfrak{a}_1$. Similar structures also appear for $N>2$, however, they are in general accompanied by more complicated terms containing more involved sums probing the structure of the $U(N)$ gauge group. The latter, however, can still be expressed in terms of the generating functions of multiple divisor sums. The appearance of these latter functions might hint to a better understanding of the full non-perturbative partition function and free energy of the LSTs, potentially also their algebraic properties.

This paper is organised as follows: section~\ref{Sect:Review} provides a review of the partition function $\mathcal{Z}_{N,1}$ as well as the free energy $F_{N,1}$. After having established our notation, in an attempt to render this paper more readable, section~\ref{Sect:SummaryofResults} provides a summary of the results obtained or conjectured in the remainder of this paper. These shall be worked out explicitly in the subsequent sections: Section~\ref{Sect:CaseN2} discusses the case $N=2$, which (due to the moderate complexity of the partition function) allows to study the free energy up to order $Q_R^3$ (and partially even $Q_R^4$), which from a field theoretic perspective corresponds to the instanton level. We particularly argue that the free energy can be re-written in terms of the generating functions of multiple divisor sums mentioned above. Sections~\ref{Sect:CaseN3} and \ref{Sect:CaseN4} repeat the same analysis (albeit to less orders in $Q_R$ owing to the increased complexity of the free energy) for $N=3$ and partially for $N=4$ respectively. Section~\ref{Sect:Conclusion} contains our conclusions and an outlook to future work. Finally, additional reviews on (quasi)modular forms, the generating functions \`a la \cite{Bachmann:2013wba} as well as several lists of expansion parameters, which have been deemed too lengthy for the main text, have been relegated to three appendices.

%%%%%%%%%%%%%%%%%%%%%%%%%%%%%%%%%%%%%%
\section{M-brane Partition Functions and its Symmetries}\label{Sect:Review}
In an attempt to make this paper minimally self-contained, we shall provide a lightning review of some crucial concepts, as well as our notation for the BPS partition function of $\mathcal{Z}_{N,1}$ of a system of $N$ parallel M5-branes on a circle probing a transverse $\mathbb{R}^4$. The latter can also be computed as the topological string partition function on a class of toric Calabi-Yau manifolds $X_{N,1}$. For more information, we refer the reader to the companion paper \cite{Companion1} as well as the original literature, which we shall reference throughout this section.
%%%%%%%%%%%%%%%%%%%%%%%%%%%%%%%
\subsection{Partition Function and Free Energy}
Following the notation of \cite{Bastian:2017ing}, the partition function can be expressed as
\begin{align}
&\mathcal{Z}_{N,1}(\widehat{a}_{1,\ldots,N},S,R;\epsilon_{1,2})=\sum_{\{\alpha\}}\left(\prod_{i=1}^N Q_{m_i}^{|\alpha_i|}\right)\,W^{\alpha_1,\ldots,\alpha_N}_{\alpha_{1},\ldots,\alpha_{N}}(\widehat{a}_{1,\ldots,N},S;\epsilon_{1,2})\,,\label{DefBuildingZ}
\end{align}
Here $(\widehat{a}_1\ldots,\widehat{a}_N,S,R)$ is a basis of independent K\"ahler parameters of $X_{N,1}$ (see \emph{e.g.} \cite{Companion1} for the precise definition) and $m_i=m_i(R_,S,\widehat{a}_{i1,\ldots,N})$ is a function that notably depends on $R$ and $Q_{m_i}=e^{2\pi i m_i}$. Therefore (\ref{DefBuildingZ}) is essentially a series expansion in $Q_R=e^{2\pi i R}$: indeed $\{\alpha\}$ stands for a sum over $N$ integer partitions $\alpha_{1,\ldots,N}$. Furthermore, the precise expression for the building blocks $W^{\alpha_1,\ldots,\alpha_N}_{\alpha_{1},\ldots,\alpha_{N}}(\widehat{a}_{1,\ldots,N},S;\epsilon_{1,2})$ can be found in \cite{Bastian:2017ing,Bastian:2018jlf,Companion1}. Rather than working directly with $\mathcal{Z}_{N,1}$, in this paper we prefer the free energy
\begin{align}
F_{N,1}(\widehat{a}_{1,\ldots,N},S,R;\epsilon_{1,2})=\text{PLog}\,\mathcal{Z}_{N,1}(\widehat{a}_{1,\ldots,N},S,R;\epsilon_{1,2})\,,\label{PlethLog}
\end{align}
where $\text{PLog}$ denotes the plethystic logarithm which only receives contributions from single particle BPS states \cite{Feng:2007ur}. Following the notation of \cite{Companion1}, it can be expanded in the following manner
\begin{align}
F_{N,1}(\widehat{a}_1,\ldots,\widehat{a}_N,S,R;\epsilon_{1,2})=\sum_{s_1,s_2=0}^\infty\sum_{r=0}^\infty\sum_{i_1,\ldots,i_N}^\infty\sum_{k\in\mathbb{Z}}\epsilon_{1}^{s_1-1}\epsilon_{2}^{s_2-1}f^{(s_1,s_2)}_{i_1,\ldots,i_N,k,r}\,Q_{\widehat{a}_1}^{i_1}\ldots Q_{\widehat{a}_N}^{i_N}\,Q_S^k\,Q_R^r\,,\label{TaylorFreeEnergy}
\end{align}
introducing the coefficients $f^{(s_1,s_2)}_{i_1,\ldots,i_N,k,r}$, along with the notation 
\begin{align}
&Q_{\widehat{a}_i}=e^{2\pi i\widehat{a}_i}\,,&&Q_S=e^{2\pi i S}\,,&&Q_R=e^{2\pi i R}\,,&&\text{for} &&i=1,\ldots,N\,.
\end{align}
Extending the notation of \cite{Companion1}, we introduce the expansions
\begin{align}
G_{(s_1,s_2)}^{(i_1,\ldots,i_N)}(R,S)&=\sum_{r=0}^\infty\sum_{k\in \mathbb{Z}}f^{(s_1,s_2)}_{i_1,\ldots,i_N,k,r}\, Q_S^kQ_R^r\,, &&\forall i_{1,\ldots,N}\in\mathbb{N}\cup\{0\}\,,\label{DefinitionG}\\
H_{(s_1,s_2)}^{(i_1,\ldots,i_N,r)}(\rho,S)&=\sum_{\ell=0}^\infty\sum_{k\in\mathbb{Z}} f^{(s_1,s_2)}_{i_1+\ell,i_2+\ell,\ldots,i_N+\ell,k,r}\,Q_S^k\,Q_\rho^\ell\,.\label{DefinitionH}
\end{align}
In the following we will focus on the cases of $H_{(s_1,s_2)}^{(i_1,\ldots,i_N,r)}(\rho,S)$, where at least one of the $i_{1,\ldots,N}$ vanishes. Since the coefficients $f^{(s_1,s_2)}_{i_1,\ldots,i_N,k,r}$ are invariant under cyclic rotations of $(i_1,\ldots,i_N)$, we shall in most cases assume (without restriction of generality) that $i_N=0$.

On the one hand, from the perspective of the system of parallel M5-branes with M2-branes stretched between them, the functions $G_{(s_1,s_2)}^{(i_1,\ldots,i_N)}(R,S)$ count BPS states for a configuration with fixed numbers $(i_1,\ldots,i_N)$ of M2-branes. These functions have previously been studied in the literature and several properties have been discussed (see the following subsubsection). On the other hand, the functions $H_{(s_1,s_2)}^{(i_1,\ldots,i_N,r)}(\rho,S)$ sum configurations of $(i_1+\ell,\ldots,i_N+\ell)$ M2-branes stretched between the M5-branes. From the little string sector, they count the multi string expansion of a specific charge sector. Resumming the latter, we can define
\begin{align}
B^{N,(r)}_{(s_1,s_2)}(\widehat{a}_1,\ldots,\widehat{a}_N,S)=\sum_{i_1,\ldots,i_N}^\infty\sum_{k\in\mathbb{Z}}f^{(s_1,s_2)}_{i_1,\ldots,i_N,k,r}\,Q_{\widehat{a}_1}^{i_1}\ldots Q_{\widehat{a}_N}^{i_N}\,Q_S^k\,.\label{ResummedHfunc}
\end{align}

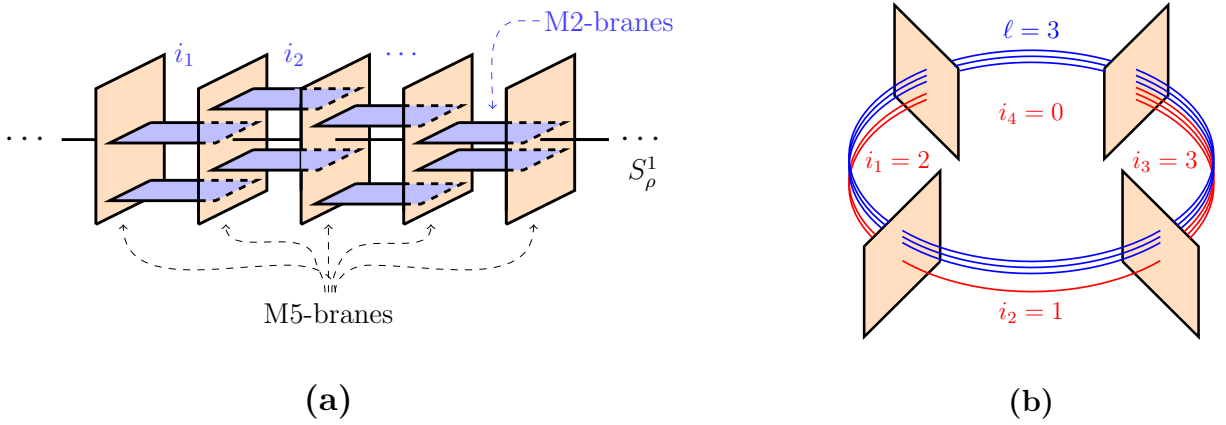
\begin{figure}[h]
\begin{center}
\scalebox{0.9}{\parbox{10.5cm}{\begin{tikzpicture}[scale = 1]
\draw[very thick,fill=orange!25!white] (0,0) -- (1,0.5) -- (1,2.5) -- (0,2) -- (0,0);
\draw[very thick,xshift=1.5cm,fill=orange!25!white] (0,0) -- (1,0.5) -- (1,2.5) -- (0,2) -- (0,0);
\draw[yshift=-0.65cm,fill=blue!25!white,very thick] (0.8,1.3) -- (0.2,1) -- (1.5,1) -- (1.5,1.3) -- (0.8,1.3);
\draw[yshift=-0.65cm,fill=blue!25!white,very thick,dashed] (1.5,1) --(1.75,1) -- (2.35,1.3) -- (1.5,1.3);

\draw[yshift=0.2cm,fill=blue!25!white,very thick] (0.8,1.3) -- (0.2,1) -- (1.5,1) -- (1.5,1.3) -- (0.8,1.3);
\draw[yshift=0.2cm,fill=blue!25!white,very thick,dashed] (1.5,1) --(1.75,1) -- (2.35,1.3) -- (1.5,1.3);
\draw[very thick] (1.5,0.2) -- (1.5,2);
%%%%%%%%%%%%%%%%%
%\draw[yshift=-0.6cm,fill=blue!25!white,very thick] (0.8,1.3) -- (0.2,1) -- (1.5,1) -- (1.5,1.3) -- (0.8,1.3);
%\draw[yshift=-0.6cm,fill=blue!25!white,very thick,dashed] (1.5,1) --(1.75,1) -- (2.35,1.3) -- (1.5,1.3);
%%%%%%%%%%%%%%%%%
\draw[yshift=0.7cm,fill=blue!25!white,very thick,xshift=1.5cm] (0.8,1.3) -- (0.2,1) -- (1.5,1) -- (1.5,1.3) -- (0.8,1.3);
\draw[very thick,xshift=3cm,fill=orange!25!white] (0,0) -- (1,0.5) -- (1,2.5) -- (0,2) -- (0,0);
\draw[yshift=0.7cm,fill=blue!25!white,very thick,dashed,,xshift=1.5cm] (1.5,1) --(1.75,1) -- (2.35,1.3) -- (1.4,1.3);
\draw[very thick,xshift=1.5cm] (1.5,0.5) -- (1.5,2);
\draw[yshift=-0.2cm,fill=blue!25!white,very thick,xshift=1.5cm] (0.8,1.3) -- (0.2,1) -- (1.5,1) -- (1.5,1.3) -- (0.8,1.3);
\draw[yshift=-0.2cm,fill=blue!25!white,very thick,dashed,xshift=1.5cm] (1.5,1) --(1.75,1) -- (2.35,1.3) -- (1.5,1.3);
%%%%%%%%%%%%%%%%%
\draw[very thick,xshift=4.5cm,fill=orange!25!white] (0,0) -- (1,0.5) -- (1,2.5) -- (0,2) -- (0,0);
\draw[yshift=-0.7cm,fill=blue!25!white,very thick,xshift=3cm] (0.8,1.3) -- (0.2,1) -- (1.5,1) -- (1.5,1.3) -- (0.8,1.3);
\draw[yshift=-0.7cm,fill=blue!25!white,very thick,dashed,xshift=3cm] (1.5,1) --(1.75,1) -- (2.35,1.3) -- (1.5,1.3);
\draw[yshift=0.45cm,fill=blue!25!white,very thick,xshift=3cm] (0.8,1.3) -- (0.2,1) -- (1.5,1) -- (1.5,1.3) -- (0.8,1.3);
\draw[yshift=0.45cm,fill=blue!25!white,very thick,dashed,xshift=3cm] (1.5,1) --(1.75,1) -- (2.35,1.3) -- (1.5,1.3);
%%%%%%%%%%%%%%%%%
\draw[very thick,xshift=6cm,fill=orange!25!white] (0,0) -- (1,0.5) -- (1,2.5) -- (0,2) -- (0,0);
%\draw[yshift=0.6cm,fill=blue!25!white,very thick,xshift=4.5cm] (0.8,1.3) -- (0.2,1) -- (1.5,1) -- (1.5,1.3) -- (0.8,1.3);
%\draw[yshift=0.6cm,fill=blue!25!white,very thick,dashed,xshift=4.5cm] (1.5,1) --(1.75,1) -- (2.35,1.3) -- (1.5,1.3);
%
\draw[yshift=0.2cm,fill=blue!25!white,very thick,xshift=4.5cm] (0.8,1.3) -- (0.2,1) -- (1.5,1) -- (1.5,1.3) -- (0.8,1.3);
\draw[yshift=0.2cm,fill=blue!25!white,very thick,dashed,xshift=4.5cm] (1.5,1) --(1.75,1) -- (2.35,1.3) -- (1.5,1.3);
\draw[yshift=-0.2cm,fill=blue!25!white,very thick,xshift=4.5cm] (0.8,1.3) -- (0.2,1) -- (1.5,1) -- (1.5,1.3) -- (0.8,1.3);
\draw[yshift=-0.2cm,fill=blue!25!white,very thick,dashed,xshift=4.5cm] (1.5,1) --(1.75,1) -- (2.35,1.3) -- (1.5,1.3);
%%%%%%%%%%%%%%%%%
\node at (-1,1.25) {\large $\cdots$}; 
\draw[very thick] (-0.5,1.25) -- (0,1.25);
%\draw[very thick] (0.5,1.25) -- (1.5,1.25);
\draw[very thick,xshift=1.5cm] (0.5,1.25) -- (1.5,1.25);
\draw[very thick,xshift=3cm] (0.5,1.25) -- (1.5,1.25);
%\draw[very thick,xshift=4.5cm] (0.5,1.25) -- (1.5,1.25);
\draw[very thick,xshift=6cm] (0.5,1.25) -- (1.5,1.25);
\draw[very thick,xshift=1.5cm] (5,1.25) -- (6,1.25);
%%%%
\node at (3.4,-1.3) {\text{M5-branes}};
\draw[dashed,->] (3.4,-1) -- (3.4,0);
\draw[dashed,->] (3.45,-1) to [out=90,in=240] (3.6,-0.5) to [out=60,in=270] (4.9,0);
\draw[dashed,->] (3.5,-1) to [out=90,in=180] (4,-0.6) to [out=0,in=270] (6.4,0);
\draw[dashed,->] (3.35,-1) to [out=90,in=285] (3.2,-0.5) to [out=105,in=270] (1.9,0);
\draw[dashed,->] (3.5,-1) to [out=90,in=0] (2.8,-0.6) to [out=180,in=270] (0.4,0);
\node at (8,1.25) {\large $\cdots$}; 
\node[blue!75!white] at (1.3,2.5) {$i_1$};
\node[blue!75!white] at (2.9,2.5) {$i_2$};
\node[blue!75!white] at (4.5,2.5) {$\cdots$};
%\node[blue!75!white] at (5.8,2.5) {$i_4$};
\draw[dashed,<-,blue!75!white] (5.8,1.7) to [out=90,in=180] (6.2,3) to [out=0,in=180] (6.5,3);
\node[blue!75!white] at (7.5,3) {\text{M2-branes}};
\draw[very thick,xshift=3cm] (1.5,0.2) -- (1.5,2);
\draw[very thick,xshift=4.5cm] (1.5,0.2) -- (1.5,2);
\node at (8,0.75) {$S^1_\rho$};
\node at (3.4,-2.6) {\large\text{\bf{(a)}}};
\end{tikzpicture}}}
%%%%%%%%%%%%%%%%%%%%%%%%%%%
\hspace{1.5cm}
%%%%%%%%%%%%%%%%%
\scalebox{0.8}{\parbox{6.1cm}{\begin{tikzpicture}[scale = 1]
%\draw (0,0) ellipse (3cm and 1.75cm);
\draw[very thick,fill=orange!25!white] (1.5,-1.25) -- (2.75,-2.5) -- (2.75,-1) -- (1.5,0.25) -- (1.5,-1.25);
\draw[very thick,fill=orange!25!white] (-1.5,-1.25) -- (-2.75,-2.5) -- (-2.75,-1) -- (-1.5,0.25) -- (-1.5,-1.25);
\draw[very thick,fill=orange!25!white] (1.2,1.95) -- (2.25,3) -- (2.25,1.5) -- (1.2,0.45) -- (1.2,1.95);
\draw[very thick,fill=orange!25!white] (-1.2,1.95) -- (-2.25,3) -- (-2.25,1.5) -- (-1.2,0.45) -- (-1.2,1.95);
%%%%%%
\draw [thick,red,domain=-30:55] plot ({3*cos(\x)}, {1.75*sin(\x)});
\draw [thick,red,domain=-31:55,yshift=0.1cm] plot ({3*cos(\x)}, {1.75*sin(\x)});
\draw [thick,red,domain=-33:55,yshift=0.2cm] plot ({3*cos(\x)}, {1.75*sin(\x)});
\draw [thick,blue,domain=-35:55,yshift=0.3cm] plot ({3*cos(\x)}, {1.75*sin(\x)});
\draw [thick,blue,domain=-37:55,yshift=0.4cm] plot ({3*cos(\x)}, {1.75*sin(\x)});
\draw [thick,blue,domain=-39:55,yshift=0.5cm] plot ({3*cos(\x)}, {1.75*sin(\x)});
\node[red] at (2.2,0.4) {$i_3=3$}; 
\draw [thick,red,domain=-135:-45] plot ({3*cos(\x)}, {1.75*sin(\x)});
\draw [thick,blue,domain=-135:-45,yshift=0.3cm] plot ({3*cos(\x)}, {1.75*sin(\x)});
\draw [thick,blue,domain=-135:-45,yshift=0.4cm] plot ({3*cos(\x)}, {1.75*sin(\x)});
\draw [thick,blue,domain=-135:-45,yshift=0.5cm] plot ({3*cos(\x)}, {1.75*sin(\x)});
\node[red] at (0,-2.1) {$i_2=1$}; 
\draw [thick,red,domain=125:209] plot ({3*cos(\x)}, {1.75*sin(\x)});
\draw [thick,red,domain=125:211,yshift=0.1cm] plot ({3*cos(\x)}, {1.75*sin(\x)});
\draw [thick,blue,domain=125:215,yshift=0.3cm] plot ({3*cos(\x)}, {1.75*sin(\x)});
\draw [thick,blue,domain=125:217,yshift=0.4cm] plot ({3*cos(\x)}, {1.75*sin(\x)});
\draw [thick,blue,domain=125:219,yshift=0.5cm] plot ({3*cos(\x)}, {1.75*sin(\x)});
\node[red] at (-2.2,0.4) {$i_1=2$}; 
%
%\draw [thick,domain=66:113] plot ({3*cos(\x)}, {1.75*sin(\x)});
\draw [thick,blue,domain=66:113,yshift=0.3cm] plot ({3*cos(\x)}, {1.75*sin(\x)});
\draw [thick,blue,domain=66:115,yshift=0.4cm] plot ({3*cos(\x)}, {1.75*sin(\x)});
\draw [thick,blue,domain=64:116,yshift=0.5cm] plot ({3*cos(\x)}, {1.75*sin(\x)});
\node[red] at (0,1.2) {$i_4=0$}; 
\node[blue] at (0,2.6) {$\ell=3$}; 
%%%%%%%%%%%%%%%%%
\node at (0,-3.6) {\large\text{\bf{(b)}}};
\end{tikzpicture}}}
\end{center} 
\caption{{\it (a) Configuration of M5-branes (drawn in orange) with $(i_1,\ldots,i_N)$ M2-branes (drawn in blue) stretched between them. Here the direction normal to the M5-branes is compactified on the circle $S_\rho^1$. (b) Schematic representation of the configuration of $N=4$ with $(i_1,i_2,i_3,i_4)=(2,1,3,0)$ and $\ell=3$. The blue lines represent the 'common' number of 3 M2-branes stretched between each pair of M5-branes, while the red lines represent 'additional' M2-branes leading to the configuration $(2,1,3,0)$. To obtain $H_{(s_1,s_2)}^{(2,1,3,0,r)}$ in (\ref{DefinitionH}) a summation over similar configurations with all possible $\ell\in\mathbb{N}$ is required.}}
\label{fig:PartM5braneConfig}
\end{figure}

%%%%%%%%%%%%%%%%%%%%%%%%%%%%%%
\subsection{Symmetries and Modular Properties}\label{Sect:ReviewSyms}
The function $G_{(s_1,s_2)}^{(i_1,\ldots,i_N)}(R,S)$ in (\ref{DefinitionG}) transforms like a quasi-Jacobi form with respect to a congruence subgroup\footnote{We shall be more specific for a particular case in section~\ref{Sect:N2HighOrder}.} of $SL(2,\mathbb{Z})_R$ that acts in the following manner \cite{Companion1}:
\begin{align}
&SL(2,\mathbb{Z})_R:&&(R,S,\rho)\longrightarrow \left(\frac{aR+b}{cR+d},\frac{S}{cR+d},\rho-\frac{c N S^2}{cR+d}\right)\,,\label{SL2Raction}
\end{align}
More concretely, $G_{(s_1,s_2)}^{(i_1,\ldots,i_N)}(R,S)$ is a quasi-Jacobi form of index $K=\sum_{a=1}^Ni_a$ and weight $s_1+s_2-2$ so it can be written in the following manner\footnote{For our convention of modular objects, see appendix~\ref{App:ModularReview}.}
\begin{align}
G_{(s_1,s_2)}^{(i_1,\ldots,i_N)}(R,S)=\sum_{u=0}^{K}g_u^{(s_1,s_2)}(R)\,(\phi_{0,1}(R,S))^{K-u}\,(\phi_{-2,1}(R,S))^{u}\,.\label{ExpansionG}
\end{align}
Here $g_u^{(s_1,s_2)}(R)$ is a quasi-modular form with weight $s_1+s_2+2(u-1)$, \emph{i.e.} it is a polynomial in the Eisenstein series $\{E_2(p_i R),E_4(p_i R),E_6(p_iR)\}$ for all the prime factors $p_i$ appearing in the prime factorisation of $K$. Notice that the presence of $E_2(\rho)$ leads to the fact the $G_{(s_1,s_2)}^{(i_1,\ldots,i_N)}$ do not only transform with a weight factor under modular transformations, but in general also produce a shift term (see appendix~\ref{Sect:QuasiJacobiForms} for more information on quasi-Jacobi forms). 

Furthermore, a priori, the function $H_{(s_1,s_2)}^{(i_1,\ldots,i_N,r)}(\rho,S)$ in (\ref{DefinitionH}) is not a modular object at all. However, both $G_{(s_1,s_2)}^{(i_1,\ldots,i_N)}(R,S)$ and $H_{(s_1,s_2)}^{(i_1,\ldots,i_N,r)}(\rho,S)$ are invariant under the action of the group $\widetilde{\mathbb{G}}(N)\cong\mathbb{G}(N)\times \mathcal{S}_N$, which acts in the following manner on the coefficients in (\ref{DefinitionG}) and (\ref{DefinitionH})
\begin{align}
&f^{(s_1,s_2)}_{i_1,\ldots,i_N,k,r}=f^{(s_1,s_2)}_{i'_1,\ldots,i'_N,k',r'}&&\text{for}&&\left\{\begin{array}{l}(i'_1,\ldots,i'_N,k',r')^T=G^T\cdot (i_1,\ldots,i_N,k,r)^T\,, \\ G\in \widetilde{\mathbb{G}}(N)\,;\hspace{0.3cm}s_1,s_2\in\mathbb{N}\cup \{0\}\,.\end{array}\right.\label{DihedralAction}
\end{align}
Here $\mathcal{S}_N\subset S_N$ acts by shuffling the $\widehat{Q}_{1,\ldots, N}$ in (\ref{TaylorFreeEnergy}). Since the M5-brane setup can be thought of as a regular $N$-gone, whose symmetry group is $S_N$, we have $\mathcal{S}_N\cong \text{Dih}_N$ (which can be indeed checked by studying explicit expressions for $f^{(s_1,s_2)}_{i_1,\ldots,i_N,k,r}$).

In the following we will mostly focus on studying the functions $H_{(s_1,s_2)}^{(i_1,\ldots,i_N,r)}(\rho,S)$ for the values $N=2,3,4$, which will reveal a number of interesting patterns that shall allow us to make a conjecture of their generic form.

%%%%%%%%%%%%%%%%%%%%%%%%%%%%%%%%%%%%%
\section{Summary of Results}\label{Sect:SummaryofResults}
After having introduced the main actors of this paper in the previous section, we are now ready to analyse in detail the $H_{(s_1,s_2)}^{(i_1,\ldots,i_N,r)}(\rho,S)$. However, due to the complexity of some of the results obtained or conjectured below, we shall summarise the most important patterns that we have found before presenting the technical analysis in the subsequent sections. We hope that this increases the readability of this work. Furthermore, while these patterns appear in their cleanest and most tangible form (which shall also allow us to make contact to many other structures previously encountered in the literature) for the case $N=2$, they also extend to $N>2$. However, in these cases all expressions tend to become more complicated and involved. 

Moreover, we stress that the above mentioned patterns shall appear from studying expansions of the free energy in various parameters. Since these, in most cases, do not enjoy modular properties, most results do not lend themselves to (simple) rigorous proofs to all orders and therefore have to be considered conjectures. We note, however, that below we present an overwhelming number of different cases, all following the same type of pattern, thus giving us reasonable confidence that our conjectures actually hold true.

As mentioned above, the main object we shall study throughout this work is the sector of the free energy $H_{(s_1,s_2)}^{(i_1,\ldots,i_N,r)}(\rho,S)$ defined in (\ref{DefinitionH}), or $B^{N,(r)}_{(s_1,s_2)}(\widehat{a}_1,\ldots,\widehat{a}_N,S)$ given in (\ref{ResummedHfunc}), where various contributions have been summed up. While not strictly speaking modular objects (with respect to a group $SL(2,\mathbb{Z})_\rho$ acting on $\rho$), explicit expansions (shown below) suggest that they can be written in the form
\begin{align}
H_{(s_1,s_2)}^{(i_1,\ldots,i_N,r)}(\rho,S)=\sum_{u=1}^{Nr+1}g_{(s_1,s_2)}^{u,(i_1,\ldots,i_N,r)}(\rho)\,\phi_{-2,1}^{Nr+1-u}(\rho,S)\,\phi_{0,1}^{u-1}(\rho,S)\,.\label{ExpansionHNgen}
\end{align}
This is compatible with the group action $\mathbb{G}(N)$ that acts on the Fourier coefficients $f^{(s_1,s_2)}_{i_1,\ldots,i_N,k,r}$ in (\ref{TaylorFreeEnergy}). The $g_{(s_1,s_2)}^{u,(i_1,\ldots,i_N,r)}(\rho)$ can be characterised as series expansions in $Q_\rho=e^{2\pi i\rho}$. While in general not (quasi)modular forms, they nevertheless exhibit interesting patterns: in general, they are not constrained by $\mathbb{G}(N)$, however, the group $\mathcal{S}_N\subset\widetilde{\mathbb{G}}(N)$ (together with the fact that at least one of the $i_{1,\ldots,N}$ has to be chosen to be 0) leaves a finite number of distinct classes of $H_{(s_1,s_2)}^{(i_1,\ldots,i_N,r)}(\rho,S)$ for fixed $N$ (see \emph{e.g.} (\ref{PatternsN4}) below for $N=4$). While their explicit form strongly depends on the indices $(i_1,\ldots,i_N)$ (or more precisely the class under consideration), they can schematically be written in the following form 
\begin{align}
g_{(s_1,s_2)}^{u,(i_1,\ldots,i_N,r)}(\rho)=\sum_\ell\frac{p_\ell(i_1,\ldots,i_N;E_{2}(\rho),E_{4}(\rho),E_{6}(\rho);s_{1,2};u)\,Q_\rho^{t_\ell(i_1,\ldots,i_N;s_{1,2};u)}}{\prod_{a=1}^p (1-Q_\rho^{q_{\ell,a}(i_1,\ldots,i_N;s_{1,2};u)})}\,.\label{GenFormCoefsFreeEnergy}
\end{align} 
Here $p\in\mathbb{N}$, the $q_{\ell,a}$ and $t_\ell$ are linear combinations of $i_{1,\ldots,N}$, while the $p_\ell$ are homogeneous polynomials in (differences of) the $i_{1,\ldots,N}$ and polynomials of the Eisenstein series with fixed weight. Based on the explicit expressions studied in the subsequent sections, we find the following pattern: let $w$ be the (combined) weight of all Eisenstein series in $p_\ell$ and $d$ the degree of the polynomial in $i_{1,\ldots,N}$, then
\begin{align}
2u-2Nr+w+d+p=s_1+s_2\,.
\end{align}
The functions $H_{(s_1,s_2)}^{(i_1,\ldots,i_N,r)}(\rho,S)$ are only certain terms appearing in the expansion of the free energy in a Fourier series. They can be partially resummed in the sense of (\ref{ResummedHfunc}) to form the $B^{N,(r)}_{(s_1,s_2)}(\widehat{a}_1,\ldots,\widehat{a}_N,S)$. The form of (\ref{GenFormCoefsFreeEnergy}), in particular the denominator, suggests that the the latter can be expressed as combinations of generating functions of multiple divisor sums, called $T(X_1,\ldots, X_p;\rho)$, which were first introduced in \cite{Bachmann:2013wba} and whose definition is given in (\ref{GenFuncBracket}). Here $X_{1,\ldots,p}$ are suitable linear combinations of $\widehat{a}_{1,\ldots,N-1}$ and $\rho$. This becomes intuitively clear, since the $Q_\rho$-expansion of $B^{N,(r)}_{(s_1,s_2)}(\widehat{a}_1,\ldots,\widehat{a}_N,S)$ can in general be written as (multiple) divisor sums (decorated sums over (part of the) root lattices of the gauge algebra $\mathfrak{a}_{N-1}$). While this is most transparent in the case of $N=2$ (see \emph{e.g.} section~\ref{Sect:N2RootLattice}), similar patterns also appear for cases $N>2$. However, they generically involve summations over several copies of the root lattices. The generating functions $T(X_1,\ldots, X_p;\rho)$ are not Jacobi forms themselves, but form a natural generalisation thereof (see appendix~\ref{App:MultiDivisorSums}). In the case of $N=2$, we can furthermore show that the appearing combinations naturally combine into the generalised Eisenstein series defined in (\ref{GenEisenstein}) (see \cite{Weil} for further information). For $N>2$, the explicit combinations of $T(X_1,\ldots, X_p;\rho)$ that appear seem more complicated and it is not obvious that they can be recombined into a simpler object. 

Furthermore, the explicit expressions obtained for $B^{N,(r)}_{(s_1,s_2)}(\widehat{a}_1,\ldots,\widehat{a}_N,S)$ lend themselves for further analysis. Indeed, combining the latter with $H_{(s_1,s_2)}^{(0,\ldots,0,r)}(\rho,S)$, which is a quasi-Jacobi form, certain cancellations among them occur: these reduce the depth of the combined expression, \emph{i.e.} the maximal power of $E_2(\rho)$ that is appearing. Finally, we have also verified the relation
\begin{align}
H_{(s_1,0)}^{(\overbrace{\text{\scriptsize $0,\ldots,0$}}^{N-\text{times}},r)}(\rho,S)&+B_{(s_1,0)}^{N,r}(\tfrac{\rho}{N},\ldots,\tfrac{\rho}{N},S)=NH_{(s_1,0)}^{(0,r)}(\tfrac{\rho}{N},S)\,,
\end{align}
which was first conjectured in \cite{Hohenegger:2016eqy}. This not only serves as a strong check of the correctness of our results presented here, but also the work available in the literature. 

In the following we shall substantiate the above conjectures (in particular the explicit form (\ref{GenFormCoefsFreeEnergy})) for the examples $N=2,3,4$ and for low orders in $Q_R$.

%%%%%%%%%%%%%%%%%%%%%%%%%%%%%%%%%%%

%%%%%%%%%%%%%%%%%%%%%%%%%%%%%%%%%%%
%%%%%%%%%%%%%%%%%%%%%%%%%%%%%%%%%%%
%%%%%%%%%%%%%%%%%%%%%%%%%%%%%%%%%%%
%%%%%%%%%%%%%%%%%%%%%%%%%%%%%%%%%%%
\section{Case $N=2$}\label{Sect:CaseN2}\label{Sect:CaseN2}
The simplest non-trivial example to study is the case $N=2$.\footnote{The case $N=1$ is trivial from the point of view of (\ref{DefinitionH}), since for fixed $r$ (and $s_{1,2}$) there is only a single non-trivial $H_{(s_1,s_2)}^{(i,r)}$ (namely $i=0$), which is a quasi-Jacobi form and has already been studied extensively in the literature (see \emph{e.g.} \cite{Hohenegger:2015btj}).} In the following we discuss different orders of $Q_R$ (\emph{i.e.} different values of $r\geq 1$ in eq.~(\ref{DefinitionH})) and specify the functions appearing in (\ref{ExpansionHNgen}) and (\ref{GenFormCoefsFreeEnergy}) respectively.
%%%%%%%%%%%%%%
\subsection{Order $Q_R^1$}
\subsubsection{Explicit Contributions to the Free Energy}
For $N=2$ and at order $Q_R^1$, the distinct functions (\ref{DefinitionH}) (for given $(s_1,s_2)$) are characterised by a single integer $n$:
\begin{align}
H_{(s_1,s_2)}^{(n,0,1)}(\rho,S)&=\sum_{\ell=0}^\infty\sum_{k\in\mathbb{Z}} f^{(s_1,s_2)}_{n+\ell,\ell,k,1}\,Q_S^k\,Q_\rho^\ell\,,&&\forall\,n\geq 1\,.\label{N2HfunctionSimple}
\end{align}
While the case $n=0$ is also well defined, it is in fact part of the reduced free energy that was studied in \cite{Companion1} and we shall therefore not discuss it in detail at present. As remarked above, while the object (\ref{N2HfunctionSimple}) is not in general modular, the coefficients $f^{(s_1,s_2)}_{n+\ell,\ell,k,1}$ are still symmetric with respect to $\mathbb{G}(2)$ (see appendix~\ref{App:ModularReview}). The latter imposes in fact infinitely many relations among the Fourier coefficients: \emph{e.g.} acting on $f^{(s_1,s_2)}_{n+\ell,\ell,k,1}$ with $(\mathcal{G}_2(2)\cdot \mathcal{B})^\kappa$ we find
\begin{align}
&f^{(s_1,s_2)}_{n+\ell,\ell,k,1}=f^{(s_1,s_2)}_{n+\ell+\kappa k+2\kappa^2,\ell+\kappa k+2\kappa^2,k+4\kappa,1}&&\text{for} &&\left\{\begin{array}{l}s_1,s_2,\ell,\kappa\in\mathbb{N}\cup\{0\}\,, \\ n\in\mathbb{N}\,,\\ k\in\mathbb{Z}\,.\end{array}\right.\label{RelationCoefsN2}
\end{align}
Here $\mathcal{G}_2(2)$ is one of the generators of $\mathbb{G}(2)$. Indeed, for generic $N$, we have
\begin{align}
\mathbb{G}(N)\cong\left\langle\{\mathcal{G}_2(N),\mathcal{G}'_2(N)\big|(\mathcal{G}(N))^2=(\mathcal{G}'(N))^2=(\mathcal{G}(N)\cdot\mathcal{G}'(N))^n=1\!\!1_{(N+2)\times (N+2)}\}\right\rangle\,,
\end{align} 
Specifically, as $(N+2)\times (N+2)$ matrices, we have
\begin{align}
&\mathcal{G}_2(N)=\left(\begin{array}{ccccc} & & & 0 & 0 \\ & 1\!\!1_{N\times N} & & \vdots & \vdots \\ & & & 0 & 0 \\ 1 & \cdots & 1 & -1 & 0 \\ N & \cdots & N & -2N & 1 \end{array}\right)\,,&&\text{and}&&\mathcal{G}'_2(N)=\left(\begin{array}{ccccc} & & & -2 & 1 \\ & 1\!\!1_{N\times N} & & \vdots & \vdots \\ & & & -2 & 1 \\ 0 & \cdots & 0 & -1 & 1 \\ 0 & \cdots & 0 & 0 & 1 \end{array}\right)\,.\label{DefGinfGeneral}
\end{align}
Furthermore, $\mathcal{B}$ is defined as
\begin{align}
\mathcal{B}=\text{diag}(1,1,-1,1)\in \mathbb{Z}_2\,,\label{Defbacker}
\end{align}
which realises the symmetry $f^{(s_1,s_2)}_{i_1,i_2,k,1}=f^{(s_1,s_2)}_{i_1,i_2,-k,1}$. Using relation (\ref{RelationCoefsN2}) as well as studying the explicit coefficients stemming from (\ref{PlethLog}) we find the following form\footnote{Since in this way we are only capable of checking a finite number of coefficients and since $H_{(s_1,s_2)}^{(n,0,1)}(\rho,S)$ does not enjoy modular properties, eq.~(\ref{Hfunc21}) should be considered a conjecture. We checked the latter up to order $Q_\rho^{15}$ in eq.~(\ref{FuncsExpansionNsimple}).} (see (\ref{ExpansionHNgen})).
\begin{align}
H_{(s_1,s_2)}^{(n,0,1)}(\rho,S)=g^{1,(n,r=1)}_{(s_1,s_2)}(\rho)\,\phi^2_{-2,1}(\rho,S)+g^{2,(n,r=1)}_{(s_1,s_2)}(\rho)\,\phi_{0,1}(\rho,S)\,\phi_{-2,1}(\rho,S)+g^{3,(n,r=1)}_{(s_1,s_2)}(\rho)\,\phi^2_{0,1}(\rho,S)\,,\label{Hfunc21}
\end{align}
where the functions in (\ref{GenFormCoefsFreeEnergy})
\begin{align}
&g^{i,(n,r=1)}_{(s_1,s_2)}(\rho)=\sum_{k=0}^\infty c_{k}^{i,(s_1,s_2)}(n,r=1)\,Q_\rho^k\,,&&c_{k}^{i,(s_1,s_2)}(n,r=1)\in\mathbb{Z}\,,&&\forall i=1,2\,,\label{FuncsExpansionNsimple}
\end{align}
are a priori generic series in $Q_\rho$. Expansion up to order $Q_\rho^{20}$ suggests that for $(s_1,s_2)=(0,0)$ the latter are in fact given by\footnote{In the following we shall assume that $|Q_\rho|<1$.}
\begin{align}
&g^{1,(n,1)}_{(0,0)}(\rho)=-\frac{2n}{1-Q_\rho^n}\,,&&\text{and} &&g^{2,(n,1)}_{(0,0)}(\rho)=g^{3,(n,1)}_{(0,,0)}(\rho)=0\,, 
\end{align}
such that 
\begin{align}
H_{(0,0)}^{(n,0,1)}(\rho,S)&=-\frac{2n}{1-Q_\rho^n}\,\phi^2_{-2,1} (\rho,S)\,.
\end{align}
We can treat other values of $(s_1,s_2)$ in the same fashion, leading to the following conjectures of closed form expressions of $H_{(s_1,s_2)}^{(n,0,1)}(\rho,S)$
{\allowdisplaybreaks
\begin{align}
g^{1,(n,1)}_{(2,0)}&=\frac{n\,\left(4 n^2- E_2\right)}{12\,(1-Q_\rho^n)}\,, \hspace{3.2cm}g^{2,(n,1)}_{(2,0)}=\frac{n}{24\,(1-Q_\rho^n)}\,,\hspace{3.2cm}g^{3,(n,1)}_{(2,0)}=0\,,\nonumber\\[12pt]
g^{1,(n,1)}_{(4,0)}&=\frac{-96 n^5+80 E_2 n^3-10 E_2^2 n-13 E_4 n}{5760 \left(1-Q_\rho^n\right)}\,, \hspace{3cm}g^{2,(n,1)}_{(4,0)}=-\frac{n(4 n^2-E_2)}{576 \left(1-Q_\rho^n\right)}\,,\nonumber\\
g^{3,(n,1)}_{(4,0)}&=-\frac{n}{8\cdot 24^2 \left(1-Q_\rho^n\right)}\,,\nonumber\\[12pt]
g^{1,(n,1)}_{(6,0)}&=\frac{n \left(1152 n^6-2016 E_2 n^4+84 \left(10 E_2^2+13 E_4\right) n^2-70 E_2^3-273 E_2 E_4-92 E_6\right)}{210\cdot 24^3
   \left(1-Q_\rho^n\right)}\,, \nonumber\\
g^{2,(n,1)}_{(6,0)}&=\frac{n \left(48 n^4-40 E_2 n^2+5 \left(E_2^2+E_4\right)\right)}{10\cdot 24^3 \left(1-Q_\rho^n\right)}\,,\hspace{3cm}
g^{3,(n,1)}_{(6,0)}=\frac{n(4 n^2-E_2)}{8\cdot 24^3 (1-Q_\rho^n)}\,,\nonumber\\[12pt]
g^{1,(n,1)}_{(1,1)}&=-\frac{n \left(4 n^2-E_2\right) }{6 \left(1-Q_\rho^n\right)}\,, \hspace{4cm}g^{2,(n,1)}_{(1,1)}=0\,,\hspace{4cm}g^{3,(n,1)}_{(1,1)}=0\,,\nonumber\\[12pt]
g^{1,(n,1)}_{(3,1)}&=\frac{n \left(-40 n^2 E_2+5 E_2^2+2E_4+48 n^4\right)}{30\cdot 24 \left(1-Q_\rho^n\right)}\,, \hspace{1cm}g^{2,(n,1)}_{(3,1)}=\frac{n(4 n^2- E_2)}{12\cdot 24 \left(1-Q_\rho^n\right)}\,,\hspace{1cm}g^{3,(n,1)}_{(3,1)}=0\,,\nonumber\\[12pt]
g^{1,(n,1)}_{(5,1)}&=\frac{n \left(2016 n^4 E_2 -84 n^2 \left(10 E_2 ^2+7 E_4 \right)+70 E_2 ^3+147
   E_2  E_4 +32 E_6 -1152 n^6\right)}{35\cdot 24^3 \left(1-Q_\rho^n\right)}\,, \nonumber\\
g^{2,(n,1)}_{(5,1)}&=\frac{n \left(40 n^2 E_2 -5 E_2 ^2-2 E_4 -48 n^4\right)}{60\cdot 24^2 \left(1-Q_\rho^n\right)}\,,\hspace{4cm}g^{3,(n,1)}_{(5,1)}=-\frac{n \left(4 n^2-E_2 \right)}{4\cdot 24^3 \left(1-Q_\rho^n\right)}\,,\nonumber\\[12pt]
g^{1,(n,1)}_{(2,2)}&=-\frac{n \left(96 n^4-80 E_2 n^2+10 E_2^2+9 E_4\right)}{40\cdot 24 \left(1-Q_\rho^n\right)}\,, \hspace{0.1cm}g^{2,(n,1)}_{(2,2)}=-\frac{n \left(4 n^2-E_2\right)}{12\cdot 24 \left(1-Q_\rho^n\right)}\,,\hspace{0.1cm}g^{3,(n,1)}_{(2,2)}=\frac{n}{4\cdot 24^2 \left(1-Q_\rho^n\right)}\,,\nonumber\\[12pt]
g^{1,(n,1)}_{(4,2)}&=-\frac{n \left(10080 n^4 E_2 -84 n^2 \left(50 E_2 ^2+33 E_4 \right)+350 E_2 ^3+693
   E_2  E_4 +300 E_6 -5760 n^6\right)}{70\cdot 24^3 \left(1-Q_\rho^n\right)}\,,\nonumber\\ 
g^{2,(n,1)}_{(4,2)}&=\frac{n \left(-280 n^2 E_2 +35 E_2 ^2+11 E_4 +336 n^4\right)}{10\cdot 24^3 \left(1-Q_\rho^n\right)}\,,\hspace{3cm}g^{3,(n,1)}_{(4,2)}=-\frac{n \left(4 n^2-E_2 \right)}{110592 \left(1-Q_\rho^n\right)}\,,\nonumber\\[12pt]
g^{1,(n,1)}_{(3,3)}&=\frac{n \left(2016 n^4 E_2 -84 n^2 \left(10 E_2 ^2+7 E_4 \right)+70 E_2 ^3+147E_2  E_4 +32 E_6 -1152 n^6\right)}{252\cdot 24^2 \left(1-Q_\rho^n\right)}\,,\nonumber\\ 
g^{2,(n,1)}_{(3,3)}&=\frac{n \left(40 n^2 E_2 -5 E_2 ^2-2 E_4 -48 n^4\right)}{30\cdot 24^2 \left(1-Q_\rho^n\right)}\,,\hspace{4cm}g^{3,(n,1)}_{(3,3)}=\frac{n (4n^2-E_2 )}{2\cdot 24^3 \left(1-Q_\rho^n\right)}\,.\label{N2ExplicitgFuncs}
\end{align}}
Notice that we also have the symmetry $H_{(s_1,s_2)}^{(n,0,1)}(\rho,S)=H_{(s_2,s_1)}^{(n,0,1)}(\rho,S)$. 
%%%%%%%%%%%%%%%%%%%%%%%%%
\subsubsection{(Generalised) Eisenstein Series}
In order to compute $B^{N=2,(r)}_{(s_1,s_2)}(\widehat{a}_1,\widehat{a}_2,S)$ (which was defined in (\ref{ResummedHfunc})) we have to sum the contributions in (\ref{N2ExplicitgFuncs}) over $n\in \mathbb{N}$, weighted by $(Q_{\widehat{a}_1})^n+\left(\tfrac{Q_\rho}{Q_{\widehat{a}_1}}\right)^n$:\footnote{We recall that $\rho=\widehat{a}_1+\widehat{a}_2$.} 
\begin{align}
B_{(s_1,s_2)}^{N=2,1}(\rho,\widehat{a}_1,S)=\sum_{n=1}^\infty H_{(s_1,s_2)}^{(n,0,1)}(\rho,S)\left(Q_{\widehat{a}_1}^n+\tfrac{Q_\rho^n}{Q_{\widehat{a}_1}^n}\right)\,.\label{ResummedN2R1}
\end{align}
Taking into account the form of the $g^{i,(n,1)}_{(s_1,s_2)}$ in (\ref{N2ExplicitgFuncs}), the relevant sum to study is of the form
\begin{align}
&\mathcal{I}_\alpha=\sum_{n=0}^\infty\frac{n^{2\alpha+1}}{1-Q_\rho^n}\,\left(Q_{\widehat{a}_1}^n+\tfrac{Q_\rho^n}{Q_{\widehat{a}_1}^n}\right)=D_{\widehat{a}_1}^{2\alpha}\sum_{n=1}^\infty\frac{n}{1-Q_\rho^n}\,\left(Q_{\widehat{a}_1}^n+\tfrac{Q_\rho^n}{Q_{\widehat{a}_1}^n}\right)=D_{\widehat{a}_1}^{2\alpha}\,\mathcal{I}_0\,,&&\text{for} &&\alpha\in \mathbb{N}\cup \{0\}\,,\label{DefIalpha}
\end{align}
where $D_{\widehat{a}_1}=\frac{1}{2\pi i}\,\tfrac{\partial}{\partial \widehat{a}_1}=Q_{\widehat{a}_1}\tfrac{\partial}{\partial Q_{\widehat{a}_1}}$. Upon writing
{\allowdisplaybreaks
\begin{align}
\sum_{n=1}^\infty\frac{n}{1-Q_\rho^n}\,Q_{\widehat{a}_1}^{n}&=\sum_{n=1}^\infty\sum_{k=0}^\infty n\,Q_\rho^{nk}\,Q_{\widehat{a}_1}^{n}=\sum_{n=1}^\infty n\,Q_{\widehat{a}_1}^{n}+\sum_{n=1}^\infty\sum_{k=1}^\infty n\,Q_\rho^{nk}\,Q_{\widehat{a}_1}^{n}\,,\nonumber\\
\sum_{n=1}^\infty\frac{nQ_\rho^n}{1-Q_\rho^n}\,Q_{\widehat{a}_1}^{-n}&=%\sum_{n=1}^\infty\sum_{k=0}^\infty n\,Q_\rho^{n(k+1)}\,Q_{\widehat{a}_1}^{-n}=
\sum_{n=1}^\infty\sum_{k=1}^\infty n\,Q_\rho^{nk}\,Q_{\widehat{a}_1}^{-n}\,,\label{ManipulationDivisor}
\end{align}}
the generating function $\mathcal{I}_0$ is
\begin{align}
\mathcal{I}_0=D_{\widehat{a}_1}\frac{Q_{\widehat{a}_1}}{1-Q_{\widehat{a}_1}}+\sum_{n=1}^\infty\sum_{k=1}^\infty n\,Q_\rho^{nk}\left(Q_{\widehat{a}_1}^n+Q_{\widehat{a}_1}^{-n}\right)\,.\label{ResumGenEisen}
\end{align}
The first term can be rewritten as follows
\begin{align}
D_{\widehat{a}_1}\frac{Q_{\widehat{a}_1}}{1-Q_{\widehat{a}_1}}=\frac{Q_{\widehat{a}_1}}{(1-Q_{\widehat{a}_1})^2}=\frac{1}{\left(e^{\pi i \widehat{a}_1}-e^{-i\pi \widehat{a}_1}\right)^2}=-\frac{1}{4\sin^2\pi \widehat{a}_1}=-\frac{1}{4\pi^2}\,e_2(\widehat{a}_1)\,,
\end{align}
where we used the notation of \cite{Weil}, reviewed in appendix~\ref{App:WeilEisen}, specifically eq.~(\ref{DefTrigWeil}). We therefore find with eq.~(\ref{SumGenEisenFourier}) for $|Q_\rho|<|Q_{\widehat{a}_1}|<|Q_\rho|^{-1}$
\begin{align}
\mathcal{I}_0=-\frac{1}{4\pi^2}\left[e_2(\widehat{a}_1)-4\pi^2\sum_{n=1}^\infty\sum_{k=1}^\infty k\,Q_{\rho}^{nk}\left(Q_{\widehat{a}_1}^k+Q_{\widehat{a}_1}^{-k}\right)\right]=\frac{1}{(2\pi i)^2}\,\mathcal{E}_2(\widehat{a}_1;\rho)\,.
\end{align}
Here $\mathcal{E}_2$ is a generalised Eisenstein series, as defined in (\ref{GenEisenstein}). Following \cite{Weil}, the latter can also be expressed in terms of the Weierstrass function $\wp$ (see (\ref{GenEisenWeierstrass})), such that
\begin{align}
\mathcal{I}_0=\frac{1}{(2\pi i )^2}\left[G_2(\rho)+\wp(\widehat{a}_1;\rho)\right]\,,\label{DefI0Weierstrass}
\end{align}
where $G_2(\rho)$ is the Eisenstein series defined in (\ref{NormEisenstein}). Furthermore, using the recursive relation (\ref{RecursiveGenEisenstein}), we have
\begin{align}
&\mathcal{I}_\alpha=\frac{(2\alpha+1)!}{(2\pi i)^{2\alpha+2}}\,\,\mathcal{E}_{2\alpha+2}(\widehat{a}_1;\rho)\,,&&\forall \alpha\in\mathbb{N}\cup\{0\}\,.\label{RelIAlphaEisen}
\end{align}
Using this result, we can write for example
\begin{align}
B_{(0,0)}^{N=2,1}(\rho,\widehat{a}_1,S)&=-\frac{2\,\mathcal{E}_2(\widehat{a}_1;\rho)}{(2\pi i)^2}\,\phi_{-2,1}^2(\rho,S)\,,
\end{align}
and similarly for the remaining examples (for simplicity, we do not display explicitly all arguments)
{\allowdisplaybreaks
\begin{align}
B_{(2,0)}^{2,1}&=-\frac{6\,\mathcal{E}_4+\pi^2 E_2\,\mathcal{E}_2}{12\pi^2(2\pi i)^2}\,\phi_{-2,1}^2+\frac{\mathcal{E}_2}{24(2\pi i)^2}\,\phi_{-2,1}\,\phi_{0,1}\,,\nonumber\\[12pt]
B_{(4,0)}^{2,1}&=-\frac{\left(720\, \mathcal{E}_6+120 \pi ^2\, E_2\, \mathcal{E}_4+\pi ^4 \left(10\, E_2^2+13 \,E_4\right) \mathcal{E}_2\right) \phi_{-2,1}^2}{10\cdot 24^2 \pi ^4 (2\pi i)^2 }+\frac{\left(6\, \mathcal{E}_4+\pi ^2\, E_2\, \mathcal{E}_2\right) \phi _{0,1} \phi _{-2,1}}{24^2 \pi^2 (2\pi i)^2}\nonumber\\
   &\hspace{0.5cm}-\frac{\mathcal{E}_2\, \phi _{0,1}^2}{8\cdot 24^2 (2\pi i)^2}\,,\nonumber\\[12pt]
B_{(6,0)}^{2,1}&=\frac{\left(90720 \mathcal{E}_8+15120 \pi ^2 E_2 \mathcal{E}_6+126 \pi ^4 \left(10 E_2^2+13 E_4\right) \mathcal{E}_4+\pi ^6 \left(70E_2^3+273 E_4 E_2+92 E_6\right) \mathcal{E}_2\right) \phi _{-2,1}^2}{1260\cdot 24^2 \pi ^4 (2\pi i)^4}\nonumber\\
&\hspace{0.5cm}-\frac{\left(72 \mathcal{E}_6+12 \pi ^2 E_2 \mathcal{E}_4+\pi ^4 \left(E_2^2+E_4\right) \mathcal{E}_2\right) \phi _{0,1} \phi _{-2,1}}{12\cdot 24^2 \pi ^2 (2\pi i)^4}+\frac{\left(6\, \mathcal{E}_4+\pi ^2\, E_2\, \mathcal{E}_2\right) \phi _{0,1}^2}{2\cdot 24^3 (2\pi i)^4}\,,\nonumber\\[12pt]
B_{(1,1)}^{2,1}&=-\frac{2\left(6\,\mathcal{E}_4+\pi^2 E_2\,\mathcal{E}_2\right)}{3(2\pi i )^4}\, \phi _{-2,1}^2\,,\nonumber\\[10pt]
B_{(3,1)}^{2,1}&=\frac{\left(\pi ^2 \mathcal{E}_2 E_2+6 \mathcal{E}_4\right)\phi _{-2,1} \phi _{0,1}}{3\cdot 24 (2\pi i )^4}-\frac{\phi _{-2,1}^2 \left(60
   \pi ^2 \mathcal{E}_4 E_2+\pi ^4 \mathcal{E}_2 \left(5 E_2^2+2 E_4\right)+360
   \mathcal{E}_6\right)}{180 \pi ^2 (2\pi i)^4}\,,\nonumber\\
B_{(5,1)}^{2,1}&=\frac{ \left(15120 \pi ^2 \mathcal{E}_6 E_2+126 \pi ^4 \mathcal{E}_4 \left(10 E_2^2+7
   E_4\right)+\pi ^6 \mathcal{E}_2 \left(70 E_2^3+147 E_4 E_2+32
   E_6\right)+90720 \mathcal{E}_8\right) \phi _{-2,1}^2}{-210 \cdot 24^2 \pi ^4 (2\pi i)^4}\nonumber\\
&\hspace{0.5cm}+\frac{\left(60 \pi ^2 \mathcal{E}_4
   E_2+\pi ^4 \mathcal{E}_2 \left(5 E_2^2+2 E_4\right)+360 \mathcal{E}_6\right)\phi _{0,1} \phi _{-2,1} }{15\cdot 24^2 \pi
   ^2(2\pi i)^4}+\frac{ \left(-\pi ^2 \mathcal{E}_2 E_2-6 \mathcal{E}_4\right)\phi _{0,1}^2}{24^3(2\pi i)^4}\,,\nonumber\\[10pt]
B_{(2,2)}^{2,1}&= \frac{\left(720 \mathcal{E}_6+120 \pi ^2 E_2 \mathcal{E}_4+\pi ^4 \left(10 E_2^2+9 E_4\right) \mathcal{E}_2\right) \phi _{-2,1}^2}{-40\cdot 24
   \pi ^4 (2\pi i)^2}+\frac{\left(6 \mathcal{E}_4+\pi ^2 E_2 \mathcal{E}_2\right) \phi _{0,1} \phi _{-2,1}}{12\cdot 24 \pi ^2 (2\pi i)^2}+\frac{\mathcal{E}_2 \phi
   _{0,1}^2}{4\cdot 24^2 (2\pi i)^2}\,,\nonumber\\[10pt]
B_{(4,2)}^{2,1}&=\frac{ \left(75600 \pi ^2 \mathcal{E}_6 E_2+126 \pi ^4 \mathcal{E}_4 \left(50 E_2^2+33
   E_4\right)+\pi ^6 \mathcal{E}_2 \left(350 E_2^3+693 E_4 E_2+300
   E_6\right)+453600 \mathcal{E}_8\right)\phi _{-2,1}^2}{420\cdot 24^2 \pi ^4(2\pi i)^4}\nonumber\\
   &\hspace{0.5cm}-\frac{\left(420 \pi ^2 \mathcal{E}_4
   E_2+\pi ^4 \mathcal{E}_2 \left(35 E_2^2+11 E_4\right)+2520 \mathcal{E}_6\right)\phi _{0,1} \phi _{-2,1} }{60\cdot 24^2 \pi
   ^2(2\pi i)^4}-\frac{ \left(6\,\mathcal{E}_4+\pi^2 E_2\,\mathcal{E}_2\right)\phi _{0,1}^2}{2\cdot 24^3(2\pi i)^4}\,,\nonumber\\
B_{(3,3)}^{2,1}&=-\frac{ \left(15120 \pi ^2 \mathcal{E}_6 E_2+126 \pi ^4 \mathcal{E}_4 \left(10 E_2^2+7
   E_4\right)+\pi ^6 \mathcal{E}_2 \left(70 E_2^3+147 E_4 E_2+32
   E_6\right)+90720 \mathcal{E}_8\right)\phi _{-2,1}^2}{63\cdot 24^2 \pi ^4(2\pi i)^4}\nonumber\\
&\hspace{0.5cm}+\frac{\phi _{0,1} \phi _{-2,1} \left(60 \pi ^2 \mathcal{E}_4
   E_2+\pi ^4 \mathcal{E}_2 \left(5 E_2^2+2 E_4\right)+360 \mathcal{E}_6\right)}{180\cdot 24 \pi
   ^2(2\pi i)^4}+\frac{ \left(6\,\mathcal{E}_4+\pi^2 E_2\,\mathcal{E}_2\right)\phi _{0,1}^2}{12\cdot 24^2(2\pi i)^4}\,.\label{EisensteinExpressionsN2}
\end{align}}

%%%%%%%%%%%%%%%%%%%%%%%%%
\subsubsection{Root Lattices}\label{Sect:N2RootLattice}
Finally, another way of writing $\mathcal{I}_\alpha$ (which will generalise to later cases) is to exchanged the order of the two summations in the last term of (\ref{ResumGenEisen}) 
\begin{align}
\mathcal{I}_\alpha=D_{\widehat{a}_1}^{2\alpha}\frac{Q_{\widehat{a}_1}}{(1-Q_{\widehat{a}_1})^2}+\sum_{n=1}^\infty Q_\rho^n\sum_{k|n}k^{2\alpha+1}\left(Q_{\widehat{a}_1}^k+Q_{\widehat{a}_1}^{-k}\right)\,.\label{DivisorFirst}
\end{align}
While the first term has a second order pole at $\widehat{a}_1=0$, the limit of the second term in fact yields the holomorphic Eisenstein series $E_2(\rho)$
\begin{align}
\lim_{\widehat{a}_1\to 0}\left[\mathcal{I}_\alpha-D_{\widehat{a}_1}^{2\alpha}\frac{Q_{\widehat{a}_1}}{(1-Q_{\widehat{a}_1})^2}\right]=2\sum_{n=1}^\infty Q_\rho^n\sum_{k|n}k^{2\alpha+1}=2\sum_{n=1}^\infty \sigma_{2\alpha+1}(n)\,Q_\rho^n=\frac{B_{2\alpha+2}}{2\alpha+2}\left[1-E_{2\alpha+2}(\rho)\right]\,,\nonumber
\end{align}
where $B_{2\alpha}=(-1)^\alpha\,\frac{2\zeta(2\alpha+2)\,(2\alpha+1)!}{(2\pi)^{2\alpha+2}}$ (for $\alpha\in\mathbb{N}\cup\{0\}$) are the Bernoulli numbers.

The form (\ref{DivisorFirst}) of the generating function $\mathcal{I}_\alpha$ (which is the building block for the free energy) can be written in another suggestive fashion:
\begin{align}
\mathcal{I}_\alpha=D_{\widehat{a}_1}^{2\alpha+1}\sum_{n=1}^\infty \sum_{\ell\in\Delta_{\mathfrak{a}_1}^+}e^{2\pi i n \ell}+\sum_{n=1}^\infty Q_\rho^n\sum_{k|n}k^{2\alpha+1}\sum_{\ell\in \Delta_{\mathfrak{a}_1}}\,e^{2\pi i \ell k}\,,\label{DivisorRootSum}
\end{align}
where $\Delta_{\mathfrak{a}}$ is the set of all roots of the Lie algebra $\mathfrak{a}_1$, \emph{i.e.} the set $\{-\widehat{a}_1,\widehat{a}_1\}$ and $\Delta_{\mathfrak{a}}^+$ the set of positive roots of $\mathfrak{a}_1$ (\emph{i.e.} the set $\{\widehat{a}_1\}$. While a seemingly trivial rewriting of (\ref{DivisorFirst}), the presentation (\ref{DivisorRootSum}) is very similar to the Fourier expansion of the Eisenstein series in (\ref{NormEisenstein}), except that the divisor sigma $\sigma_{2k-1}(n)$ is replaced by a summation over the root lattice of $\mathfrak{a}_1$. Notice in this regards that also 
\begin{align}
\sum_{n=1}^\infty \sum_{\ell\in\Delta_{\mathfrak{a}_1}^+}e^{2\pi i n \ell}=\frac{1}{1-Q_{\widehat{a}_1}}\,,
\end{align} 
can in a sense be thought of as a generating function of Riemann zeta functions as can be seen from (\ref{GeneratingFunctionaleta}). From this perspective, it is clear that the generating functions $\mathcal{I}_\alpha$ encode how the modular parameter $\rho$ is coupled together with the roots of the $\mathfrak{su}(2)$ gauge algebra in the free energy, to make both the modular and gauge structure manifest. As we shall see below, similar patterns appear also for $N>2$. Furthermore, in the following subsubsection, we shall give another perspective on the $\mathcal{I}_\alpha$, which also focuses on generating functions of divisor sigmas, as advocated in \cite{Bachmann:2013wba}.

%%%%%%%%%%%%%%%%%%%%%%%%%
\subsubsection{Generating Functions of Multiple Zeta Values}\label{Sect:N2MutliZeta}
In view of generalising to the cases $N>2$, we provide another way of organising the $B_{(s_1,s_2)}^{N=2,1}$, which formalises (\ref{DivisorFirst}) and introduces generating functions of multiple-divisor sums and multiple zeta values, as reviewed in appendix~\ref{App:MultiDivisorSums}. Indeed, using the generating function $T(\widehat{a}_1,\ldots,\widehat{a}_\ell;\rho)$ (first defined in \cite{Bachmann:2013wba} and reviewed in (\ref{GenFuncBracket})) of brackets of length $\ell=1$, we can write for $\mathcal{I}_0$ in  (\ref{DefIalpha})
\begin{align}
\mathcal{I}_0=D_{\widehat{a}_1}\, \left[T(\widehat{a}_1-\rho;\rho)-T(-\widehat{a}_1;\rho)\right]=D_{\widehat{a}_1}\, \wprim (\widehat{a}_1;\rho)\,,\label{IntroTcal}
\end{align}
where for later convenience, we have introduced the combination
\begin{align}
\wprim (\widehat{a}_1;\rho):=T(\widehat{a}_1-\rho;\rho)-T(-\widehat{a}_1;\rho)\,, \label{Defwprim}
\end{align}
such that \emph{e.g.}
\begin{align}
B_{(0,0)}^{N=2,1}(\rho,\widehat{a}_1,S)&=-2\,\wprim(\widehat{a}_1;\rho)\,\phi_{-2,1}^2(\rho,S)\,,\label{ExamplereN2T}
\end{align}
and similarly for all other terms in (\ref{SummedCoefficientsN2R2}). Following the discussion of (\ref{DefI0Weierstrass}) and the form (\ref{IntroTcal}), $\wprim$ can be understood as a primitive of the Weierstrass elliptic function.

\begin{figure}[h]
\begin{center}
\scalebox{0.95}{\parbox{15.8cm}{\begin{tikzpicture}[scale = 1]
\draw[->] (-3.4,0) -- (3,0);
\node at (3.25,0) {$u$};
\draw[->] (0,-3) -- (0,3);
\node at (0,3.25) {$v$};
\draw[thick,red,->] (0,0) -- (0.725,0.525);
\node at (0.8,0.6) {$\bullet$};
\node[red] at (0.35,0.65) {$\widehat{a}_1$};
\node at (-3.2,-2.4) {$\bullet$};
\node at (-2.2,-2.4) {$\bullet$};
\node at (-1.2,-2.4) {$\bullet$};
\node at (-0.2,-2.4) {$\bullet$};
\node at (0.8,-2.4) {$\bullet$};
\node at (1.8,-2.4) {$\bullet$};
\node at (2.8,-2.4) {$\bullet$};
\node at (-3.2,-1.4) {$\bullet$};
\node at (-2.2,-1.4) {$\bullet$};
\node at (-1.2,-1.4) {$\bullet$};
\node at (-0.2,-1.4) {$\bullet$};
\node at (0.8,-1.4) {$\bullet$};
\node at (1.8,-1.4) {$\bullet$};
\node at (2.8,-1.4) {$\bullet$};
\node at (-3.2,-0.4) {$\bullet$};
\node at (-2.2,-0.4) {$\bullet$};
\node at (-1.2,-0.4) {$\bullet$};
\node at (-0.2,-0.4) {$\bullet$};
\node at (0.8,-0.4) {$\bullet$};
\node at (1.8,-0.4) {$\bullet$};
\node at (2.8,-0.4) {$\bullet$};
\node at (-3.2,0.6) {$\bullet$};
\node at (-2.2,0.6) {$\bullet$};
\node at (-1.2,0.6) {$\bullet$};
\node at (-0.2,0.6) {$\bullet$};
\node at (1.8,0.6) {$\bullet$};
\node at (2.8,0.6) {$\bullet$};
\node at (-3.2,1.6) {$\bullet$};
\node at (-2.2,1.6) {$\bullet$};
\node at (-1.2,1.6) {$\bullet$};
\node at (-0.2,1.6) {$\bullet$};
\node at (0.8,1.6) {$\bullet$};
\node at (1.8,1.6) {$\bullet$};
\node at (2.8,1.6) {$\bullet$};
\node at (-3.2,2.6) {$\bullet$};
\node at (-2.2,2.6) {$\bullet$};
\node at (-1.2,2.6) {$\bullet$};
\node at (-0.2,2.6) {$\bullet$};
\node at (0.8,2.6) {$\bullet$};
\node at (1.8,2.6) {$\bullet$};
\node at (2.8,2.6) {$\bullet$};
\node at (0,-3.75) {{\bf (a)}};
%%%%%%%%%%%%%%%%
\begin{scope}[xshift=9cm]
\draw[->] (-3.4,0) -- (3,0);
\node at (3.25,0) {$u$};
\draw[->] (0,-3) -- (0,3);
\node at (0,3.25) {$v$};
\draw[thick,red,->] (0,0) -- (0.725,0.525);
\node[green!75!black] at (0.8,0.6) {$\bullet$};
\node[red] at (0.35,0.65) {$\widehat{a}_1$};
\node[orange!75!black] at (-3.2,-2.4) {$\bullet$};
\node[orange!75!black] at (-2.2,-2.4) {$\bullet$};
\node[orange!75!black] at (-1.2,-2.4) {$\bullet$};
\node[orange!75!black] at (-0.2,-2.4) {$\bullet$};
\node[orange!75!black] at (0.8,-2.4) {$\bullet$};
\node[orange!75!black] at (1.8,-2.4) {$\bullet$};
\node[orange!75!black] at (2.8,-2.4) {$\bullet$};
\node[orange!75!black] at (-3.2,-1.4) {$\bullet$};
\node[orange!75!black] at (-2.2,-1.4) {$\bullet$};
\node[orange!75!black] at (-1.2,-1.4) {$\bullet$};
\node[orange!75!black] at (-0.2,-1.4) {$\bullet$};
\node[orange!75!black] at (0.8,-1.4) {$\bullet$};
\node[orange!75!black] at (1.8,-1.4) {$\bullet$};
\node[orange!75!black] at (2.8,-1.4) {$\bullet$};
\node[orange!75!black] at (-3.2,-0.4) {$\bullet$};
\node[orange!75!black] at (-2.2,-0.4) {$\bullet$};
\node[orange!75!black] at (-1.2,-0.4) {$\bullet$};
\node[orange!75!black] at (-0.2,-0.4) {$\bullet$};
\node[orange!75!black] at (0.8,-0.4) {$\bullet$};
\node[orange!75!black] at (1.8,-0.4) {$\bullet$};
\node[orange!75!black] at (2.8,-0.4) {$\bullet$};
\node[green!75!black] at (-3.2,0.6) {$\bullet$};
\node[green!75!black] at (-2.2,0.6) {$\bullet$};
\node[green!75!black] at (-1.2,0.6) {$\bullet$};
\node[green!75!black] at (-0.2,0.6) {$\bullet$};
\node[green!75!black] at (1.8,0.6) {$\bullet$};
\node[green!75!black] at (2.8,0.6) {$\bullet$};
\node[blue!75!black] at (-3.2,1.6) {$\bullet$};
\node[blue!75!black] at (-2.2,1.6) {$\bullet$};
\node[blue!75!black] at (-1.2,1.6) {$\bullet$};
\node[blue!75!black] at (-0.2,1.6) {$\bullet$};
\node[blue!75!black] at (0.8,1.6) {$\bullet$};
\node[blue!75!black] at (1.8,1.6) {$\bullet$};
\node[blue!75!black] at (2.8,1.6) {$\bullet$};
\node[blue!75!black] at (-3.2,2.6) {$\bullet$};
\node[blue!75!black] at (-2.2,2.6) {$\bullet$};
\node[blue!75!black] at (-1.2,2.6) {$\bullet$};
\node[blue!75!black] at (-0.2,2.6) {$\bullet$};
\node[blue!75!black] at (0.8,2.6) {$\bullet$};
\node[blue!75!black] at (1.8,2.6) {$\bullet$};
\node[blue!75!black] at (2.8,2.6) {$\bullet$};
\node at (0,-3.75) {{\bf (b)}};
\end{scope}
\end{tikzpicture}}}
\end{center} 
\caption{{\it (a) Summation over the whole lattice $W$, shifted by $\widehat{a}_1$ as in (\ref{EisensteinLatticeInit}). (b) Split of the summation in three parts according to the first line of (\ref{E2IntoT}): the green points correspond to the summation over $n\in\mathbb{Z}$ in the first term, while the blue and orange points are summed over in the second and third term respectively.}}
\label{fig:SeparationLatticeEisenstein}
\end{figure}
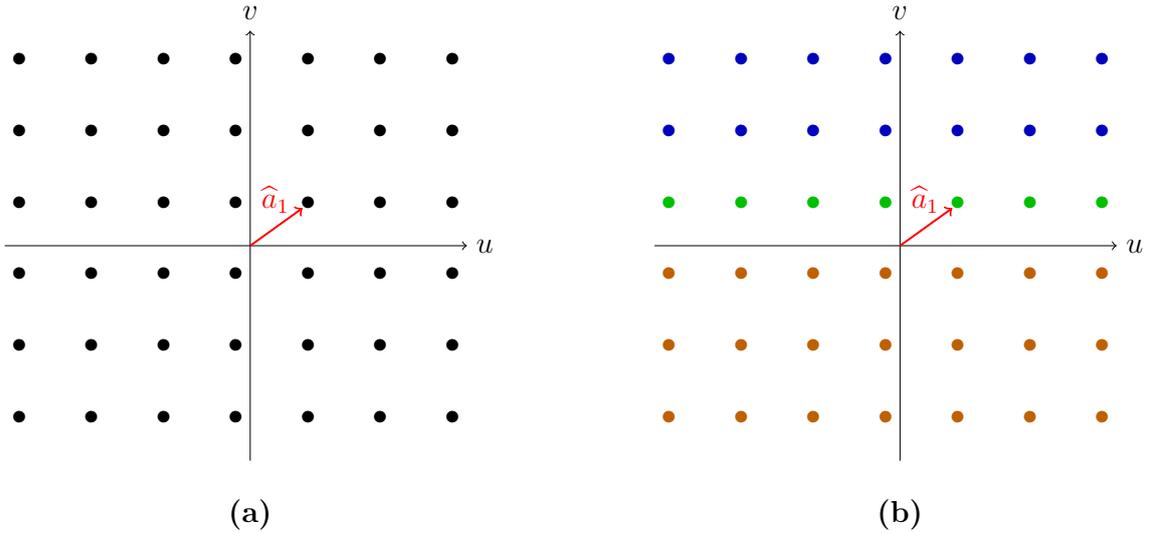

While (\ref{IntroTcal}) is evident by comparing the form of (\ref{N2ExplicitgFuncs}) to the definition (\ref{GenFuncBracket}), the relation between $T(\widehat{a}_1;\rho)$ and the generalised Eisenstein series discussed previously can also be understood in a different fashion. To this end, we consider
\begin{align}
\mathcal{E}_2(\widehat{a}_1;\rho)=\sum_{w\in W}\frac{1}{(\widehat{a}_1+w)^2}\,,\label{EisensteinLatticeInit}
\end{align}
where $W$ is a two-dimensional lattice with generators $(u,v)=(1,\rho)$, as in appendix~\ref{App:WeilEisen}. Using the notation of \cite{BachmannMaster} we write
\begin{align}
\mathcal{E}_2(\widehat{a}_1;\rho)&=\sum_{n\in\mathbb{Z}}\frac{1}{(\widehat{a}_1+n)^2}+\sum_{m=1}^\infty \left(\sum_{n\in\mathbb{Z}}\frac{1}{(\widehat{a}_1+m\tau +n)^2}\right)+\sum_{m=-\infty}^{-1} \left(\sum_{n\in\mathbb{Z}}\frac{1}{(\widehat{a}_1+m\tau+ n)^2}\right)\nonumber\\
&=e_2(\widehat{a}_1)+\sum_{m=1}^\infty \left(\sum_{n\in\mathbb{Z}}\frac{1}{(\widehat{a}_1+m\tau +n)^2}\right)+\sum_{m=1}^\infty \left(\sum_{n\in\mathbb{Z}}\frac{1}{(-\widehat{a}_1+m\tau+ n)^2}\right)\,.\label{E2IntoT}
\end{align}
Using the Lipschitz summation formula
\begin{align}
&\sum_{d\in\mathbb{Z}}\frac{1}{(\rho+d)^k}=\frac{(-2\pi i)^k}{(k-1)!}\sum_{m=1}^\infty m^{k-1}\,Q_\rho^m\,,&& \forall k\in\mathbb{N}\,,
\end{align}
we can also write
\begin{align}
\sum_{m=1}^\infty \left(\sum_{n\in\mathbb{Z}}\frac{1}{(\widehat{a}_1+m\tau +n)^2}\right)=\sum_{m=1}^\infty(-2\pi i)^2 \sum_{k=1}^\infty k\,e^{2\pi i \widehat{a}_1 k}\,Q_\rho^{km}=D_{\widehat{a}_1}\sum_{m=1}^\infty(-2\pi i)^2\sum_{k=1}^\infty Q_{\widehat{a}_1}^k Q_\rho^{km}\,.\nonumber
\end{align}
Here we have differentiated under the summation in the last expression (assuming convergence of the sum\footnote{As remarked in appendix~\ref{App:WeilEisen}, we implicitly use the Eisenstein summation prescription (see eq.~(\ref{DefEisensteinPrescription})).}). Using the presentation (\ref{TEquivalent}) we have
\begin{align}
\sum_{m=1}^\infty \left(\sum_{n\in\mathbb{Z}}\frac{1}{(\widehat{a}_1+m\tau +n)^2}\right)=-4\pi^2D_{\widehat{a}_1}\sum_{n>0}Q_{\widehat{a}_1}^n\sum_{k>0}Q_\rho^{nk}=-4\pi^2 D_{\widehat{a}_1}\,T(\widehat{a}_1;\rho)\,.
\end{align}
Therefore (\ref{E2IntoT}) can be written as
\begin{align}
\mathcal{E}_2(\widehat{a}_1;\rho)=-4\pi^2D_{\widehat{a}_1}\left[T(\widehat{a}_1;\rho)-T(-\widehat{a}_1;\rho)\right]-\frac{d}{d\widehat{a}_1}\,e_1(\widehat{a}_1)\,,
\end{align}
and with (\ref{RecursionSmalle}) we have
\begin{align}
\mathcal{E}_2(\widehat{a}_1;\rho)=-4\pi^2D_{\widehat{a}_1}\left[T(\widehat{a}_1;\rho)-T(-\widehat{a}_1;\rho)+\frac{2\pi i}{4\pi^2}\,e_1(x)+c\right]\,,
\end{align}
for $c\in\mathbb{C}$ a constant. Finally, with (\ref{e1Def}) as well as
\begin{align}
\frac{2\pi i}{4\pi^2}\,\frac{\pi\cos\pi \,\widehat{a}_1}{\sin\pi\, \widehat{a}_1}=\frac{1}{2}-\frac{1}{1-e^{-2\pi i \widehat{a}_1}}\,,
\end{align}
the behaviour of $T(\widehat{a}_1;\rho)$ under elliptic transformations (\ref{EllipticTrafoAllArgs}) finally implies
\begin{align}
\mathcal{E}_2(\widehat{a}_1;\rho)=-4\pi^2D_{\widehat{a}_1}\left[T(\widehat{a}_1-\rho;\rho)-T(-\widehat{a}_1;\rho)\right]\,,
\end{align}
which indeed yields (\ref{IntroTcal}). The crucial step in the above computation is (\ref{E2IntoT}): intuitively, (\ref{EisensteinLatticeInit}) corresponds to a summation over the whole lattice $W$, shifted by $x$ (notice that $x\notin W$ in general) as shown in \figref{fig:SeparationLatticeEisenstein}(a). The expression (\ref{E2IntoT}) corresponds to a separation of this summation into three different contributions, as in \figref{fig:SeparationLatticeEisenstein}(b). 
The summation over the blue and green points ultimately combine into $D_{\widehat{a}_1}T(\widehat{a}_1-\rho;\rho)$, while the summation over the orange lattice points translates into $D_{\widehat{a}_1}T(-\widehat{a}_1;\rho)$. In the following, we will see that the separation into various different sublattices according to (\ref{ExamplereN2T}) continues to higher orders in $Q_R$, as well as higher values of $N$, however, with increased complexity.
%%%%%%%%%%%%%%%%%%%%%%%%
\subsubsection{Free Energy and Self-Similarity}
So far we have only considered the contributions $n>0$ in (\ref{ResummedN2R1}). Indeed, the sector $n=0$ behaves rather differently (for one, it transforms under a modular symmetry acting on $\rho$) and was discussed in detail in \cite{Companion1}. It is, however, an interesting question to combine it with the present results (\ref{ResummedN2R1}). 

To leading order in $\epsilon_{1,2}$ the contribution $H_{(0,0)}^{(0,0,1)}(\rho,S)$ is given in (\ref{OrbifoldSectorContributions}) such that we have
\begin{align}
H_{(0,0)}^{(0,0,1)}(\rho,S)&+B_{(0,0)}^{N=2,1}(\rho,\widehat{a}_1,S)\nonumber\\
&=-2\phi_{-2,1}^2(\rho,S)\,D_{\widehat{a}_1}\, \wprim (\widehat{a}_1;\rho)-\frac{1}{12}\phi_{-2,1}(\rho,S)\left[\phi_{0,1}(\rho,S)+2E_2(\rho)\,\phi_{-2,1}(\rho,S)\right]\nonumber\\
%&=\frac{2}{(2\pi i )^2}\,\phi_{-2,1}^2(\rho,S)\,\wp(\widehat{a}_1)+\frac{1}{12}\,\phi_{-2,1}(\rho,S)\phi_{0,1}(\rho,S)\,,\nonumber\\
&=-\frac{1}{12}\,\phi_{-2,1}(\rho,S)\left[\phi_{0,1}(\rho,S)-\frac{6}{\pi^2}\,\phi_{-2,1}(\rho,S)\,\wp(\widehat{a}_1)\right]\,,
\end{align}
which does not depend on $E_2(\rho)$. This expression is therefore invariant under modular transformations (\ref{SL2Raction}), while at the same time holomorphic. 

Similar cancellations also take place for higher orders in $\epsilon_{1,2}$, however, in general only the highest power in $E_2(\rho)$ cancels exactly. Various contributions $H_{(s_1,s_2)}^{(0,0,1)}(\rho,S)$ for low values of $s_{1,2}$ are given in appendix~\ref{App:OrbifoldSector}. Indeed, we can therefore extract the following terms containing the highest power $\ell$ in $E_2(\rho)$
\begin{center}
\begin{tabular}{c|c|l|l}
$(s_1,s_2)$ & $\ell$ & $H_{(s_1,s_2)}^{(0,0,1)}(\rho,S)$ & $B_{(s_1,s_2)}^{N=2,1}(\rho,\widehat{a}_1,S)$\\[4pt]\hline\hline
&&&\\[-16pt]
$(0,0)$ & $ 1 $ & $-\frac{1}{6}\, E_2\, \phi _{-2,1}^2+\ldots$ & $\frac{1}{6}\, E_2\, \phi _{-2,1}^2+\ldots$\\[4pt]\hline
&&&\\[-16pt]
$(2,0)$ & $ 2 $ & $-\frac{1}{144}\, E_2^2\, \phi _{-2,1}^2+\ldots$ & $\frac{1}{144}\, E_2^2\, \phi _{-2,1}^2+\ldots$\\[4pt]\hline
&&&\\[-16pt]
$(4,0)$ & $ 3 $ & $-\frac{1}{12\cdot 24^2}\,E_2^3\, \phi _{-2,1}^2+\ldots$ & $\frac{1}{12\cdot 24^2}\,E_2^3\, \phi _{-2,1}^2+\ldots$\\[4pt]\hline
&&&\\[-16pt]
$(6,0)$ & $ 4 $ & $-\frac{1}{36\cdot 24^3}\,E_2^4\, \phi _{-2,1}^2+\ldots$ & $\frac{1}{36\cdot 24^3}\,E_2^4\, \phi _{-2,1}^2+\ldots$\\[4pt]\hline
&&&\\[-16pt]
$(1,1)$ & $ 2 $ & $\frac{1}{72}\, E_2^2\, \phi _{-2,1}^2+\ldots$ & $-\frac{1}{72}\, E_2^2\, \phi _{-2,1}^2+\ldots$\\[4pt]\hline
&&&\\[-16pt]
$(3,1)$ & $ 3 $ & $\frac{1}{3\cdot 24^2}\,E_2^3 \phi _{-2,1}^2+\ldots$ & $-\frac{1}{3\cdot 24^2}\,E_2^3 \phi _{-2,1}^2+\ldots$\\[4pt]\hline
&&&\\[-16pt]
$(5,1)$ & $ 4 $ & $\frac{1}{6\cdot 24^3}\,E_2^4 \phi _{-2,1}^2+\ldots$ & $-\frac{1}{6\cdot 24^3}\,E_2^4 \phi _{-2,1}^2+\ldots$\\[4pt]\hline
&&&\\[-16pt]
$(2,2)$ & $ 3 $ & $-\frac{1}{2\cdot 24^2}\,E_2^3 \phi _{-2,1}^2+\ldots$ & $\frac{1}{2\cdot 24^2}\,E_2^3 \phi _{-2,1}^2+\ldots$\\[4pt]\hline
&&&\\[-16pt]
$(4,2)$ & $ 4 $ & $-\frac{5 }{12\cdot 24^3}\,E_2^4 \phi _{-2,1}^2+\ldots$ & $\frac{5 }{12\cdot 24^3}\,E_2^4 \phi _{-2,1}^2+\ldots$\\[4pt]\hline
&&&\\[-16pt]
$(3,3)$ & $ 4 $ & $\frac{5 }{9\cdot 24^3}\,E_2^4 \phi _{-2,1}^2+\ldots$ & $-\frac{5 }{9\cdot 24^3}\,E_2^4 \phi _{-2,1}^2+\ldots$\\[4pt]
\end{tabular}
\end{center}
Here the dots indicate terms containing powers of $E_2(\rho)$ that are lower than $\ell$ and which generically do not cancel between $H_{(s_1,s_2)}^{(0,0,1)}(\rho,S)$ and $B_{(s_1,s_2)}^{N=2,1}(\rho,\widehat{a}_1,S)$.

With the combination of $H_{(s_1,s_2)}^{(0,0,1)}(\rho,S)$ and $B_{(s_1,s_2)}^{N=2,1}(\rho,\widehat{a}_1,S)$ we can in fact perform a check to confirm (\ref{EisensteinExpressionsN2}) for $s_2=0$. In \cite{Hohenegger:2016eqy} it was observed that in a particular region in the moduli space and in the Nekrasov-Shatashvili limit (\emph{i.e.} in the limit of vanishing $\epsilon_2$), the free energy enjoys a so-called \emph{self-similarity}. In our notation, the latter amounts to the relation
\begin{align}
H_{(s_1,0)}^{(0,0,1)}(\rho,S)&+B_{(s_1,0)}^{2,1}(\rho,\tfrac{\rho}{2},S)=2H_{(s_1,0)}^{(0,1)}(\tfrac{\rho}{2},S)\,.\label{SelfSimilarityGeneral}
\end{align}
The coefficients $H_{(s_1,0)}^{(0,1)}(\rho,S)$ are given in appendix~\ref{App:CoefsHN1} and the particular region in the moduli space corresponds to $\widehat{a}_1=\widehat{a}_2=\frac{\rho}{2}$. To compute the left hand side of (\ref{SelfSimilarityGeneral}), we recall the definition of $\mathcal{I}_\alpha$ in (\ref{DefIalpha}), which for $\widehat{a}_1=\frac{\rho}{2}$ takes the following form
\begin{align}
I_\alpha\big|_{\widehat{a}_1=\frac{\rho}{2}}=2\sum_{n=1}^\infty\frac{n^{2\alpha+1}Q^n}{1-Q^{2n}}\,,
\end{align}
where we have introduced $Q^2=Q_\rho$ and also define $\rho=2\rho'$ (such that $Q=e^{2\pi i \rho'}$). Using the relation (for $z\in\mathbb{C}$)
\begin{align}
\frac{\theta'_4(z;\rho)}{\theta_4(z;\rho)}=4\pi \sum_{n=1}^\infty\frac{Q^n\,\sin(2\pi nz) }{1-Q^{2n}}\,,
\end{align}
where $\theta_4$ is a Jacobi theta function, we can write
\begin{align}
I_\alpha\big|_{\widehat{a}_1=\frac{\rho}{2}}=-\frac{1}{(2\pi i)^{2\alpha+2}}\,\frac{d^{2\alpha+1}}{dz^{2\alpha+1}}\left(\frac{\theta'_4(z;\rho)}{\theta_4(z;\rho)}\right)\bigg|_{z=0}\,.\label{IalphaThetaFct}
\end{align}
With this, we indeed find
\begin{align}
H_{(0,0)}^{(0,0,1)}(\rho,S)+B_{(0,0)}^{2,1}(\rho,\tfrac{\rho}{2},S)&=-\frac{1}{12}\phi_{-2,1}(2\rho',S)\left[\phi_{0,1}(2\rho')+2E_2(2\rho')\,\phi_{-2,1}(2\rho',S)\right]\nonumber\\
&\hspace{1cm}-4\phi_{-2,1}^2(2\rho',S)\sum_{n=1}^\infty n\frac{Q^n}{1-Q^{2n}}\nonumber\\
&=-2\phi_{-2,1}(\rho',S)\,.\label{SelfSimilarityOrd0}
\end{align}
Here the last equality can be established by comparing the first few coefficients in an expansion in $Q$: Indeed, (\ref{IalphaThetaFct}) implies that each term on the left-hand side of (\ref{SelfSimilarityOrd0}) is a quasi-Jacobi form of weight $-2$ and index $1$ of the congruence subgroup $\Gamma_0(2)\subset SL(2,\mathbb{Z})_{\rho'}$, which acts in the following fashion
\begin{align}
SL(2,\mathbb{Z})_{\rho'}:&&(\rho,S)\longrightarrow \left(\frac{a\rho'+b}{c\rho'+d},\frac{S}{c\rho'+d}\right)\,.\label{SL2ActionRhop}
\end{align}
The precise combination appearing in (\ref{SelfSimilarityOrd0}) is in fact a Jacobi form of weight $-2$ and index $1$ of $SL(2,\mathbb{Z})_{\rho'}$. Comparing with the coefficient $H_{(0,0)}^{(0,1)}(\tfrac{\rho}{2},S)$ given in (\ref{CoeffsH1NS}), eq.~(\ref{SelfSimilarityOrd0}) indeed agrees with (\ref{SelfSimilarityGeneral}). In the same fashion we have also verified (\ref{SelfSimilarityGeneral}) for $s_1\in\{2,4,6\}$. This is a highly non-trivial check indicating that our results are compatible with \cite{Hohenegger:2016eqy}.
%%%%%%%%%%%%%%%%%%%%%%%
\subsubsection{Non-holomorphicity and Differential Equations}
Before moving to cases $N>2$, we stop to comment on a further detail with regards to the explicit expressions (\ref{N2ExplicitgFuncs}). Indeed, upon close examination, the same structures (\emph{i.e.} combinations of powers of $n$ and Eisenstein series) appear in various different $g_{(s_1,s_2)}^{i,(n,1)}$. Furthermore, we also have (with $x_{1,2,3},y_{1,2}\in\mathbb{N}$)
\begin{align}
n \frac{\partial}{\partial E_2} &\left[-70 E_2^3+84 n^2 \left(10 E_2^2+x_1 E_4\right)-273 E_2
   E_4-2016 E_2 n^4+x_2 E_6+x_3 n^6\right]\nonumber\\
   &=-21 n \left(10 E_2^2-80 E_2 n^2+13 E_4+96 n^4\right)\,,\nonumber\\
n\frac{\partial}{\partial E_2}&\left[10 E_2^2-80 E_2 n^2+y_1 E_4+y_2 n^4\right]   =-20 n \left(4 n^2-E_2\right)\,.
\end{align}
This suggests certain (differential) relations between $g_{(s_1,s_2)}^{i,(n,1)}$. For example, based on the expressions we find evidence for the following relations
{\allowdisplaybreaks
\begin{align}
\frac{\partial^2}{\partial E_2^2}g_{(s_1,0)}^{1,(n,1)}+8\frac{\partial}{\partial E_2}g_{(s_1,0)}^{2,(n,1)}+48\,g_{(s_1,0)}^{3,(n,1)}&=\chi_{s_1-5}\,\frac{(-1)^{s_1+1} n^{2s_1-5} E_4}{40\cdot 24^3 (2 s_1-5)!}\nonumber\\
&\hspace{-6.5cm}+\chi_{s_1-7}\,\frac{(-1)^{s_1} n^{2s_1-7} (21 E_2 E_4+5 E_6)}{35\cdot 24^5 (2 s_1-7)!}+\chi_{s_1-9}\,\frac{(-1)^{s_1+1}n^{s_1-9} \left(420 E_4 E_2^2+200 E_6 E_2+273
   E_4^2\right)}{1400\cdot 24^6 (2 s_1-9)!}\nonumber\\
&\hspace{-6.5cm}+\chi_{s_1-11}  \frac{(-1)^{s_1} n^{s_1-11} \left(1540 E_4 E_2^3+1100 E_6 E_2^2+3003 E_4^2
   E_2+1894 E_4 E_6\right)}{15400\cdot 24^7 (2 s_1-11)!}+\ldots\nonumber\\
\frac{\partial^2}{\partial E_2^2}g_{(s_1,1)}^{1,(n,1)}+4\frac{\partial}{\partial E_2}g_{(s_1,1)}^{2,(n,1)}+16\,g_{(s_1,1)}^{3,(n,1)}&=\chi_{s_1-5}\,\frac{(-1)^{s_1+1} n^{2s_1-5} E_4}{20\cdot 24^3 (2 s_1-5)!}\nonumber\\
&\hspace{0.5cm}+\chi_{s_1-7}\,\frac{2(-1)^{s_1} n^{2s_1-7} (21 E_2 E_4+5 E_6)}{35\cdot 24^5 (2 s_1-7)!}+\ldots\nonumber\\
%&\hspace{-5cm}+\chi_{s_1-9}\,\frac{(-1)^{s_1+1}n^{s_1-9} \left(210 E_4 E_2^2+200 E_6 E_2+273E_4^2\right)}{1400\cdot 24^6 (2 s_1-9)!}\nonumber\\
%   &\hspace{-5cm}+\chi_{s_1-11}  \frac{(-1)^{s_1} n^{s_1-11} \left(1540 E_4 E_2^3+1100 E_6 E_2^2+3003 E_4^2 E_2+1894 E_4 E_6\right)}{7700\cdot 24^7 (2 s_1-11)!}+\ldots\nonumber\\
%
\frac{\partial^2}{\partial E_2^2}g_{(s_1,2)}^{1,(n,1)}+4\frac{\partial}{\partial E_2}g_{(s_1,2)}^{2,(n,1)}+16\,g_{(s_1,2)}^{3,(n,1)}&=\chi_{s_1-5}\,\frac{(-1)^{s_1} s_1 n^{2s_1-5} E_4}{10\cdot 24^3 (2 s_1-5)!}\nonumber\\
&\hspace{0.5cm}-\chi_{s_1-7}\,\frac{4(-1)^{s_1}s_1 n^{2s_1-7} (21 E_2 E_4+5 E_6)}{35\cdot 24^5 (2 s_1-7)!}+\ldots\nonumber\\
%&\hspace{-5cm}+\chi_{s_1-9}\,\frac{(-1)^{s_1+1}n^{s_1-9} \left(210 E_4 E_2^2+200 E_6 E_2+273 E_4^2\right)}{1400\cdot 24^6 (2 s_1-9)!}\nonumber\\
%&\hspace{-5cm}+\chi_{s_1-11}  \frac{(-1)^{s_1} n^{s_1-11} \left(1540 E_4 E_2^3+1100 E_6 E_2^2+3003 E_4^2 E_2+1894 E_4 E_6\right)}{7700\cdot 24^7 (2 s_1-11)!}+\ldots\nonumber\\   
\frac{\partial^2}{\partial E_2^2}g_{(s_1,3)}^{1,(n,1)}+4\frac{\partial}{\partial E_2}g_{(s_1,3)}^{2,(n,1)}+16\,g_{(s_1,3)}^{3,(n,1)}&=\chi_{s_1-3}\,\frac {(-1)^{s_ 1} n^{s_1-3} (s_1 + 1) (2 s_1 + 1) E_4} {20\cdot 24^3 (2 s_ 1 - 3)!}\nonumber\\
&\hspace{-0.5cm}-\chi_{s_1-5}\,\frac{(-1)^{s_1} n^{s_1-5} (2 s_1+2)! (21E_2 E_4+5E_6)}{70\cdot 24^5 s_1 (2 s_1-5)! (2 s_1-1)!}+\ldots
\end{align}}
where the dots indicate lower powers in $n$, while
\begin{align}
\chi_{s}=\left\{\begin{array}{lcl} 1 & \text{if} & s\geq 0\,, \\ 0 & \text{if} &s<0\,.\end{array}\right.
\end{align}
These equations generalise the holomorphic anomaly equation, \emph{i.e.} similar relations found in \cite{Haghighat:2013gba,Haghighat:2013tka}.

%%%%%%%%%%%%%%%%%%%%%%%%
\subsection{Order $Q_R^2$}
\subsubsection{Generating Functions of Multiple Divisor Sums}
To order $Q_R^2$, the $H_{(s_1,s_2)}^{(n,0,1)}(\rho,S)$ can be decomposed in a form similar to (\ref{Hfunc21})
\begin{align}
&H_{(s_1,s_2)}^{(n,0,r=2)}(\rho,S)=g^{1,(n,2)}_{(s_1,s_2)}(\rho)\,\phi^4_{-2,1}(\rho,S)+g^{2,(n,2)}_{(s_1,s_2)}(\rho)\,\phi_{0,1}(\rho,S)\,\phi^3_{-2,1}(\rho,S)\nonumber\\
&\hspace{0.2cm}+g^{3,(n,2)}_{(s_1,s_2)}(\rho)\,\phi_{0,1}^2(\rho,S)\,\phi^2_{-2,1}(\rho,S)+g^{4,(n,2)}_{(s_1,s_2)}(\rho)\,\phi_{0,1}^3(\rho,S)\,\phi_{-2,1}(\rho,S)+g^{5,(n,2)}_{(s_1,s_2)}(\rho)\,\phi_{0,1}^4(\rho,S)\,,\nonumber
\end{align}
where $g^{i,(n,r=2)}_{(s_1,s_2)}(\rho)$ have an integer series expansion in $Q_\rho$. For simplicity, we shall only discuss the case $(s_1,s_2)=(0,0)$ in the following.\footnote{Explicit computations become very difficult for higher orders in $\epsilon_{1,2}$. However, we still expect similar results to also hold in the cases $(s_1,s_2)\neq (0,0)$: for example, the case $(2,0)$ is exhibited in appendix~\ref{App:CoefsgR2N2}.} Repeating the computation of the previous subsection to order $\mathcal{O}(Q_R^2)$, we find a pattern, which suggests the following expressions
{\allowdisplaybreaks
\begin{align}
g^{1,(n,2)}_{(0,0)}&= -\frac{n^5}{24(1-Q_\rho^n)}-\frac{n}{12(1-Q_\rho^n)}\,E_4(\rho)+\left\{\begin{array}{lcl}0 & \text{if} & \text{gcd}(n,2)=1\,,\\[4pt]\frac{n}{72(1-Q^n_\rho)}\,\psi_2^2\, & \text{if} & \text{gcd}(n,2)>1\,.\end{array}\right.\\[10pt]
g^{2,(n,2)}_{(0,0)}&=-\frac{n^3}{12(1-Q_\rho^n)}-\left\{\begin{array}{lcl} 0 & \text{if} & \text{gcd}(n,2)=1\,,\\[4pt]\frac{n}{72(1-Q_\rho^n)}\,\psi_2\, & \text{if} & \text{gcd}(n,2)>1\,,\end{array}\right.\\[10pt]
g^{3,(n,2)}_{(0,0)}&= -\frac{6n}{24^2(1-Q_\rho^n)} +\left\{\begin{array}{lcl} 0 & \text{if} & \text{gcd}(n,2)=1\,,\\[4pt] \frac{2n}{24^2(1-Q_\rho^n)} & \text{if} & \text{gcd}(n,2)>1\,,\end{array}\right.\\[10pt]
g^{4,(n,2)}_{(0,0)}&=g^{5,(n,2)}_{(0,0)}=0\,.
\end{align}}
Here we have used the notation
\begin{align}
\psi_2(\rho)=\theta_3^4(\rho,0)+\theta_4^4(\rho,0)=-2(E_2(\rho)-2E_2(2\rho))\,,
\end{align}
which is in fact a weight 2 Eisenstein series of the congruence subgroup $\Gamma_0(2)$,\footnote{See \cite{Lang,Stein} (see also \cite{Gaberdiel:2010ca}) as well as citations therein for a basis of modular forms for congruence subgroups of the type $\Gamma_0(N)$.} \emph{i.e.} despite being composed from $E_2$, it transforms without shift-term under modular transformations with respect to the latter (see \cite{Lang,Stein,Gaberdiel:2010ca}). Notice, we prefer to distinguish the cases $\text{gcd}(n,2)=1$ and $\text{gcd}(n,2)>1$, rather than $n\in\mathbb{N}_\text{even}$ or $n\in\mathbb{N}_\text{odd}$. We will provide non-trivial evidence that the former is the correct prescription that generalises to higher orders in $Q_R$ as well in section~\ref{Sect:N2HighOrder}. With these coefficients, we can compute
\begin{align}
&B_{(0,0)}^{N=2,2}(\rho,\widehat{a}_1,S)=\sum_{n=1}^\infty H_{(0,0)}^{(n,0,2)}(\rho,S)\left(Q_{\widehat{a}_1}^n+\tfrac{Q_\rho^n}{Q_{\widehat{a}_1}^n}\right)\nonumber\\
&\hspace{0.2cm}=\mathfrak{g}^{1,(2)}_{(0,0)}(\rho,\widehat{a}_1)\,\phi^4_{-2,1}(\rho,S)+\mathfrak{g}^{2,(2)}_{(0,0)}(\rho,\widehat{a}_1)\,\phi_{0,1}(\rho,S)\,\phi^3_{-2,1}(\rho,S)+\mathfrak{g}^{3,(2)}_{(0,0)}(\rho,\widehat{a}_1)\,\phi_{0,1}^2(\rho,S)\,\phi^2_{-2,1}(\rho,S)\nonumber\\
&\hspace{0.8cm}+\mathfrak{g}^{4,(2)}_{(0,0)}(\rho,\widehat{a}_1)\,\phi_{0,1}^3(\rho,S)\,\phi_{-2,1}(\rho,S)+\mathfrak{g}^{5,(2)}_{(0,0)}(\rho,\widehat{a}_1)\,\phi_{0,1}^4(\rho,S)\,,\nonumber
\end{align}
which can be expressed in terms of the generating functions (\ref{GenFuncBracket}). More concretely, we have
{\allowdisplaybreaks
\begin{align}
\mathfrak{g}^{1,(2)}_{(0,0)}(\rho,\widehat{a}_1)&=\sum_{n=1}^\infty g^{1,(n,2)}_{(0,0)}\left(Q_{\widehat{a}_1}^n+\tfrac{Q_\rho^n}{Q_{\widehat{a}_1}^n}\right)=-\frac{1}{72}D_{\widehat{a}_1}\left[3(D_{\widehat{a}_1}^4+2E_4(\rho))\wprim (\widehat{a}_1;\rho)-\psi_2^2\,\wprim (2\widehat{a}_1;2\rho)\right]\,,\nonumber\\
\mathfrak{g}^{2,(2)}_{(0,0)}(\rho,\widehat{a}_1)&=\sum_{n=1}^\infty g^{2,(n,2)}_{(0,0)}\left(Q_{\widehat{a}_1}^n+\tfrac{Q_\rho^n}{Q_{\widehat{a}_1}^n}\right)=-\frac{1}{72}D_{\widehat{a}_1}\left[6 D_{\widehat{a}_1}^2\wprim (\widehat{a}_1;\rho)+\psi_2\,\wprim (2\widehat{a}_1;2\rho)\right]\,,\nonumber\\
\mathfrak{g}^{3,(2)}_{(0,0)}(\rho,\widehat{a}_1)&=\sum_{n=1}^\infty g^{3,(n,2)}_{(0,0)}\left(Q_{\widehat{a}_1}^n+\tfrac{Q_\rho^n}{Q_{\widehat{a}_1}^n}\right)=-\frac{2}{24^2}D_{\widehat{a}_1}\left[3\wprim (\widehat{a}_1;\rho)-\wprim (2\widehat{a}_1;2\rho)\right]\,,\nonumber\\
\mathfrak{g}^{4,(2)}_{(0,0)}(\rho,\widehat{a}_1)&=\mathfrak{g}^{5,(2)}_{(0,0)}(\rho,\widehat{a}_1)=0\,,\label{SummedCoefficientsN2R2}
\end{align}}
where $\wprim$ was defined in (\ref{Defwprim}). The novel element is the appearance of $T(2\widehat{a}_1;2\rho)$. 

To analyse the generating functions $T(x_1,\ldots,x_\ell;\rho)$ (for $x_{1,\ldots,\ell}\in \mathbb{R}$ and $\ell\in\mathbb{N}$) with arguments multiplied by a fixed $p\in\mathbb{N}$, we can follow closely the discussion of \cite{Bachmann:2013wba}
{\allowdisplaybreaks
\begin{align}
T(px_1,\ldots,px_\ell;p\rho)&=\sum_{n_1,\ldots,n_\ell>0}\prod_{j=1}^\ell\frac{e^{2\pi i n_j p x_j}\,Q_\rho^{p(n_1+\ldots+n_j)}}{1-Q_\rho^{p(n_1+\ldots+n_j)}}\nonumber\\
&=\sum_{n_1,\ldots,n_\ell>0}\prod_{j=1}^\ell\sum_{k_j\geq 0}\frac{(pn_j)^{k_j}\,(2\pi i x_j)^{k_j}}{k_j!}\,\sum_{v_j>0}Q_\rho^{pv_j(n_1+\ldots+n_j)}\,.
\end{align}}
Introducing $u_j=v_j+\ldots +v_\ell$ we find \cite{Bachmann:2013wba}
{\allowdisplaybreaks
\begin{align}
T(px_1,&\ldots,px_\ell;p\rho)=\sum_{k_1,\ldots,k_\ell\geq 0}\left(\sum_{{u_1>\ldots>u_\ell>0}\atop n_1,\ldots,n_\ell>0}\frac{(pn_1)^{k_1}\ldots (pn_\ell)^{k_\ell}}{k_1!\ldots k_\ell!}Q_\rho^{p(u_1n_1+\ldots +u_\ell n_\ell)}\right)(2\pi i x_1)^{k_1}\ldots (2\pi i x_\ell)^{k_\ell}\nonumber\\
&=\sum_{k_1,\ldots,k_\ell\geq 0}\left(\sum_{n>0}Q_\rho^{np}\sum_{{u_1v_1+\ldots+u_\ell v_\ell=n}\atop{u_1>\ldots>u_\ell>0}}\frac{(pv_1)^{k_1}\ldots (pv_\ell)^{k_\ell}}{k_1!\ldots k_\ell!}\right)(2\pi i x_1)^{k_1}\ldots (2\pi i x_\ell)^{k_\ell}\nonumber\\
&=\sum_{k_1,\ldots,k_\ell\geq 0}\left(\sum_{n>0}Q_\rho^{np}\sum_{{u_1v_1+\ldots+u_\ell v_\ell=np}\atop{u_1>\ldots>u_\ell>0}}\frac{v_1^{k_1}\ldots v_\ell^{k_\ell}}{k_1!\ldots k_\ell!}\right)(2\pi i x_1)^{k_1}\ldots (2\pi i x_\ell)^{k_\ell}\nonumber\\
&=\sum_{k_1,\ldots,k_\ell\geq 0}\sum_{n>0}Q_\rho^{np}\,\frac{\sigma_{k_1,\ldots,k_\ell}(np)}{k_1!\ldots k_\ell!}\,(2\pi i x_1)^{k_1}\ldots (2\pi i x_\ell)^{k_\ell}\nonumber\\
&=\sum_{s_1,\ldots,s_\ell> 0}[s_1,\ldots,s_\ell;p\rho]_p\,(2\pi i x_1)^{k_1}\ldots (2\pi i x_\ell)^{k_\ell}\,,
\end{align}}
where the multiple divisor sum $\sigma_{k_1,\ldots,k_\ell}(n)$ was introduced in (\ref{DefMultiDivisor}) and 
\begin{align}
[s_1,\ldots,s_\ell;\rho]_p=\sum_{n>0}Q_\rho^{n}\,\frac{\sigma_{s_1-1,\ldots,s_\ell-1}(np)}{(s_1-1)!\ldots (s_\ell-1)!}\,,\label{GenBracketFourier}
\end{align}
which generalises the bracket $[s_1,\ldots,s_\ell;\rho]=[s_1,\ldots,s_\ell;\rho]_1$ introduced in (\ref{DefBracket}). Notice, when interpreting (\ref{GenBracketFourier}) in the sense of a Fourier expansion, \emph{i.e.}
\begin{align}
[s_1,\ldots,s_\ell;\rho]_p=\sum_{n>0}d_{p}^{\,(s_1-1,\ldots,s_\ell-1)}(n)\,Q_\rho^{n}\,,&&\text{with} &&d_{p}^{\,(s_1-1,\ldots,s_\ell-1)}(n)=\frac{\sigma_{s_1-1,\ldots,s_\ell-1}(np)}{(s_1-1)!\ldots (s_\ell-1)!}\in\mathbb{Q}\,,\nonumber
\end{align}
we have
\begin{align}
&d_{p}^{\,(s_1-1,\ldots,s_\ell-1)}(n)=d_{1}^{\,(s_1-1,\ldots,s_\ell-1)}(np)\,,&&\forall s_1,\ldots,s_\ell>0 &&\text{and} &&\forall n\in\mathbb{N}\,.\label{HeckeSpirit}
\end{align}
This relation is very similar in spirit to eq.~(15) of \cite{Ahmed:2017hfr} (at least at the level of the Fourier coefficients), which was interpreted as a particular type of Hecke structure in the $H_{(0,0)}^{(n,0,2)}(\rho,S)$ (see also \cite{Companion1}). In the current context (due to the fact that $T(x_1,\ldots,x_\ell;\rho)$ is not a (quasi-) Jacobi form), we do not find the action of a Hecke operator, however, a similar relation at the level of the Fourier coefficients. We leave further study of this relation (and a potential generalisation and extension of the work \cite{Ahmed:2017hfr}) for the future.

Finally, we remark that with (\ref{Defwprim}) and (\ref{RelIAlphaEisen}), we can re-write (\ref{SummedCoefficientsN2R2}) in terms of generalised Eisenstein series. However, in order to keep our presentation compact, we refrain from writing the explicit expressions.
%%%%%%%%%%%%%%%%%%%%%%%%%%%%%%%%%%%%%%%%%%%
%%%%%%%%%%%%%%%%%%%%%%%%%%%%%%%%%%%%%%%%%%%
\subsubsection{Free Energy and Self-Similarity}
Finally, for completeness, we shall also attempt to combine the results above in (\ref{SummedCoefficientsN2R2}) with the contribution $H_{(0,0)}^{(0,0,2)}(\rho,S)$, which is explicitly given in (\ref{OrbifoldSectorContributionR2}). To order $Q_R$, we have seen that there are certain cancellations that  reduce the depth of the complete free energy (to this order in $Q_R$) as a quasi-Jacobi form. In the following we shall see, that similar cancellations also occur to order $Q_R^2$: indeed, expanding (\ref{OrbifoldSectorContributionR2}) in powers of $\widehat{a}_1$ (using (\ref{IntroTcal}) along with (\ref{DefI0Weierstrass})) we can write
{\allowdisplaybreaks
\begin{align}
\mathfrak{g}^{1,(2)}_{(0,0)}(\rho,\widehat{a}_1)&=\frac{1}{432}\left[E_2(2\rho)\psi_2^2-3 E_2(\rho)E_4(\rho)\right]+\ldots\nonumber\\
\mathfrak{g}^{2,(2)}_{(0,0)}(\rho,\widehat{a}_1)&=\frac{1}{432}\,\psi_2\,E_2(2\rho)+\ldots\nonumber\\
\mathfrak{g}^{3,(2)}_{(0,0)}(\rho,\widehat{a}_1)&=\frac{1}{6\cdot 24^2}\left[3E_2(\rho)-2E_2(2\rho)\right]+\ldots\,,\nonumber
\end{align}}
where the dots denote terms depending explicitly on $\widehat{a}_1$ that are Jacobi forms of the congruence subgroup $\Gamma_0(2)$.\footnote{This means they depend on $E_2$ only through the combination $\psi_2$.} Comparing with (\ref{OrbifoldSectorContributionR2}), we see that these terms do not completely cancel, however, add up to 
\begin{align}
&B_{(0,0)}^{N=2,2}(\rho,\widehat{a}_1,S)+H_{(0,0)}^{(0,0,2)}(\rho,S)\nonumber\\
&=-\frac{\psi_2}{6\cdot 24^2}\, \phi_{0,1}^2 \phi_{-2,1}^2 -\frac{1}{3\cdot 24^2}\,(3E_4(\rho)-2\psi_2)\, \phi_{0,1} \phi_{-2,1}^3 +\frac{ \psi_2}{576}(5E_4(\rho)-\psi_2^2)\,\phi_{-2,1}^4+\ldots\,.\nonumber
\end{align}
which is a Jacobi form of $\Gamma_0(2)$.

Furthermore, we have verified that 
\begin{align}
H_{(0,0)}^{(0,0,2)}(\rho,S)&+B_{(0,0)}^{2,2}(\rho,\tfrac{\rho}{2},S)=2H_{(0,0)}^{(0,2)}(\tfrac{\rho}{2},S)\,,
\end{align}
where the contribution $H_{(0,0)}^{(0,2)}(\rho,S)$ is given in (\ref{CoeffsH2NS}), thus confirming (\ref{SelfSimilarityGeneral}) also to order $Q_R^2$.

%%%%%%%%%%%%%%
\subsection{Order $Q_R^3$}
We can provide further support for the fact that similar structures that we found in the previous subsections in fact persist to all orders in $Q_R$, by analysing the case $Q_R^3$. Since computations are exceedingly difficult in this case, we once more focus on $(s_1,s_2)=(0,0)$.

Expanding the free energy to order $Q_R^3$, the results are compatible with the following presentation of $H_{(s_1,s_2)}^{(n,0,3)}$
\begin{align}
H_{(s_1,s_2)}^{(n,0,3)}(\rho,S)=\sum_{u=1}^7g^{u,(n,3)}_{(0,0)}(\rho)\,\phi^{7-u}_{-2,1}(\rho,S)\,\phi^{u-1}_{0,1}(\rho,S)\,,\label{DecompositionHN2R3}
\end{align}
with the following coefficients
{\allowdisplaybreaks
\begin{align}
g^{1,(n,3)}_{(0,0)}&=-\frac{n^9}{7560(1-Q_\rho^n)}-\frac{7n^5\,E_4(\rho)}{1080(1-Q_\rho^n)}+\frac{11n^3\,E_6(\rho)}{1134(1-Q_\rho^n)}-\frac{7n\,E_8(\rho)}{864(1-Q_\rho^n)}\nonumber\\[6pt]
&\hspace{1cm}+\left\{\begin{array}{lcl} 0 & \text{if} & \text{gcd}(n,3)=1\,,\\[12pt]\frac{n}{2048(1-Q_\rho^n)}\,\psi_3^4 & \text{if} & \text{gcd}(n,3)>1 \end{array}\right.\nonumber\\[10pt]
g^{2,(n,3)}_{(0,0)}&=-\frac{n^7}{810(1-Q_\rho^n)}-\frac{n^3\,E_4(\rho)}{90(1-Q_\rho^n)}+\frac{n\,E_6(\rho)}{216(1-Q_\rho^n)}+\left\{\begin{array}{lcl}0  & \text{if} & \text{gcd}(n,3)=1\,,\\[4pt]\frac{9n}{24^3(1-Q_\rho^n)}\,\psi_3^3\,, & \text{if} & \text{gcd}(n,3)>1\,,\end{array}\right.\nonumber\\[10pt]
g^{3,(n,3)}_{(0,0)}&=-\frac{n^5}{432(1-Q_\rho^n)}-\frac{n\,E_4(\rho)}{576(1-Q_\rho^n)}+\left\{\begin{array}{lcl} 0 & \text{if} & \text{gcd}(n,3)=1\,,\\[4pt]\frac{9n}{2\cdot 24^3(1-Q_\rho^n)}\,\psi_3^2\,, & \text{if} & \text{gcd}(n,3)>1\,,\end{array}\right.\nonumber\\[10pt]
g^{4,(n,3)}_{(0,0)}&=-\frac{32n^3}{3\cdot 24^3(1-Q_\rho^n)}+\left\{\begin{array}{lcl} 0 & \text{if} & \text{gcd}(n,3)=1\,,\\[4pt]\frac{n}{24^3(1-Q_\rho^n)}\,\psi_3\,, & \text{if} & \text{gcd}(n,3)>1\,,\end{array}\right.\nonumber\\[10pt]
g^{5,(n,3)}_{(0,0)}&=-\frac{n}{3\cdot 24^3(1-Q_\rho^n)}+\left\{\begin{array}{lcl} 0 & \text{if} & \text{gcd}(n,3)=1\,,\\[4pt]\frac{n}{12\cdot 24^3(1-Q_\rho^n)}\, & \text{if} & \text{gcd}(n,3)>1\,,\end{array}\right.\nonumber\\[10pt]
g^{6,(n,3)}_{(0,0)}&=g^{7,(n,3)}_{(0,0)}=0\,.\label{CoeffsN2R3}
\end{align}}
where we used the shorthand notation
\begin{align}
\psi_3=E_2(\rho)-3E_2(3\rho)\,.
\end{align}
In a similar fashion as (\ref{DecompositionHN2R3}), we can also decompose $B_{(0,0)}^{2,3}(\rho,\widehat{a}_1,S)$
\begin{align}
B_{(0,0)}^{2,3}(\rho,\widehat{a}_1,S)=\sum_{u=1}^7\mathfrak{g}^{u,(3)}_{(0,0)}(\rho,\widehat{a}_1)\,\phi^{7-u}_{-2,1}(\rho,S)\,\phi^{u-1}_{0,1}(\rho,S)\,,\label{DecompositionBN2R3}
\end{align}
where the coefficients (\ref{CoeffsN2R3}) lead to
{\allowdisplaybreaks
\begin{align}
\mathfrak{g}^{1,(3)}_{(0,0)}(\rho,\widehat{a}_1)&=\sum_{n=1}^\infty g_5^{(3)}\left(Q_{\widehat{a}_1}^n+\tfrac{Q_\rho^n}{Q_{\widehat{a}_1}^n}\right)=-\frac{1}{420\cdot 24^3}D_{\widehat{a}_1}\big[64(12D_{\widehat{a}_1}^8+588E_4(\rho)D_{\widehat{a}_1}^4\nonumber\\
&\hspace{1cm}-880E_6(\rho)D_{\widehat{a}_1}^2+735 E_8(\rho))\wprim (\widehat{a}_1;\rho)-2835\,\psi_3^4\,\wprim (3\widehat{a}_1;3\rho)\big]\,,\nonumber\\
\mathfrak{g}^{2,(3)}_{(0,0)}(\rho,\widehat{a}_1)&=\sum_{n=1}^\infty g_4^{(3)}\left(Q_{\widehat{a}_1}^n+\tfrac{Q_\rho^n}{Q_{\widehat{a}_1}^n}\right)\nonumber\\
&=-\frac{1}{15\cdot 24^3}D_{\widehat{a}_1}\left[64(4D_{\widehat{a}_1}^6+36E_4(\rho)D_{\widehat{a}_1}^2-15 E_6(\rho))\wprim (\widehat{a}_1;\rho)-135\,\psi_3^3\,\wprim (3\widehat{a}_1;3\rho)\right]\,,\nonumber\\
\mathfrak{g}^{3,(3)}_{(0,0)}(\rho,\widehat{a}_1)&=\sum_{n=1}^\infty g_3^{(3)}\left(Q_{\widehat{a}_1}^n+\tfrac{Q_\rho^n}{Q_{\widehat{a}_1}^n}\right)=-\frac{1}{2\cdot 24^3}D_{\widehat{a}_1}\left[16(4D_{\widehat{a}_1}^4+3E_4(\rho))\wprim (\widehat{a}_1;\rho)-9\,\psi_3^2\,\wprim (3\widehat{a}_1;3\rho)\right]\,,\nonumber\\
\mathfrak{g}^{4,(3)}_{(0,0)}(\rho,\widehat{a}_1)&=\sum_{n=1}^\infty g_2^{(3)}\left(Q_{\widehat{a}_1}^n+\tfrac{Q_\rho^n}{Q_{\widehat{a}_1}^n}\right)=-\frac{1}{3\cdot 24^3}D_{\widehat{a}_1}\left[32D_{\widehat{a}_1}^2\wprim (\widehat{a}_1;\rho)-3\,\psi_3\,\wprim (3\widehat{a}_1;3\rho)\right]\,,\nonumber\\
\mathfrak{g}^{5,(3)}_{(0,0)}(\rho,\widehat{a}_1)&=\sum_{n=1}^\infty g_1^{(3)}\left(Q_{\widehat{a}_1}^n+\tfrac{Q_\rho^n}{Q_{\widehat{a}_1}^n}\right)=-\frac{1}{12\cdot 24^3}D_{\widehat{a}_1}\left[4\wprim (\widehat{a}_1;\rho)-\wprim (3\widehat{a}_1;3\rho)\right]\,,\nonumber\\
\mathfrak{g}^{6,(3)}_{(0,0)}(\rho,\widehat{a}_1)&=\mathfrak{g}^{7,(3)}_{(0,0)}(\rho,\widehat{a}_1)=0\,.\label{SummedCoefficientsN2R3}
\end{align}}

%%%%%%%%%%%%%%%%%%%%%%%%%%%%%%%%%%%%%%%%%%%
%%%%%%%%%%%%%%%%%%%%%%%%%%%%%%%%%%%%%%%%%%%
\subsubsection{Free Energy and Self-Similarity}
Following the approach of the previous sections, we combine the results above in (\ref{SummedCoefficientsN2R3}) with the contribution $H_{(0,0)}^{(0,0,3)}(\rho,S)$, as exhibited in  (\ref{OrbifoldSectorContributionR3}). Expanding the latter in powers of $\widehat{a}_1$ (using (\ref{IntroTcal}) along with (\ref{DefI0Weierstrass})) we find
\begin{align}
&\mathfrak{g}^{1,(3)}_{(0,0)}(\rho,\widehat{a}_1)=\frac{7 E_2(\rho) E_8(\rho)}{18\cdot 24^2}-\frac{\psi_3^4 E_2(3\rho)}{8192}+\ldots\,,&&\mathfrak{g}^{2,(3)}_{(0,0)}(\rho,\widehat{a}_1)=-\frac{\psi_3^3 E_2(3\rho)}{256\cdot 24}-\frac{E_2(\rho) E_6(\rho)}{108\cdot 24}+\ldots\,,\nonumber\\
&\mathfrak{g}^{3,(3)}_{(0,0)}(\rho,\widehat{a}_1)=\frac{16 E_2(\rho) E_4(\rho)-9 \psi_3^2 E_2(3\rho)}{8\cdot 24^3}+\ldots\,,&&\mathfrak{g}^{4,(3)}_{(0,0)}(\rho,\widehat{a}_1)=-\frac{\psi_3 E_2(3\rho)}{4\cdot 24^3}+\ldots\nonumber\\
&\mathfrak{g}^{5,(3)}_{(0,0)}(\rho,\widehat{a}_1)=\frac{4 E_2(\rho)-3 E_2(3\rho)}{6\cdot 24^4}+\ldots\,,
\end{align}
where as before the dots denote terms depending explicitly on $\widehat{a}_1$ that are Jacobi forms of the congruence subgroup $\Gamma_0(3)$. Comparing with (\ref{OrbifoldSectorContributionR3}), we see that these terms combine to
\begin{align}
&B_{(0,0)}^{N=2,3}(\rho,\widehat{a}_1,S)+H_{(0,0)}^{(0,0,3)}(\rho,S)=\frac{\psi_3}{3\cdot 24^4}\, \phi_{0,1}^4 \phi_{-2,1}^2 +\frac{15 \psi_3^2-38 E_4(\rho)+108 E_4(3\rho)}{15\cdot 24^4}\, \phi_{0,1}^3 \phi_{-2,1}^3 \nonumber\\
&\hspace{1cm}+\frac{81 \psi_3^3-12 \psi_3 E_4(\rho)+224 E_6(\rho)}{6\cdot 24^4}\, \phi_{0,1}^2 \phi_{-2,1}^4+\frac{81 \psi_3^4-16 \psi_3 E_6(\rho)-436 E_8(\rho)}{3\cdot 24^4}\, \phi_{0,1} \phi_{-2,1}^5\nonumber\\
&\hspace{1cm}-\frac{243 \psi_3^5-48 \psi_3 E_8(\rho)+2560 E_{10}(\rho)}{12\cdot 24^4}\, \phi_{-2,1}^6+\ldots\,.\nonumber
\end{align}
which is a Jacobi form of $\Gamma_0(3)$.

Finally, we have verified that 
\begin{align}
H_{(0,0)}^{(0,0,3)}(\rho,S)&+B_{(0,0)}^{2,3}(\rho,\tfrac{\rho}{2},S)=2H_{(0,0)}^{(0,3)}(\tfrac{\rho}{2},S)\,,
\end{align}
where the contribution $H_{(0,0)}^{(0,3)}(\rho,S)$ is explicitly exhibited in (\ref{CoeffsH3NS}), thus confirming (\ref{SelfSimilarityGeneral}) also to order $Q_R^3$.

%%%%%%%%%%%%%%%%%%

\subsection{Higher Orders in $Q_R$}\label{Sect:N2HighOrder}
While in general, orders higher than $Q_R^3$ are extremely difficult to analyse directly by expanding the free energy (\ref{TaylorFreeEnergy}), there are other methods that we can use to extract certain information. Indeed, as discussed in \cite{Hohenegger:2015btj} and reviewed briefly in section~\ref{Sect:ReviewSyms}, the $G_{(0,0)}^{(i_1,i_2)}(R,S)$ (as introduced in (\ref{DefinitionG}) are quasi-Jacobi forms of weight $-2$ and index $i_1+i_2$ under the congruence subgroup $\Gamma_0(\subg(\text{gcd}(i_1,i_2)))\subset SL(2,\mathbb{Z})_R$. Here we are using the notation introduced in \cite{Companion1}: 
\begin{align}
&\subg(n)=\prod_{i=1}^\ell p_i&&\text{for} &&n=\prod_{i=1}^\ell p_i^{s_i}\,, &&\text{with} &&\begin{array}{l}\ell\in\mathbb{N}\,, \\ p_i\in\mathbb{N}_{\text{prime}}\,, \\  s_{1,\ldots,\ell}\in\mathbb{N}\end{array}\,,\label{PrimeFunct}
\end{align}
In order to determine the explicit expansion (\ref{ExpansionG}) only a finite number of coefficients are required, which (at least for low values of $i_{1,2}$) are provided by the results up to order $Q_R^3$. We exhibit a few explicit expressions in appendix~\ref{Sect:FuncGN2}. In turn, upon assuming the general decomposition
\begin{align}
H_{(0,0)}^{(n,0,r)}(\rho,S)=\sum_{u=1}^{2r+1}g_{(0,0)}^{u,(n,r)}(\rho)\,\phi_{-2,1}^{2r+1-u}(\rho,S)\,\phi_{0,1}^{u-1}(\rho,S)\,,\label{ExpansionHN2higher}
\end{align}
the former allow us to determine generically 2 coefficients in a series expansion of $g_{(0,0)}^{u,(n,0,r)}(\rho)$ in $Q_\rho$. While by far insufficient for generic $u$, there are two\footnote{We do not count the trivial vanishing $g_{(0,0)}^{2r,(n,r)}(\rho)=g_{(0,0)}^{2r+1,(n,r)}(\rho)=0$.} classes of coefficients for which we can make non-trivial statements:
\begin{itemize}
\item coefficients $g_{(0,0)}^{2r-1,(n,r)}(\rho)$\\
These are the simplest non-vanishing coefficients, which, following the pattern arising from the results for order up to $Q_R^3$, should be of the form $g^{2r-1,(n,r)}_{(0,0)}(\rho)\sim \frac{n}{(1-Q_\rho^n)}$ up to some numerical prefactor.\footnote{In particular, it should not contain an Eisenstein series $E_{2}(\rho)$, which, since we can check two coefficients in an expansion in $Q_\rho$, constitutes a non-trivial check.} Analysing the latter up to order $r=50$ and $n=4$ we conjecture
\begin{align}
g^{2r-1,(n,r)}_{(0,0)}(\rho)=-\frac{2n}{24^{2r-2}(1-Q_\rho^n)}\,\sigma_1(r)+\left\{\begin{array}{lcl}  0 & \text{if} & \text{gcd}(n,r)=1\,,\\[4pt] \frac{2n}{24^{2r-2}(1-Q_\rho^n)}\,\sigma_1(r,n) & \text{if} & \text{gcd}(n,r)>1\,,\end{array}\right.
\end{align}
where we defined
\begin{align}
\sigma_1(r,n)=\sum_{{d|r}\atop{\text{gcd}(r/d,n)>1}}d\,.
\end{align} 
Notice that the appearance of (parts of) divisor sums might indicate that similar structures that we have found when analysing the free energy as a function of $\rho$, may also apply when we think of $F_{2,1}(\rho,\widehat{a}_1,S,R;\epsilon_{1,2})$ as a function of $R$. It would therefore be interesting to analyse the latter from the point of view of genus two generalisations of the generating functions (\ref{GenFuncBracket}).
\item coefficients $g_{(0,0)}^{2r-2,(n,r)}(\rho)$\\
Following the pattern of previous sections, the structure of this coefficient should schematically be
\begin{align}
g^{2r-2,(n,r)}_{(0,0)}(\rho)\sim c_1(n,r)\,\frac{n^3}{(1-Q_\rho^n)}+c_2(n,r)\,\frac{n\,P(E_2)}{(1-Q_\rho^n)}\sim \frac{\sum_{i=1}^\infty d_i(n,r) Q_\rho^i}{1-Q_\rho^n}\,,
\end{align}
where $c_{1,2},d_i\in\mathbb{Q}$, while $P$ is a linear function in Eisenstein series with various arguments. While we do not have enough information to fully determine $P$, we find (by considering up to $r=10$) for $r>1$
\begin{align}
&d_1(n,r)\neq0&&\text{and} &&d_2(n,r)\left\{\begin{array}{lcl}=0 & \text{if} & n=1 \\ \neq 0 & \text{if} & n=2\end{array}\right.
\end{align}
While not a lot of information, this indicates that the coefficients $g^{2r-2,(n,r)}_{(0,0)}(\rho)$ behave differently depending on $\text{gcd}(n,r)=1$ or $\text{gcd}(n,r)\neq 1$. Notice, if the deciding condition would have been $r|n$, $d_2(1,r)=0$ would have implied $d_2(2,r)=0$.
\end{itemize}

%%%%%%%%%%%%%%%%%%%%%%%%%%%%%%%%%%%
%%%%%%%%%%%%%%%%%%%%%%%%%%%%%%%%%%%
%%%%%%%%%%%%%%%%%%%%%%%%%%%%%%%%%%%
\section{Case $N=3$}\label{Sect:CaseN3}
After analysing the free energy for $N=2$ to various orders in $Q_R$, we now switch to $N=3$. The main difference compared to the previous case is the fact that the $H_{(s_1,s_2)}^{(i_1,i_2,i_3,r)}(\rho,S)$ are no longer characterised by a single integer. Indeed, taking into account that the coefficients $f^{(s_1,s_2)}_{i_1,i_2,i_3,k,r}$ are invariant under $\text{Dih}_3$, which acts through permutations of $(i_1,i_2,i_3)$, there are three a priori distinct classes of functions
\begin{align}
H_{(s_1,s_2)}^{(n,0,0,r)}(\rho,S)&=\sum_{r=0}^\infty\sum_{k\in\mathbb{Z}}f^{(s_1,s_2)}_{n+\ell,\ell,\ell,k,r}\,Q_\rho^\ell\,Q_S^k\,,&&\forall\,n\geq 1\,,\nonumber\\
H_{(s_1,s_2)}^{(n,n,0,r)}(\rho,S)&=\sum_{\ell=0}^\infty\sum_{k\in\mathbb{Z}}f^{(s_1,s_2)}_{n+\ell,n+\ell,\ell,k,r}\,Q_\rho^\ell\,Q_S^k\,,&&\forall\,n\geq 1\,,\nonumber\\
H_{(s_1,s_2)}^{(n_1+n_2,n_1,0,r)}(\rho,S)&=\sum_{\ell=0}^\infty\sum_{k\in\mathbb{Z}}f^{(s_1,s_2)}_{n_1+n_2+\ell,n_1+\ell,\ell,k,r}\,Q_\rho^\ell\,Q_S^k\,,&&\forall\,n_{1,2}\geq 1\,.
\end{align}
Due to the complexity of the computations, we content ourselves with a discussion to order $Q_R^1$ (\emph{i.e.} $r=1$) and partially $Q_R^2$ (\emph{i.e.} $r=2$) , as well as leading order in $\epsilon_{1,2}$ (\emph{i.e.} $(s_1,s_2)=(0,0)$). Within the limits of this restriction, we shall find similar structures as in the case $N=2$, albeit much more complicated.

%%%%%%%%%%%%%%
\subsection{Order $Q_R^1$}
\subsubsection{Explicit Contributions to the Free Energy}
Analysing the the coefficients $f^{(0,0)}_{i_1,i_2,i_3,k,1}$ up to order $i_1+i_2+i_3=30$, suggests the following expressions
{\allowdisplaybreaks
\begin{align}
&H_{(0,0)}^{(n,0,0,1)}(\rho,S)=-\left[\frac{n E_2(\rho)}{6(1-Q_\rho^n)}+\frac{n^2 Q_\rho^n}{(1-Q_\rho^n)^2}\right]\,\phi_{-2,1}^3(\rho,m)-\frac{n}{12(1-Q_\rho^n)}\,\phi_{-2,1}^2(\rho,m)\phi_{0,1}(\rho,m)\,,\label{CoefsN3QR10}\\
&H_{(0,0)}^{(n,n,0,1)}(\rho,S)=-\left[\frac{n E_2(\rho)}{6(1-Q_\rho^n)}+\frac{n^2 }{(1-Q_\rho^n)^2}\right]\,\phi_{-2,1}^3(\rho,m)-\frac{n}{12(1-Q_\rho^n)}\,\phi_{-2,1}^2(\rho,m)\phi_{0,1}(\rho,m)\,,\\
&H_{(0,0)}^{(n_1+n_2,n_1,0,1)}(\rho,S)=-\left[\frac{n_2(n_2+2n_1)}{(1-Q_\rho^{n_1})(1-Q_\rho^{n_2})}+\frac{n_1^2-n_2^2 }{(1-Q_\rho^{n_1})(1-Q_\rho^{n_1+n_2})}\right]\,\phi_{-2,1}^3(\rho,m)\,.\label{CoefsN3QR1}
\end{align}}
These objects are not modular forms in any sense, such that the analysis of a finite number of coefficients (\emph{i.e.} such as $i_1+i_2+i_3=30$) is not sufficient to proof the relations above. However, since we find structures that (at least superficially) look very similar to those we have encountered in the previous section, we conjecture that (\ref{CoefsN3QR10})--(\ref{CoefsN3QR1}) are in fact correct. 
%%%%%%%%%%%%%%
\subsubsection{Root Lattices}
After having established a conjecture for the $H_{(0,0)}^{(i_1,i_2,i_3,1)}(\rho,S)$, the next step is to compute $B_{(0,0)}^{3,1}(\rho,\widehat{a}_1,\widehat{a}_2,S)$ defined in (\ref{ResummedHfunc}). We shall present two different approaches, generalising the computations for $N=2$ in sections~\ref{Sect:N2RootLattice} and \ref{Sect:N2MutliZeta} respectively. We shall start with the former, using summations over the root lattice of $\mathfrak{a}_2$.

The full contribution of $B_{(0,0)}^{3,1}(\rho,\widehat{a}_1,\widehat{a}_2,S)$ is given by
\begin{align}
B_{(0,0)}^{3,1}&(\rho,\widehat{a}_1,\widehat{a}_2,S)=\sum_{n=1}^\infty H_{(0,0)}^{(n,0,0,1)}(\rho,S)\left(Q_{\widehat{a}_1}^n+Q_{\widehat{a}_2}^n+\frac{Q_\rho^n}{Q_{\widehat{a}_1}^nQ_{\widehat{a}_2}^n}\right)\nonumber\\
&+\sum_{n=1}^\infty H_{(0,0)}^{(n,n,0,1)}(\rho,S)\left(Q_{\widehat{a}_1}^n Q_{\widehat{a}_2}^n+\frac{Q_\rho^n}{Q_{\widehat{a}_1}^n}+\frac{Q_\rho^n}{Q_{\widehat{a}_2}^n}\right)\nonumber\\
&+\sum_{n_1,n_2=1}^\infty H_{(0,0)}^{(n_1+n_2,n_1,0,1)}(\rho,S)\left(Q_{\widehat{a}_1}^{n_1+n_2} Q_{\widehat{a}_2}^{n_1}+\frac{Q_{\widehat{a}_1}^{n_2} Q_{\rho}^{n_1}}{Q_{\widehat{a}_2}^{n_1}}+\frac{Q_\rho^{n_1+n_2}}{Q_{\widehat{a}_1}^{n_1+n_2}Q_{\widehat{a}_2}^{n_2}}+(\widehat{a}_1\leftrightarrow \widehat{a}_2)\right)\,.
\end{align}
Here we choose to regroup $B_{(0,0)}^{3,1}$ into four different contributions
{\allowdisplaybreaks
\begin{align}
L_1&=\frac{\phi_{-2,1}^2}{12}\,(\phi_{0,1}-2 E_2\phi_{-2,1})\,\lambda_1\nonumber\\
&=\frac{\phi_{-2,1}^2}{12}\,(\phi_{0,1}-2 E_2\phi_{-2,1})\sum_{n=1}^\infty \frac{n}{1-Q_\rho^n}\big[\left(Q_{\widehat{a}_1}^n+Q_{\widehat{a}_2}^n+(Q_{\widehat{a}_1}Q_{\widehat{a}_2})^n\right)\nonumber\\
&\hspace{5.5cm}+Q_\rho^n \left(Q_{\widehat{a}_1}^{-n}+Q_{\widehat{a}_2}^{-n}+(Q_{\widehat{a}_1}Q_{\widehat{a}_2})^{-n}\right)\big]\,,\nonumber\\
L_2&=-\phi_{-2,1}^3\lambda_2=-\phi_{-2,1}^3\,\sum_{n=1}^\infty \frac{n^2}{(1-Q_\rho^n)^2}\left[(Q_{\widehat{a}_1}Q_{\widehat{a}_2})^n+Q_\rho^n \left(Q_{\widehat{a}_1}^n+Q_{\widehat{a}_2}^n+Q_{\widehat{a}_1}^{-n}+Q_{\widehat{a}_2}^{-n}\right)+Q_\rho^{2n}(Q_{\widehat{a}_1}Q_{\widehat{a}_2})^{-n}\right]\nonumber\\
%&\hspace{5.5cm}+Q_\rho^n \left(Q_{\widehat{a}_1}^{-n}+Q_{\widehat{a}_2}^{-n}+(Q_{\widehat{a}_1}Q_{\widehat{a}_2})^{-n}\right)\big]\,,\nonumber\\
%
L_3%&=-\phi_{-2,1}^3\lambda_3^{(1)}-\phi_{-2,1}^3\lambda_3^{(2)}\nonumber\\
&=-\phi_{-2,1}^3\sum_{n_{1,2}=1}^\infty\frac{1}{1-Q_\rho^{n_1}}\left[\frac{n_2(n_2+2n_1)}{1-Q_\rho^{n_2}}\,\left(Q_{\widehat{a}_1}^{n_1+n_2} Q_{\widehat{a}_2}^{n_1}+\frac{Q_\rho^{n_1+n_2}}{Q_{\widehat{a}_1}^{n_1+n_2}Q_{\widehat{a}_2}^{n_2}}\right)+\frac{n_1^2-n_2^2}{1-Q_\rho^{n_1+n_2}}\frac{Q_{\widehat{a}_1}^{n_2} Q_{\rho}^{n_1}}{Q_{\widehat{a}_2}^{n_1}}\right] \nonumber\\
&\phantom{=} -\phi_{-2,1}^3\sum_{n_{1,2}=1}^\infty\frac{1}{1-Q_\rho^{n_1}} \left[\frac{n_2(n_2+2n_1)}{1-Q_\rho^{n_2}}\frac{Q_{\widehat{a}_1}^{n_2} Q_{\rho}^{n_1}}{Q_{\widehat{a}_2}^{n_1}}+\frac{n_1^2-n_2^2}{1-Q_\rho^{n_1+n_2}}\,\left(Q_{\widehat{a}_1}^{n_1+n_2} Q_{\widehat{a}_2}^{n_1}+\frac{Q_\rho^{n_1+n_2}}{Q_{\widehat{a}_1}^{n_1+n_2}Q_{\widehat{a}_2}^{n_2}}\right)\right] \,,\nonumber\\[20pt]
L_4&=L_3\big|_{\widehat{a}_1\leftrightarrow \widehat{a}_2}\,.
\end{align}}
Following similar steps as in (\ref{ManipulationDivisor}), the summation appearing in $L_1$ can be re-arranged as\footnote{Throughout these manipulations we assume that the expressions are in fact convergent.} 
{\allowdisplaybreaks
\begin{align}
\lambda_1&=\sum_{n=1}^\infty n\left(Q_{\widehat{a}_1}^n+Q_{\widehat{a}_2}^n+(Q_{\widehat{a}_1}Q_{\widehat{a}_2})^n\right)\nonumber\\
&\hspace{1cm}+\sum_{n=1}^\infty n\sum_{k=1}^\infty Q_\rho^{nk}\left[Q_{\widehat{a}_1}^n+Q_{\widehat{a}_2}^n+(Q_{\widehat{a}_1}Q_{\widehat{a}_2})^n+Q_{\widehat{a}_1}^{-n}+Q_{\widehat{a}_2}^{-n}+(Q_{\widehat{a}_1}Q_{\widehat{a}_2})^{-n}\right]\nonumber\\
&=\sum_{n=1}^\infty n\sum_{\ell\in \Delta_{\mathfrak{a}_2}^+} e^{2\pi i n \ell}+\sum_{n=1}^\infty Q_\rho^n\sum_{k|n} k\sum_{\ell\in\Delta_{\mathfrak{a}_2}}e^{2\pi i k \ell}\,.
\end{align}}
Here $\Delta_{\mathfrak{a}_2}$ is the set of roots of $\mathfrak{a}_2$, while $\Delta_{\mathfrak{a}_2}$ is the set of positive roots of $\mathfrak{a}_2$, \emph{i.e.}
\begin{align}
&\Delta_{\mathfrak{a}_2}=\{\widehat{a}_1,\widehat{a}_2,\widehat{a}_1+\widehat{a}_2,-\widehat{a}_1,-\widehat{a}_2,-\widehat{a}_1-\widehat{a}_2\}\,,&&\text{and} &&\Delta_{\mathfrak{a}_2}=\{\widehat{a}_1,\widehat{a}_2,\widehat{a}_1+\widehat{a}_2\}\,.
\end{align}
In order to treat the summation in $L_2$, we use the following identities (with $\mathfrak{d}:=Q_\rho\frac{\partial }{\partial Q_\rho}$)
\begin{align}
\frac{n}{(1-Q_\rho^n)^2}&=\mathfrak{d}\left(\frac{Q_\rho^{-n}}{1-Q_\rho^n}\right)+\frac{n\,Q_\rho^{-n}}{1-Q_\rho^n}\,,&& \frac{n\,Q_\rho^n}{(1-Q_\rho^n)^2}=\mathfrak{d}\left(\frac{1}{1-Q_\rho^n}\right)\,,\nonumber\\
\frac{n\, Q_\rho^{2n}}{(1-Q_\rho^n)^2}&=\mathfrak{d}\left(\frac{Q_\rho^n}{1-Q_\rho^n}\right)-\frac{n\,Q_\rho^n}{1-Q_\rho^n}\,.\label{DdiffIdentitiesQ}
\end{align}
such that we have
{\allowdisplaybreaks
\begin{align}
\lambda_2=&\,\mathfrak{d}\left[\sum_{n=1}^\infty\frac{n}{1-Q_\rho^n}\left[Q_\rho^{-n}(Q_{\widehat{a}_1}Q_{\widehat{a}_2})^n+\left(Q_{\widehat{a}_1}^n+Q_{\widehat{a}_2}^n+Q_{\widehat{a}_1}^{-n}+Q_{\widehat{a}_2}^{-n}\right)+Q_\rho^{n}(Q_{\widehat{a}_1}Q_{\widehat{a}_2})^{-n}\right]\right]\nonumber\\
&+\sum_{n=1}^\infty\frac{n^2}{1-Q_\rho^n}\left[Q_\rho^{-n}(Q_{\widehat{a}_1}Q_{\widehat{a}_2})^n-Q_\rho^{n}(Q_{\widehat{a}_1}Q_{\widehat{a}_2})^{-n}\right]\nonumber\\
=&\,\mathfrak{d}\left[\sum_{n=1}^\infty n\left[ (Q_\rho^{-n}+1) (Q_{\widehat{a}_1}Q_{\widehat{a}_2})^n+ \left(Q_{\widehat{a}_1}^n+Q_{\widehat{a}_2}^n+Q_{\widehat{a}_1}^{-n}+Q_{\widehat{a}_2}^{-n}\right)\right]\right]\nonumber\\
&+\mathfrak{d}\left[\sum_{n=1}^\infty Q_\rho^n \sum_{k|d} k \left[(Q_{\widehat{a}_1}Q_{\widehat{a}_2})^k+Q_{\widehat{a}_1}^k+Q_{\widehat{a}_2}^k+Q_{\widehat{a}_1}^{-k}+Q_{\widehat{a}_2}^{-k}+(Q_{\widehat{a}_1}Q_{\widehat{a}_2})^{-k}\right]\right]\nonumber\\
&+\sum_{n=1}^\infty n^2\left[(Q_\rho^{-n}+1)(Q_{\widehat{a}_1}Q_{\widehat{a}_2})^n\right]+\sum_{n=1}^\infty Q_\rho^n\sum_{k|n} k^2\left[(Q_{\widehat{a}_1}Q_{\widehat{a}_2})^n-(Q_{\widehat{a}_1}Q_{\widehat{a}_2})^{-n}\right]
\end{align}}
where we have used similar steps as in (\ref{ManipulationDivisor}). We can further simplify the expression to 
{\allowdisplaybreaks
\begin{align}
\lambda_2=&\mathfrak{d}\left[\sum_{n=1}^\infty Q_\rho^n \sum_{k|d} k \sum_{\ell\in\Delta_{\mathfrak{a}_2}}e^{2\pi i k \ell}\right]+\sum_{n=1}^\infty Q_\rho^n\sum_{k|n} k^2\left[(Q_{\widehat{a}_1}Q_{\widehat{a}_2})^n-(Q_{\widehat{a}_1}Q_{\widehat{a}_2})^{-n}\right]
+\sum_{n=1}^\infty n^2\left[(Q_{\widehat{a}_1}Q_{\widehat{a}_2})^n\right]
\end{align}}
The summation appearing in $L_3$ and $L_4$ is more complicated and we shall not write the complete explicit expression for reasons of brevity. We remark however, that they can no longer be written as summations generalising the divisor sigma appearing in conventional Eisenstein series. Instead, applying similar methods as before, we find structures which are schematically of the form
\begin{align}
\sum_{u>1} Q_\rho^u\sum_{n_1>n_2>0}\sum_{n_1 k_1+n_2k_2=u\atop k_{1,2}>0}\,\sum_{\ell_1\in\Delta_{\mathfrak{a}_2}}\sum_ {\ell_2\in \Delta_{\mathfrak{a}_2}/ \{\ell_1,-\ell_1\}} e^{2\pi i k_1\ell_1+2\pi i k_2\ell_2}\,,\label{StrucGen3}
\end{align}
Notice in particular the appearance of two summations over (part of) the root lattice of $\mathfrak{a}_2$. Terms of this form are more akin to sigmas of length $2$, as introduced in (\ref{DefMultiDivisor}) for $\ell=2$. The full free energy $N=3$ at order $Q_R^1$ can therefore be treated more compactly using the generating functions defined in (\ref{GenFuncBracket}), as we shall do in the following subsubsection. It remains, however, an interesting question to speculate, whether the structures (\ref{StrucGen3}) are a $\mathfrak{a}_2$-generalisation of a large class of modular objects than the Eisenstein series. We will leave such speculations for future work.

%%%%%%%%%%%%%%%%%%%%%%%%%%%%%%%%%%%%%%%%
\subsubsection{Generating Functions of Multiple Divisor Sums}
In order to write the $B_{(0,0)}^{3,1}(\rho,\widehat{a}_1,\widehat{a}_2,S)$ in terms of the functions appearing in (\ref{GenFuncBracket}), we choose a slightly different decomposition:
\begin{align}
B_{(0,0)}^{3,1}(\rho,\widehat{a}_1,\widehat{a}_2,S)=F_1+F_2+F_3
\,.\label{ResummedN3R1}
\end{align}
with the explicit expressions
{\allowdisplaybreaks
\begin{align}
F_1&=\sum_{n=1}^\infty H_{(0,0)}^{(n,0,0,1)}(\rho,S)\left(Q_{\widehat{a}_1}^n+Q_{\widehat{a}_2}^n+\frac{Q_\rho^n}{Q_{\widehat{a}_1}^nQ_{\widehat{a}_2}^n}\right)\nonumber\\
&=-\frac{\phi_{-2,1}^3}{12}\bigg[2(E_2(\rho)+6\mathfrak{d})\left[D_{\widehat{a}_1}T(\widehat{a}_1-\rho;\rho)+D_{\widehat{a}_2}T(\widehat{a}_2-\rho;\rho)\right]\nonumber\\
&\hspace{2.5cm}-(D_{\widehat{a}_1}+D_{\widehat{a}_2})\left[3(D_{\widehat{a}_1}+D_{\widehat{a}_2})+E_2(\rho)+6\mathfrak{d}\right]T(-\widehat{a}_1-\widehat{a}_2;\rho)\bigg]\nonumber\\
&-\frac{\phi_{-2,1}^2\phi_{0,1}}{24}\left[2(D_{\widehat{a}_1}T(\widehat{a}_1-\rho;\rho)+D_{\widehat{a}_2}T(\widehat{a}_2-\rho;\rho))-(D_{\widehat{a}_1}+D_{\widehat{a}_2})T(-\widehat{a}_1-\widehat{a}_2;\rho)\right]\,,
\end{align}}
{\allowdisplaybreaks
\begin{align}
F_2&=\sum_{n=1}^\infty H_{(0,0)}^{(n,n,0,1)}(\rho,S)\left(Q_{\widehat{a}_1}^n Q_{\widehat{a}_2}^n+\frac{Q_\rho^n}{Q_{\widehat{a}_1}^n}+\frac{Q_\rho^n}{Q_{\widehat{a}_2}^n}\right)\nonumber\\
&=-\frac{\phi_{-2,1}^3}{12}\bigg[3(D_{\widehat{a}_1}+D_{\widehat{a}_2})(D_{\widehat{a}_1}+D_{\widehat{a}_2}-2\mathfrak{d})T(\widehat{a}_1+\widehat{a}_2-2\rho;\rho)+E_2(\rho)(D_{\widehat{a}_1}+D_{\widehat{a}_2})T(\widehat{a}_1+\widehat{a}_2-\rho;\rho)\nonumber\\
&\hspace{1cm}-2E_2(\rho)(D_{\widehat{a}_1}T(-\widehat{a}_1;\rho)+D_{\widehat{a}_2}T(-\widehat{a}_2;\rho))-12\mathfrak{d}(D_{\widehat{a}_1}T(-\widehat{a}_1-\rho;\rho)+D_{\widehat{a}_2}T(-\widehat{a}_2-\rho;\rho))\bigg]\nonumber\\
&-\frac{\phi_{-2,1}^2\phi_{0,1}}{24}\left[(D_{\widehat{a}_1}+D_{\widehat{a}_2})T(\widehat{a}_1+\widehat{a}_2-\rho;\rho)-2(D_{\widehat{a}_1}T(-\widehat{a}_1;\rho)+D_{\widehat{a}_2}T(-\widehat{a}_2;\rho))\right]\,,
\end{align}}
{\allowdisplaybreaks
\begin{align}
F_3&=\sum_{n_1,n_2=1}^\infty H_{(0,0)}^{(n_1+n_2,n_1,0,1)}(\rho,S)\left(Q_{\widehat{a}_1}^{n_1+n_2} Q_{\widehat{a}_2}^{n_1}+\frac{Q_{\widehat{a}_1}^{n_2} Q_{\rho}^{n_1}}{Q_{\widehat{a}_2}^{n_1}}+\frac{Q_\rho^{n_1+n_2}}{Q_{\widehat{a}_1}^{n_1+n_2}Q_{\widehat{a}_2}^{n_2}}+(\widehat{a}_1\leftrightarrow \widehat{a}_2)\right)\nonumber\\
&=-\phi_{-2,1}^3\bigg[\left(D_{\widehat{a}_2}^2-D_{\widehat{a}_1}^2\right) T(-\widehat{a}_2-\rho ,\widehat{a}_1-\rho;\rho)\nonumber\\
&\hspace{1cm}-D_{\widehat{a}_1}(D_{\widehat{a}_1}-2 D_{\widehat{a}_2})(T(\widehat{a}_1+\widehat{a}_2-2 \rho ,\widehat{a}_1-\rho;\rho)-T(-\widehat{a}_1-\rho ,-\widehat{a}_1-\widehat{a}_2;\rho))\nonumber\\
&\hspace{1cm}+(D_{\widehat{a}_1}T(-\widehat{a}_1;\rho))(D_{\widehat{a}_1}+D_{\widehat{a}_2})T(-\widehat{a}_1-\widehat{a}_2;\rho)+\frac{1}{4}T(-\widehat{a}_1;\rho)(D_{\widehat{a}_1}+D_{\widehat{a}_2})^2T(-\widehat{a}_1-\widehat{a}_2;\rho)\nonumber\\
&\hspace{1cm}-2(D_{\widehat{a}_2}T(-\widehat{a}_1;\rho))D_{\widehat{a}_1}T(\widehat{a}_1-\rho;\rho)+T(-\widehat{a}_1;\rho)D_{\widehat{a}_1}^2T(\widehat{a}_1-\rho;\rho)\nonumber\\
&\hspace{1cm}+((D_{\widehat{a}_1}+D_{\widehat{a}_2})T(\widehat{a}_1+\widehat{a}_2-\rho;\rho))D_{\widehat{a}_1}T(\widehat{a}_1-\rho;\rho)+T(\widehat{a}_1+\widehat{a}_2-\rho;\rho)D_{\widehat{a}_1}^2T(\widehat{a}_1-\rho;\rho)\bigg]\nonumber\\
&\hspace{0.5cm}+(\widehat{a}_1\leftrightarrow \widehat{a}_2)\,.
\end{align}}
Here we have used the generating function (\ref{GenFuncBracket}) along with the identities (\ref{DdiffIdentitiesQ}). Combining the terms in (\ref{ResummedN3R1}) we can write them as
{\allowdisplaybreaks
\begin{align}
B^{N=3,(1)}_{(0,0)}&(\rho,\widehat{a}_1,\widehat{a}_2,S)=-\frac{\phi_{-2,1}^2\phi_{0,1}}{24}(D_{\widehat{a}_1}+D_{\widehat{a}_2})\left[2\wprim (\widehat{a}_1;\rho)+2\wprim (\widehat{a}_2;\rho)+\wprim (\widehat{a}_1+\widehat{a}_2;\rho)\right]\nonumber\\
&-\frac{\phi_{-2,1}^3}{12}(D_{\widehat{a}_1}+D_{\widehat{a}_2})\bigg[E_2(\rho)\left[2\wprim (\widehat{a}_1;\rho)+2\wprim (\widehat{a}_2;\rho)+\wprim (\widehat{a}_1+\widehat{a}_2;\rho)\right]\nonumber\\
&\hspace{5cm}+6\,\left[2\tder (\widehat{a}_1;\rho)+2\tder (\widehat{a}_2;\rho)+\tder (\widehat{a}_1+\widehat{a}_2-\rho;\rho)\right]\nonumber\\
&\hspace{5cm}-3(D_{\widehat{a}_1}+D_{\widehat{a}_2})\left[T(-\widehat{a}_1-\widehat{a}_2;\rho)-T(\widehat{a}_1+\widehat{a}_2-2\rho;\rho)\right]\bigg]\nonumber\\
&-\frac{\phi_{-2,1}^3}{4}\bigg[4(D_{\widehat{a}_1}\wprim (\widehat{a}_1;\rho))(D_{\widehat{a}_1}+D_{\widehat{a}_2})\left[T(-\widehat{a}_1-\widehat{a}_2;\rho)-2T(\widehat{a}_2-\rho;\rho)\right]\nonumber\\
&\hspace{3cm}+\wprim (\widehat{a}_1;\rho)(D_{\widehat{a}_1}+D_{\widehat{a}_2})^2\left[T(-\widehat{a}_1-\widehat{a}_2;\rho)+4T(\widehat{a}_2-\rho;\rho)\right]\nonumber\\
&\hspace{3cm}+(D_{\widehat{a}_1}+D_{\widehat{a}_2})\left[(D_{\widehat{a}_1}T(\widehat{a}_1-\rho;\rho))T(\widehat{a}_1+\widehat{a}_2-\rho;\rho)\right]+(\widehat{a}_1\leftrightarrow \widehat{a}_2)\bigg]\nonumber\\
&+\phi_{-2,1}^3\bigg[D_{\widehat{a}_1}(D_{\widehat{a}_1}-2 D_{\widehat{a}_2})(T(\widehat{a}_1+\widehat{a}_2-2 \rho ,\widehat{a}_1-\rho;\rho)-T(-\widehat{a}_1-\rho ,-\widehat{a}_1-\widehat{a}_2;\rho))\nonumber\\
&\hspace{3cm}-\left(D_{\widehat{a}_2}^2-D_{\widehat{a}_1}^2\right) T(-\widehat{a}_2-\rho ,\widehat{a}_1-\rho;\rho)+(\widehat{a}_1\leftrightarrow \widehat{a}_2)\bigg]\,,\label{N3R1Tform}
\end{align}}
where we have defined 
\begin{align}
\tder (x;\rho):=\mathfrak{d}\left[T(x-\rho;\rho)-T(-x-\rho;\rho)\right]\,.
\end{align}
This expression can be understood as a generating functional of derivatives of Eisenstein series
\begin{align}
\tder (x;\rho)=\frac{1}{2\pi i}\sum_{k=1}^\infty \mathfrak{d}\,G_{2k}(\rho)\,x^{2k-1}\,.
\end{align}
Notice furthermore that \emph{e.g.}
\begin{align}
(D_{\widehat{a}_1}+D_{\widehat{a}_2})\left[T(-\widehat{a}_1-\widehat{a}_2;\rho)-2T(\widehat{a}_2-\rho;\rho)\right]=-2(T'(-\widehat{a}_1-\widehat{a}_2;\rho)+T'(\widehat{a}_2-\rho,\rho))\,,
\end{align}
where the prime signifies the derivative with respect to the first argument.

%%%%%%%%%%%%%%%%%%%%%%%%%%%
\subsubsection{Full Free Energy}
As in the case of $N=2$, it is an interesting question to combine the expression (\ref{N3R1Tform}) with $H_{(0,0)}^{(0,0,0,1)}(\rho,S)$, which was studied in \cite{Ahmed:2017hfr} (see also \cite{Companion1}). For the latter, the following expression was found
\begin{align}
H_{(0,0)}^{(0,0,0,1)}(\rho,S)&=-\frac{1}{8\cdot 24}\phi_{-2,1}(\rho,S)\left[\phi_{0,1}(\rho,S)+2E_2(\rho)\,\phi_{-2,1}(\rho,S)\right]\nonumber\\
&=-\frac{E_2^2\,\phi_{-2,1}^3}{48}-\frac{E_2\,\phi_{-2,1}^2\phi_{0,1}}{48}-\frac{\phi_{-2,1}\phi_{0,1}^2}{192}\,.
\end{align}
Studying and expansion of $B^{N=3,(1)}_{(0,0)}$ in (\ref{N3R1Tform}) in powers of $\widehat{a}_{1,2}$, we can extract the terms that contain the Eisenstein series $E_2(\rho)$. In this fashion we find
\begin{align}
B^{N=3,(1)}_{(0,0)}(\rho,\widehat{a}_1,\widehat{a}_2,S)=\left[\frac{E_2^2}{48}+\ldots\right]\,\phi_{-2,1}^3+\left[ E_2(\rho)\left(\frac{1}{48}+K(\rho,\widehat{a}_{1,2})\right)+\ldots\right]\,\phi_{-2,1}^2\phi_{0,1}\,.
\end{align}
Here the dots indicate contributions with lower\footnote{\emph{I.e.} order $(E_2)^1$ in the first bracket and order $(E_2)^0$ in the second.} powers of $E_2(\rho)$), while $K(\rho,\widehat{a}_{1,2})$ satisfies $K(\rho,\widehat{a}_{1,2}=0)=0$ and permits a series expansion in $\widehat{a}_{1,2}$, whose coefficients are modular forms (\emph{i.e.} polynomials in the Eisenstein series $(E_4,E_6)$). To leading order, we have for example
\begin{align}
K(\rho,\widehat{a}_{1,2})=\frac{\pi^2}{180}\,(\widehat{a}_1^2+\widehat{a}_1\widehat{a}_2+\widehat{a}_2^2)\,E_4(\rho)+\ldots\,.
\end{align}
Therefore, as in the case $N=2$, we see that the highest power in $E_2$ cancels between $B^{N=3,(1)}_{(0,0)}(\rho,\widehat{a}_1,\widehat{a}_2,S)$ and $H_{(0,0)}^{(0,0,0,1)}(\rho,S)$. Furthermore, the remaining non-holomorphicity is intimately coupled to $\widehat{a}_{1,2}$, \emph{i.e.} the remaining terms linear in $E_2(\rho)$ are accompanied by powers of $\widehat{a}_{1,2}$.

Finally, we have also verified explicitly 
\begin{align}
H_{(0,0)}^{(0,0,0,1)}(\rho,S)&+B_{(0,0)}^{3,1}(\rho,\tfrac{\rho}{3},S)=3H_{(0,0)}^{(0,1)}(\tfrac{\rho}{3},S)\,,
\end{align}
thus confirming the self-similarity first observed in \cite{Hohenegger:2016eqy}.

%%%%%%%%%%%%%%%%%%%%%%%%%%%
\subsection{Order $Q_R^2$}
To order $Q_R^2$, the expansion of the free energy is much more challenging to work out and we were only able to determine the first coefficients. Indeed, upon writing $g^{1,(n,3)}_{(0,0)}$
{\allowdisplaybreaks
\begin{align}
&H_{(0,0)}^{(n,0,0,2)}(\rho,S)=\phi_{-2,1}^2(\rho,m)\sum_{i=1}^5g^{i,(n,0,0),2}_{(0,0)}(\rho)\,\phi_{-2,1}^{5-i}(\rho,m)\phi_{0,1}^{i-1}(\rho,m)\,,&&\forall n\geq 1\,,\nonumber\\
&H_{(0,0)}^{(n,n,0,2)}(\rho,S)=\phi_{-2,1}^2(\rho,m)\sum_{i=1}^5g^{i,(n,n,0),2}_{(0,0)}(\rho)\,\phi_{-2,1}^{5-i}(\rho,m)\phi_{0,1}^{i-1}(\rho,m)\,,&&\forall n\geq 1\,,\nonumber\\
&H_{(0,0)}^{(n_1+n_2,n_1,0,2)}(\rho,S)=\phi_{-2,1}^2(\rho,m)\sum_{i=1}^5g^{i,(n_1+n_2,n_1,0),2}_{(0,0)}(\rho)\,\phi_{-2,1}^{5-i}(\rho,m)\phi_{0,1}^{i-1}(\rho,m)\,,&&\forall n_{1,2}\geq 1\,,\nonumber
\end{align}}
we have found the following patterns
\begin{align}
g^{5,(n,0,0),2}_{(0,0)}&=g^{5,(n,n,0),2}_{(0,0)}=-\frac{n}{4\cdot 24^3(1-Q_\rho^n)}+\left\{\begin{array}{lcl}0 & \text{if} & \text{gcd}(n,2)=1\,,\\[4pt] \frac{n}{12\cdot 24^3(1-Q_\rho^n)}& \text{if} & \text{gcd}(n,2)>1\,,\end{array}\right.\nonumber\\[14pt]
g^{5,(n_1+n_2,n_1,0),2}_{(0,0)}&=0\hspace{2cm}\forall n_{1,2}\in\mathbb{N}\,.
\end{align}
\begin{align}
g^{4,(n,0,0),2}_{(0,0)}&=\left\{\begin{array}{lcl}-\frac{n E_2+2n^3}{24^3(1-Q_\rho^n)}+\frac{n^2Q_\rho^n}{4\cdot 24^n(1-Q_\rho^n)^2} & \text{if} & \text{gcd}(n,2)=1 \\[4pt] -\frac{8nE_2+4n\psi_2+n^3}{12\cdot 24^3(1-Q_\rho^n)}+\frac{3n^2 Q_\rho^n}{4\cdot 24^2(1-Q_\rho^n)^2} & \text{if} & \text{gcd}(n,2)>1 \end{array}\right.\nonumber\\[10pt]
g^{4,(n,n,0),2}_{(0,0)}&=\left\{\begin{array}{lcl}-\frac{n E_2+2n^3}{24^3(1-Q_\rho^n)}+\frac{n^2}{4\cdot 24^n(1-Q_\rho^n)^2} & \text{if} & \text{gcd}(n,2)=1 \\[4pt] -\frac{8nE_2+4n\psi_2+n^3}{12\cdot 24^3(1-Q_\rho^n)}+\frac{2n^2 Q_\rho^n}{12\cdot 24^2(1-Q_\rho^n)^2} & \text{if} & \text{gcd}(n,2)>1 \end{array}\right.\nonumber\\[10pt]
g^{4,(n_1+n_2,n_1,0),2}_{(0,0)}&=\frac{1}{(1-Q_\rho^{n_2})}\left(\frac{(n_1^2-n_2^2)Q_\rho^{n_2}}{(1-Q_\rho^{n_1+n_2})}-\frac{n_1(n_1+2n_2)}{(1-Q_\rho^{n_1})}\right)\nonumber\\
&\hspace{1cm}\times  \left\{\begin{array}{lcl}6 & \text{if} & \text{gcd}(n_1,2)=1\text{ or } \text{gcd}(n_2,2)=1\,,\\[4pt] 4 & \text{if} & \text{gcd}(n_1,2)>1\text{ and }\text{gcd}(n_2,2)>1\,,\end{array}\right.\label{N3ResUnsum}
\end{align}
Higher terms (\emph{i.e.} the coefficients $g^{i,(\kappa_1,\kappa_2,0),2}_{(0,0)}$ for $i=1,2,3$) require an expansion of the free energy to much higher orders, which is currently out of computational reach. However, the results (\ref{N3ResUnsum}) display similar patterns as observed in the cases above.
%%%%%%%%%%%%%%%%%%%%%%%%%%%%%%%%%%%
%%%%%%%%%%%%%%%%%%%%%%%%%%%%%%%%%%%
\section{Case $N=4$}\label{Sect:CaseN4}
In this section we report the results of the analysis for the case $N=4$. Due to the complexity of the partition function (and subsequently also the free energy), the computations are rendered very difficult. Therefore, we limit ourselves to present the structure of the coefficients $f^{(s_1,s_2)}_{\ell_1,\ell_2,\ell_3,\ell_4,k,r}$ up to $\ell_1+\ell_2+\ell_3+\ell_4\leq 28$, which allows us to analyse the structure of $B^{N=4,(1)}_{(0,0)}(\rho,\widehat{a}_1,\widehat{a}_2,\widehat{a}_3,S)$ as a series expansion up to order $Q_\rho^5$ (or less, depending on the precise term). In the following we have determined analytic expressions matching this expansion but we cannot exclude that they do not receive corrections to higher order.

This expansion suggests the following form of the building blocks of the free energy
\begin{align}
&H_{(0,0)}^{(\ell_1,\ell_2,\ell_3,\ell_4,1)}(\rho,S)=\phi_{-2,1}^2(\rho,S)\sum_{i=1}^3 g^{i,(\ell_1,\ell_2,\ell_3,\ell_4,1)}_{(0,0)}(\rho)\,\phi_{-2,1}^{3-i}(\rho,S)\phi_{0,1}^{i-1}(\rho,S)\,,
\end{align}
where $(\ell_1,\ell_2,\ell_3,\ell_4)$ can be either of the following distinct configurations
{\allowdisplaybreaks
\begin{align}
&(n,0,0,0)&&\forall n\geq 1\nonumber\\
&(n,n,0,0)&&\forall n\geq 1\nonumber\\
&(n,0,n,0)&&\forall n\geq 1\nonumber\\
&(n,n,n,0)&&\forall n\geq 1\nonumber\\
&(n_1+n_2,n_1,0)&&\forall n_{1,2}\geq 1\nonumber\\
&(n_1+n_2,0,n_1,0)&&\forall n_{1,2}\geq 1\nonumber\\
&(n_1+n_2,n_1,n_1,0)&&\forall n_{1,2}\geq 1\nonumber\\
&(n_1+n_2,n_1,0,n_1)&&\forall n_{1,2}\geq 1\nonumber\\
&(n_1+n_2+n_3,n_1+n_2,n_1,0)&&\forall n_{1,2,3}\geq 1\nonumber\\
&(n_1+n_2+n_3,n_1,n_1+n_2,0)&&\forall n_{1,2,3}\geq 1\nonumber\\
&(n_1+n_2+n_3,n_1,0,n_1+n_2)&&\forall n_{1,2,3}\geq 1\,.\label{PatternsN4}
\end{align}}
By comparing to the series expansions, we found the following patterns
{\allowdisplaybreaks
\begin{align}
&g^{1,(n,0,0,0,1)}_{(0,0)}=\frac{-nE_2^2(\rho)}{72(1-Q_\rho^n)}-\frac{n^2 Q_\rho^n\,E_2(\rho)}{6(1-Q_\rho^n)^2}\,,&&g^{2,(n,0,0,0,1)}_{(0,0)}=-\frac{nE_2(\rho)}{72(1-Q_\rho^n)}-\frac{n^2 Q_\rho^n}{12(1-Q_\rho^n)^2}\,,\nonumber\\[4pt]
&g^{3,(n,0,0,0,1)}_{(0,0)}=-\frac{n}{288(1-Q_\rho^n)}\,,\nonumber\\[12pt]
%%%
&g^{1,(n,n,0,0,1)}_{(0,0)}=\frac{-nE_2^2(\rho)}{72(1-Q_\rho^n)}-\frac{n^2 (E_2(\rho)+Q_\rho^n)}{12(1-Q_\rho^n)^2}\,,&&g^{2,(n,n,0,0,1)}_{(0,0)}=\frac{3n^2 Q_\rho^n-nE_2(\rho)}{72(1-Q_\rho^n)}-\frac{n^2}{12(1-Q_\rho^n)^2}\,,\nonumber\\[4pt]
&g^{3,(n,n,0,0,1)}_{(0,0)}=-\frac{n}{288(1-Q_\rho^n)}\,.\nonumber\\[12pt]
%%%
&g^{1,(n,0,n,0,1)}_{(0,0)}=\frac{-4n^3\,Q_\rho^n}{(1-Q_\rho^n)^3}\,,&&g^{2,(n,0,n,0,1)}_{(0,0)}=g^{3,(n,0,n,0,1)}_{(0,0)}=0\,,\nonumber\\[12pt]
%%%
&g^{1,(n,n,n,0,1)}_{(0,0)}=\frac{-nE_2^2(\rho)}{72(1-Q_\rho^n)}-\frac{n^2 E_2(\rho)}{6(1-Q_\rho^n)^2}\,,&& g^{2,(n,n,n,0,1)}_{(0,0)}=\frac{-nE_2(\rho)}{72(1-Q_\rho^n)}-\frac{n^2}{12(1-Q_\rho^n)^2}\,,\nonumber\\[4pt]
&g^{3,(n,n,n,0,1)}_{(0,0)}=-\frac{n}{288(1-Q_\rho^n)}\,,\nonumber
\end{align}
%%%%%%%%%%
\begin{align}
&g^{1,(n_1+n_2,n_1,0,0)}_{(0,0)}=\frac{12(n_1n_2(n_1+n_2))-n_2(2n_1+n_2)E_2(\rho)}{12(1-Q_\rho^{n_1})(1-Q_\rho^{n_2})}+\frac{(n_2^2-n_1^2) E_2(\rho)}{12(1-Q_\rho^{n_1})(1-Q_\rho^{n_1+n_2})}\nonumber\\
&\hspace{1cm}-\frac{n_1n_2(n_1+n_2)}{(1-Q_\rho^{n_1})(1-Q_\rho^{n_2})(1-Q_\rho^{n_1+n_2})}\,,\nonumber\\[4pt]
&g^{2,(n_1+n_2,n_1,0,0)}_{(0,0)}=\frac{n_2^2-n_1^2}{24(1-Q_\rho^{n_1})(1-Q_\rho^{n_1+n_2})}-\frac{n_2(2n_1+n_2)}{24(1-Q_\rho^{n_1})(1-Q_\rho^{n_2})}\,,\hspace{1cm}g^{3,(n_1+n_2,n_1,0,0)}_{(0,0)}=0\,,\nonumber\\[12pt]
%%%
&g^{1,(n_1+n_2,0,n_1,0)}_{(0,0)}=\frac{-2n_1^2n_2}{(1-Q_\rho^{n_1})^2(1-Q_\rho^{n_1})}-\frac{2n_1n_2^2 Q_\rho^{n_2} }{(1-Q_\rho^{n_1})(1-Q_\rho^{n_2})^2}\,,\hspace{0.1cm}g^{2,(n_1+n_2,0,n_1,0)}_{(0,0)}=g^{3,(n_1+n_2,0,n_1,0)}_{(0,0)}=0\,,\nonumber\\[12pt]
%%%
&g^{1,(n_1+n_2,n_1,n_1,0)}_{(0,0)}=\frac{(n_2^2-n_1^2)E_2(\rho)}{12(1-Q_\rho^{n_1})(1-Q_\rho^{n_1+n_2})}-\frac{n_2(2n_1+n_2)E_2(\rho) }{12(1-Q_\rho^{n_1})(1-Q_\rho^{n_2})}\nonumber\\
&\hspace{1cm}-\frac{n_1n_2(n_1+n_2)Q_\rho^{n_2}}{(1-Q_\rho^{n_1})(1-Q_\rho^{n_2})(1-Q_\rho^{n_1+n_2})}\,.\label{ConfigRnnnn}\\[4pt]
&\hspace{0.4cm}g^{2,(n_1+n_2,n_1,n_1,0)}_{(0,0)}=\frac{-n_1(n_1+2n_2)}{24(1-Q_\rho^{n_1})(1-Q_\rho^{n_1+n_2})}-\frac{n_2(2n_1+n_2)Q_\rho^{n_2}}{24(1-Q_\rho^{n_2})(1-Q_\rho^{n_1+n_2})}\,,\hspace{0.4cm} g^{3,(n_1+n_2,n_1,n_1,0)}_{(0,0)}=0\,,\nonumber\\[12pt]
%%%
&g^{1,(n_1+n_2,n_1,0,n_1)}_{(0,0)}=\frac{-2n_1^2 n_2}{(1-Q_\rho^{n_1})^2(1-Q_\rho^{n_2})}-\frac{2n_1n_2^2\,Q_\rho^{n_2}}{(1-Q_\rho^{n_1})(1-Q_{\rho}^{n_2})^2}\,,\hspace{0.1cm}g^{2,(n_1+n_2,n_1,0,n_1)}_{(0,0)}=g^{3,(n_1+n_2,n_1,0,n_1)}_{(0,0)}=0\nonumber\\
%%%
&g^{1,(n_1+n_2+n_3,n_1+n_2,n_1,0)}_{(0,0)}=\frac{(n_1-n_2)n_3(n_1+n_2+n_3)}{(1-Q_\rho^{n_1})(1-Q_\rho^{n_2})(1-Q_\rho^{n_1+n_2+n_3})}-\frac{n_2n_3(2n_1+n_2+n_3)}{(1-Q_\rho^{n_1})(1-Q_\rho^{n_2})(1-Q_\rho^{n_3})}\nonumber\\
&\hspace{1cm}-\frac{n_2(n_1-n_3)(n_1+n_2+n_3)}{(1-Q_\rho^{n_1})(1-Q_\rho^{n_2})(1-Q_\rho^{n_1+n_2+n_3})}-\frac{(n_1-n_2)n_3(n_1+n_2+n_3)}{(1-Q_\rho^{n_1})(1-Q_\rho^{n_1+n_2})(1-Q_\rho^{n_3})}\,,\nonumber\\[4pt]
&g^{2,(n_1+n_2+n_3,n_1+n_2,n_1,0)}_{(0,0)}=g^{3,(n_1+n_2+n_3,n_1+n_2,n_1,0)}_{(0,0)}=0\,,\nonumber\\[12pt]
%%%
&g^{1,(n_1+n_2+n_3,n_1,n_1+n_2,0)}_{(0,0)}=\frac{n_1(n_1+n_2)(n_2+2n_3)}{(1-Q_\rho^{n_1})(1-Q_\rho^{n_1+n_2})(1-Q_\rho^{n_2+n_3})}-\frac{n_1n_3(n_1+2n_2+n_3)Q_\rho^{n_3}}{(1-Q_\rho^{n_1})(1-Q_\rho^{n_3})(1-Q_\rho^{n_2+n_3})}\nonumber\\
&\hspace{1cm}+\frac{(n_1+n_2)(n_1-n_2-n_3)n_3Q_\rho^{n_2+n_3}}{(1-Q_\rho^{n_1+n_2})(1-Q_\rho^{n_3})(1-Q_\rho^{n_2+n_3})}\,,\nonumber\\[4pt]
&g^{2,(n_1+n_2+n_3,n_1,n_1+n_2,0)}_{(0,0)}=g^{3,(n_1+n_2+n_3,n_1,n_1+n_2,0)}_{(0,0)}=0\,,\nonumber\\[12pt]
%%%
&g^{1,(n_1+n_2+n_3,n_1,0,n_1+n_2)}_{(0,0)}=\frac{n_1(n_3-n_1)(n_1-3n_2-n_3)Q_\rho^{2n_2+n_3}}{(1-Q_\rho^{n_1})^2(1-Q_\rho^{n_2})}-\frac{n_2(n_1+n_2)(n_1+2(n_2+n_3))Q_\rho^{n_2}}{(1-Q_\rho^{n_2})(1-Q_\rho^{n_1+n_2})(1-Q_\rho^{n_2+n_3})}\nonumber\\
&\hspace{1cm}+\frac{2(n_1-2n_3)(n_1-n_3)n_3Q_\rho^{n_2+2n_3}}{(1-Q_\rho^{n_1+n_2})(1-Q_\rho^{n_3})(1-Q_\rho^{n_1+n_2+n_3})}-\frac{(n_1+2n_2)(n_2+n_3)(n_1+n_2+n_3)}{(1-Q_\rho^{n_1+n_2})(1-Q_\rho^{n_2+n_3})(1-Q_\rho^{n_1+n_2+n_3})}\,,\nonumber\\[4pt]
&g^{2,(n_1+n_2+n_3,n_1,0,n_1+n_2)}_{(0,0)}=g^{3,(n_1+n_2+n_3,n_1,0,n_1+n_2)}_{(0,0)}=0\,.
\end{align}}
These coefficients show various new structures. While we do not attempt to explicitly convert all of them into generating functions of brackets, we remark that sum may require to consider infinite series of the latter. Indeed, consider for example the structure
\begin{align}
\mathcal{Q}(X_1,X_2)=\sum_{n_1,n_2>0}\frac{X_1^{n_1} X_2^{n_2}\,Q_\rho^{n_2}}{(1-Q_\rho^{n_1})(1-Q_\rho^{n_2})(1-Q_\rho^{n_1+n_2})}\,,\label{NewTermM1}
\end{align}
which appears for example in the resummation of the configuration $(n_1+n_2,n_1,n_1,0)$ in (\ref{ConfigRnnnn}) and $X_{1,2}$ may be suitable combinations of $Q_{\widehat{a}_{1,2}}$ and $Q_\rho$ with $X_{1,2}=e^{2\pi i x_{1,2}}$. The expression (\ref{NewTermM1}) cannot readily be written as a linear combination of generating functions of the type (\ref{GenFuncBracket}). However, we can write it in the form
\begin{align}
\mathcal{Q}(X_1,X_2)&=\sum_{n=0}^\infty\sum_{n_1,n_2>0}\frac{X_1^{n_1}X_2^{n_2}Q_\rho^{(n+1)n_2}}{(1-Q_\rho^{n_1})(1-Q_{\rho}^{n_1+n_2})}=\sum_{n=0}^\infty\sum_{n_1,n_2>0}\frac{(X_1\,Q^{-2}_\rho)^{n_1}(X_2 \,Q_\rho^n)^{n_2}\,Q_\rho^{2n_1+n_2}}{(1-Q_\rho^{n_1})(1-Q_{\rho}^{n_1+n_2})}\nonumber\\
&=\sum_{n=0}^\infty T(x_1-2\rho,x_2+n\rho;\rho)\,.
\end{align}
%%%%%%%%%%%%%%%%%%%%%%%%%%%%%%%%%%%
%%%%%%%%%%%%%%%%%%%%%%%%%%%%%%%%%%%
\section{Conclusions and Outlook}\label{Sect:Conclusion}
Based on instanton expansions in a number of examples, we have presented in this paper numerous intriguing patterns in the free energy of a class of six-dimensional little string theories of A-type. We have in particular shown that all our examples permit a decomposition in terms of generating functions of multiple divisor sums, first introduced in \cite{Bachmann:2013wba}, which we conjecture to be true in general. 

The results and conjectures presented in this work serve as another indication of the importance of symmetries in the study of the network of dual gauge theories engineered from the toric Calabi-Yau threefolds $X_{N,1}$. In the companion paper \cite{Companion1}, we have focused on the so-called reduced free energy $H_{(s_1,0)}^{(0,\ldots,0,r)}(\rho,S)$ and have unraveled more of the group structure acting on the particular sector of the BPS spectrum captured by it. The current work deals mostly with the complement of this part of the spectrum, \emph{i.e.} the remaining part of the free energy. We have shown that the latter also has numerous interesting new and unexpected structures and we have shown a class of functions (namely the generating functions of multiple divisor sigmas), which is adapted to capture the latter in a natural fashion. In the simplest case $N=2$, the latter is related to so-called generalised Eisenstein series which in turn are related to the elliptic Weierstrass function. 

The appearance of these functions is very interesting when one considers the case $N=2$ as a generalisation of $N=1$. As was argued in \cite{,Bastian:2018jlf}, the latter has as automorphism group $Sp(4,\mathbb{Z})$, which contains two $SL(2,\mathbb{Z})$ groups (called $SL(2,\mathbb{Z})_R$ and $SL(2,\mathbb{Z})_\rho$, since they act on $R$ and $\rho$ as modular parameters respectively) in a natural fashion. In the case of $N=2$, one of these ($SL(2,\mathbb{Z})_R$ in our conventions) remains more or less intact, while the second one gets modified due to the introduction of the root $\widehat{a}_1$ of an $SU(2)$ gauge group\footnote{Notice that the Weyl group $\text{Dih}_2$ of the latter is in fact part of $\widetilde{\mathbb{G}}(2)$.} in addition to the original modular parameter $\rho$. The functions we found in section~\ref{Sect:CaseN2}, notably the Eisenstein series $\mathcal{E}_{2k}$ combine the two, in a natural fashion, making both the modular properties, as well as the $SU(2)$ character of the free energy manifest. 

For the future it would be interesting to study the patterns we have found in this work from a more group theoretic perspective and analyse the full automorphism group of the BPS spectrum of the theories engineered from $X_{N,1}$. Notably the relation (\ref{HeckeSpirit}), which connects the coefficients that appear in the building blocks of the free energy to different orders in $Q_R$, is very similar in spirit to a Hecke type of relation appearing in the reduced free energy $H_{(s_1,0)}^{(0,\ldots,0,r)}(\rho,S)$. This could be seen as an indication that the whole free energy at order $Q_R^r$ can be understood as some operator acting on a certain seed function. In this regard, it would also be interesting to understand, what physical role is played by the algebra of the generating functions $T(\widehat{a}_1,\ldots,\widehat{a}_N;\rho)$ (which form the general building blocks of the free energy) that was uncovered in \cite{Bachmann:2013wba}. We leave these questions for future work. 

Furthermore, the generating functions of multiple divisor sums are also intimately related to multiple zeta values \cite{Bachmann:2013wba,BachmannMaster}. The latter have in recent years appeared in many other branches of physics, notably the study of scattering amplitudes in (supersymmetric) string theories (see \cite{Schlotterer:2012ny,Broedel:2013aza,Broedel:2013tta,Stieberger:2013wea,Broedel:2014vla,Broedel:2015hia,DHoker:2015wxz,MatthesPhD,Brown1,Brown2,Broedel:2018izr,Zerbini:2018sox,Zerbini:2018hgs} for recent work). While the latter are in general perturbative quantities, the appearance of the same type of objects in the instanton partition function (and the corresponding free energy), which is an inherently non-perturbative object, might indicate a more profound role of multiple zeta values in gauge theories also at the non-perturbative level. It will be interesting in the future to study, what type of lessons we can take away from this connection \cite{HoheneggerS}. 

Finally, based on earlier results \cite{Bastian:2018jlf} in this paper we have restricted ourselves to theories engineered from the toric Calabi-Yau manifolds $X_{N,1}$. The latter are part of the larger class of manifolds $X_{N,M}$, for which the free energy has also been studied in detail. It would be interesting to repeat the current analysis for these cases. Similarly it would be even more interesting to analyse theories that are engineered by parallel M5-branes probing various different orbifold backgrounds. Due to the similar nature of these theories one might expect to find the same type of patterns as described in the current work.

%%%%%%%%%%%%%%%%%%%%%%%%%%%%%%%%%%%
%%%%%%%%%%%%%%%%%%%%%%%%%%%%%%%%%%%
\section*{Acknowledgements}
We are very thankful to A.~Iqbal and S.J.~ Rey for many stimulating discussions and earlier collaborations on related projects. In particular, we are grateful to A.~Iqbal, for a reading of the manuscript prior to publication. SH would like to thank the organisers of the GGI workshop 'String Theory from a Worldsheet Perspective' (Galileo Galilei Institute, May 2019) and the CERN TH institute 'Topological String Theory and Related Topics' (CERN, June 2019) for creating a productive atmosphere, while part of the computations for this article were being carried out. SH would also like to thank O.~Schlotterer and P.~Vanhove for useful discussions and exchanges. 

%%%%%%%%%%%%%%%%%%%%%%%%%%%%%%%%%%%
%%%%%%%%%%%%%%%%%%%%%%%%%%%%%%%%%%%
\appendix
\section{Modular Objects}\label{App:ModularReview}
\subsection{(Quasi-)Jacobi Forms}
\subsubsection{Jacobi Forms}
For completeness, we recall the definition of a Jacobi form of weight $w$ and index $m$ as a function\footnote{Here $\mathbb{H}$ s the upper half-plane.}
\begin{align}
\phi:\,\,\mathbb{H}\times\mathbb{C}&\longrightarrow\mathbb{C}\,\nonumber\\
(\rho,z)&\longmapsto \phi(\rho;z)
\end{align}
which satisfies (for $\Gamma\subset SL(2,\mathbb{Z})$ some finite-index subgroup of the modular group)
\begin{align}
\phi\left(\frac{a\rho+b}{c\rho+d};\frac{z}{c\rho+d}\right)&=(c\rho+d)^w\,e^{\frac{2\pi i m c z^2}{c\tau+d}}\,\phi(\tau;z)\,,&&\forall\,\left(\begin{array}{cc}a & b \\ c & d\end{array}\right)\in\Gamma\,,\nonumber\\
\phi(\rho;z+\ell_1 \rho+\ell_2)&=e^{-2\pi i m(\ell_1^2\rho+2\ell_1 z)}\,\phi(\rho;z)\,,&&\forall\,\ell_{1,2}\in\mathbb{N}\,,\label{JacobiFormGen}
\end{align}
and which allows for a Fourier expansion of the form
\begin{align}
&\phi(z,\rho)=\sum_{n= 0}^\infty\sum_{\ell\in\mathbb{Z}}c(n,\ell)\,Q_\rho^n\,e^{2\pi i z \ell}\,,&&\text{with} &&c(n,\ell)=(-1)^w c(n,-\ell)\,.
\end{align}
As examples of Jacobi forms of index $1$ and weight $0$ and $-2$ respectively, we introduce\footnote{For the convenience of some numerical factors appearing in various expressions, these definitions differ by numerical prefactors ($2$ and $-1$ respectively) from those in the literature.}
\begin{align}
&\phi_{0,1}(\rho,z)=8\sum_{a=2}^4\frac{\theta_a^2(z;\rho)}{\theta_a^2(0,\rho)}\,,&&\text{and}&&\phi_{-2,1}(\rho,z)=\frac{\theta_1^2(z;\rho)}{\eta^6(\rho)}\,,
\end{align}
where $\theta_{a=1,2,3,4}(z;\rho)$ are Jacobi theta functions and $\eta(\rho)$ is the Dedekind eta function.

The space $J_{w,m}(\Gamma)$ of Jacobi forms of weight $w$ and index $m$, is finite dimensional, which means we can decompose any Jacobi form of index $m$ and weight $k$ through the choice of a suitable basis. The decompositions relevant in this work are of the following form: let $\phi$ be a  Jacobi form of weight $w\in\mathbb{N}_{\text{even}}$ and index $m>1$, then
\begin{align}
\phi(\rho,z)=\sum_{u=1}^{m+1}g^{u}(\rho)\,\phi_{-2,1}^{m+1-u}(\rho,z)\,\phi_{0,1}^{u-1}(\rho,z)\,,\label{ExpansionJacobiPhi}
\end{align}
where $g^{u}(\rho)$ is a modular form of weight $w+2(m+1-u)$. The latter can in turn be written as a homogeneous polynomial in the \emph{Eisenstein} series, which are defined by
\begin{align}
&E_{2k}(\rho)=1-\frac{4k}{B_{2k}}\sum_{n=1}^\infty \sigma_{2k-1}(n)\,Q_\rho^n\,,&&\forall\,k\in\mathbb{N}\,.\label{DefEisenStein}
\end{align}
where $B_{2k}$ are the Bernoulli numbers. We also introduce
\begin{align}
G_{2k}(\rho)=2\zeta(2k)+2\frac{(2\pi i)^{2k}}{(2k-1)!}\sum_{n=1}^\infty\sigma_{2k-1}(n)\,Q_\rho^n=2\zeta(2k)E_{2k}(\rho)\,,\label{NormEisenstein}
\end{align}
along with Weierstrass's elliptic function
\begin{align}
\wp(z;\rho)=\frac{1}{z^2}+\sum_{k=1}^\infty(2k+1)G_{2k+2}(\rho)\,z^{2k}\,.\label{DefWeierstrass}
\end{align}
%%%%%%%%%%%%%%%%%%%%%%%%%%%%%%%%%%%%%%%%%
%%%%%%%%%%%%%%%%%%%%%%%%%%%%%%%%%%%%%%%%%

\subsubsection{Quasi-Jacobi Forms}\label{Sect:QuasiJacobiForms}
Notice in the definition (\ref{DefEisenStein}), $E_2$ is not a modular form, but transforms with a shift-term
\begin{align}
&E_2\left(\frac{a\rho+b}{c\rho+d}\right)=(c\rho+d)^2\, E_2(\rho)-\frac{6i}{\pi}\,\frac{c}{c\rho+d}\,,&&\forall\,\left(\begin{array}{cc}a & b \\ c & d\end{array}\right)\in SL(2,\mathbb{Z})\,.
\end{align} 
Allowing the $g^u$ in (\ref{ExpansionJacobiPhi}) to also depend on $E_2$, in fact leads to a generalisation of Jacobi forms, called \emph{quasi-Jacobi forms}. To define the latter more rigorously, we follow \cite{Libgober} and first define an \emph{almost meromorphic Jacobi form}. Let
\begin{align}
&\lambda(z,\rho)=\frac{z-\bar{z}}{\rho-\bar{\rho}}\,,&&\text{and}&&\mu(\rho)=\frac{1}{\rho-\bar{\rho}}\,,
\end{align}
then an almost meromorphic Jacobi form of weight $w$, index zero and depth $(s,t)$ is a meromorphic function, which satisfies (\ref{JacobiFormGen}) for $m=0$ and which has degree $s$ in $\lambda$ and $t$ in $\mu$. A \emph{quasi-Jacobi form} $\psi(\rho;z)$ of depth $(s,t)$ is the constant term of an almost meromorphic Jacobi form of index zero and depth $(s,t)$ when considered as a polynomial in $\lambda$ and $\mu$. Under modular and elliptic transformations, it schematically behaves in the following manner (for $\left(\begin{array}{cc}a & b \\ c & d\end{array}\right)\in SL(2,\mathbb{Z})$ and $\ell_{1,2}\in\mathbb{Z}$)
\begin{align}
(c\rho+d)^{-w}\,\psi\left(\frac{a\rho+b}{c\rho+d};\frac{z}{c\rho+d}\right)&=\sum_{i=0}^s\sum_{j=0}^t S_{ij}(\psi)(\rho,z)\left(\frac{cz}{c\rho+d}\right)^i\left(\frac{cz}{c\rho+d}\right)^j\,,\nonumber\\
\psi(\rho;z+\ell_1\rho+\ell_2)&=\sum_{i=0}^s T_i(\psi)(\rho;z)\,\ell_1^i\,.
\end{align}
As discussed in \cite{Libgober}, the algebra of quasi-Jacobi forms is the algebra of functions on $\mathbb{H}\times \mathbb{C}$ generated by functions $\mathcal{E}_n(z;\rho)$ and $E_2$, where $\mathcal{E}_n$ is a class of generalised Eisenstein series that we shall discuss in the following subsection.

%%%%%%%%%%%%%%%%%%%%%%%%%%%%%%%%%%%
\subsection{Generalised Eisenstein Series}\label{App:WeilEisen}
In this appendix we recall parts of the review \cite{Weil}. Let $z\in\mathbb{C}$ and $n\in\mathbb{N}$
\begin{align}
e_n(z)=\sum_{\mu=-\infty}^\infty(z+\mu)^{-n}\,.\label{DefTrigWeil}
\end{align} 
$e_n$ is absolut convergent, while for $n=1$, the summation should be understood as the following \emph{Eisenstein summation}
\begin{align}
e_1=\lim_{M\to \infty}\sum_{\mu=-M}^M\frac{1}{z+\mu}=\frac{d}{dz}\,\log(\sin\pi z)\,.\label{e1Def}
\end{align}
The functions (\ref{DefTrigWeil}) satisfy the following recursive relation
\begin{align}
&\frac{d e_n}{dz}=-n\,e_{n+1}\,,&&\forall n\geq 1\,,\label{RecursionSmalle}
\end{align}
such that in particular $e_2(z)=\frac{\pi^2}{\sin^2(\pi z)}$.

The definition (\ref{DefTrigWeil}) can be generalised according to \cite{Weil} in the following manner: let $W$ be a lattice in the complexe plane, generated by two complex numbers $(u,v)$ that satisfy $v/u=\delta \rho$ (with $\delta=\pm 1$). The generalised Eisenstein series is defined as
\begin{align}
\mathcal{E}_n(z;\rho)=\sum_{w\in W}(z+w)^{-n}\,.\label{GenEisenstein}
\end{align}
We remark that throughout the main body of this paper, we prefer to work with the specific choice of generators
\begin{align}
&u=1\,,&&v=\rho\,,&&\delta=+1\,.
\end{align}
Originally studied by Eisenstein, (\ref{GenEisenstein}) is absolutely convergent for $n\geq 3$, while for $n\in\{1,2\}$, the summation is supposed to be understood as
\begin{align}
\sum_{w\in W}(z+w)^{-n}=\lim_{N\to \infty}\sum_{\nu=-N}^N\left(\lim_{M\to\infty}\sum_{\mu=-M}^M(z+\mu u+\nu v)^{-n}\right)\,.\label{DefEisensteinPrescription}
\end{align} 
Throughout the main body of this paper, we will always assume that this Eisenstein summation is used whenever appropriate. The $\mathcal{E}_n$ satisfy the following recursive relation
\begin{align}
\frac{d \mathcal{E}_n}{dz}=-n\,\mathcal{E}_{n+1}\,.\label{RecursiveGenEisenstein}
\end{align}
A particular Fourier expansion of $\mathcal{E}_n$ \cite{Weil} can be given for $n\in\mathbb{N}$
\begin{align}
\mathcal{E}_n(z;\rho)=u^{-n}\sum_{\nu=-M}^{+M}e_n\left(\frac{z+\nu v}{u}\right)+\frac{(2\pi/ iu)^n}{(n-1)!}\sum_{\nu=M+1}^\infty\sum_{d=1}^\infty d^{n-1}\,Q_\rho^{\nu d}\left[e^{2\pi i z d}+(-1)^n\,e^{-2\pi i z d}\right]\,,\label{SumGenEisenFourier}
\end{align}
where $M$ has to be chosen such that $|Q_\rho^{M+1} e^{2\pi i z}|<1$ and $|Q_\rho^{M+1} e^{-2\pi i z}|<1$. Focusing on the case $|Q_\rho|< |e^{2\pi i z}|<|Q_\rho|^{-1}$, one can choose $M=0$. Furthermore, $\mathcal{E}_2$ can also be expressed in terms of the Weierstrass function $\wp$ (defined in (\ref{DefWeierstrass}))
\begin{align}
\mathcal{E}_2(z;\rho)=G_2(\rho)+\sum_{w\in W}{}' \left[\frac{1}{(z+w)^2}-\frac{1}{w^2}\right]=G_2(\rho)+\wp(z;\rho)\,,\label{GenEisenWeierstrass}
\end{align}
where the summation does not include the origin of $W$.
%%%%%%%%%%%%%%%%%%%%%%%%%%%%%%%%%%%
%%%%%%%%%%%%%%%%%%%%%%%%%%%%%%%%%%%
\section{Generating Functions of Multiple Divisor Sums}\label{App:MultiDivisorSums}
Following \cite{Bachmann:2013wba} we define the {\it multiple divisor sum}
\begin{align}
&\sigma_{r_1,\ldots,r_\ell}(n)=\sum_{{u_1v_1+\ldots +u_\ell v_\ell=n}\atop{u_1>\ldots>u_\ell>0}}v_1^{r_1}\ldots v_\ell^{r_\ell}\,,&&\text{for} &&\begin{array}{l}r_1,\ldots,r_\ell\in \mathbb{N}\cup \{0\} \\ \ell,n\in\mathbb{N}\end{array}\label{DefMultiDivisor}
\end{align}
whose generating function
\begin{align}
&[s_1,\ldots,s_\ell;\rho]=\frac{1}{(s_1-1)!\ldots (s_\ell-1)!}\sum_{n>0}\sigma_{s_1-1,\ldots, s_\ell-1}(n)\,Q_\rho^n\,,&&\text{for} &&s_1,\ldots,s_\ell\in\mathbb{N}\,,\label{DefBracket}
\end{align}
is called a {\it bracket} of length $\ell$. The generating function of brackets was defined in \cite{Bachmann:2013wba} as follows
\begin{align}
T(x_1,\ldots,x_\ell;\rho)&=\sum_{s_1,\ldots,s_\ell>0}[s_1,\ldots,s_\ell;\rho] (2\pi ix_1)^{ s_1-1}\ldots (2\pi i x_\ell)^{ s_\ell-1}=\sum_{n_1,\ldots,n_\ell>0}\prod_{j=1}^\ell\frac{e^{2\pi i n_j x_j} \nome^{n_1+\ldots +n_j}}{1-\nome^{n_1+\ldots+n_j}}\,,\label{GenFuncBracket}
\end{align}
for $x_{1,\ldots,\ell}\in\mathbb{R}$. An equivalent expression was given in \cite{Bachmann:2013wba} by re-writing the denominator of (\ref{GenFuncBracket}) as a series expansion
\begin{align}
T(x_1,\ldots,x_\ell;\rho)=\sum_{n_1,\ldots,n_\ell>0}\prod_{j=1}^\ell e^{2\pi i n_j x_j}\sum_{v_j>0}Q_\rho^{v_j(n_1+\ldots+n_j)}\,.\label{TEquivalent}
\end{align}
%%%%%%%%%%%%%%%%%%%%%%%%%%%
While the $T(x_1,\ldots,x_\ell;\rho)$ are in general not modular objects, it was shown in \cite{Bachmann:2013wba} that the quasi-modular forms constitute a subalgebra of the algebra $\mathcal{MD}$ spanned by the brackets $[s_1,\ldots,s_\ell;\rho]$. A property we are interest in is the generalisation of elliptic transformations for the $T(x_1,\ldots,x_\ell;\rho)$: we first consider a shift of all arguments of $T(x_1,\ldots,x_\ell;\nom)$ (for $\ell\in\mathbb{N}$) of the following form
\begin{align}
T(x_1-\nom,\ldots,x_\ell-\nom;\nom)&=\sum_{n_1,\ldots,n_\ell>0}\nome^{-n_1-\ldots-n_\ell}\prod_{j=1}^\ell\frac{e^{2\pi i n_j x_j} \nome^{n_1+\ldots +n_j}}{1-\nome^{n_1+\ldots+n_j}}\nonumber\\
&=\sum_{n_1,\ldots,n_\ell>0}\prod_{j=1}^{\ell-1}\frac{e^{2\pi i n_j x_j} \nome^{n_1+\ldots +n_j}}{1-\nome^{n_1+\ldots+n_j}}\,e^{2\pi i n_\ell x_\ell}\left[\frac{\nome^{n_1+\ldots+n_\ell}}{1-\nome^{n_1+\ldots+n_\ell}}+1\right]\nonumber\\
&=T(x_1,\ldots,x_\ell;\nom)+\sum_{n_1,\ldots,n_{\ell-1}>0}\prod_{j=1}^{\ell-1}\frac{e^{2\pi i n_j x_j} \nome^{n_1+\ldots +n_j}}{1-\nome^{n_1+\ldots+n_j}}\sum_{n_\ell>0}e^{2\pi i n_\ell x_\ell}\nonumber\\
&=T(x_1,\ldots,x_\ell;\nom)-\frac{1}{1-e^{-2\pi i x_\ell}}\,T(x_1,\ldots,x_{\ell-1};\nom)\,.\label{EllipticTrafoAllArgs}
\end{align}
Here we have implicitly used the convention that $T(x_1,\ldots,x_\ell;\nom)\big|_{\ell=0}=1$. Notice, that the factor $\frac{1}{1-e^{2\pi i x_\ell}}$ in (\ref{EllipticTrafoAllArgs}) is in fact the generating function of zeta-values in the sense
\begin{align}
\frac{1}{1-e^{-2\pi ix_\ell}}=\frac{1}{2}+\frac{i}{\pi}\sum_{n=0}^\infty \zeta(2n) x_\ell^{2n-1}\,.\label{GeneratingFunctionaleta}
\end{align}
more general shifts of the arguments of $T(x_1,\ldots,x_\ell;\nom)$ lead to more complicated expressions, which, however, in general again contain (parts of) generating functions of zeta values as well as $T$ of lower length. As an example we present the case $\ell=2$
\begin{align}
T(x_1-\nom,x_2;\nom)&=\sum_{n_1,n_2>0}\frac{e^{2\pi i (n_1x_1+ n_2 x_2)}\,\nome^{n_1+n_2}}{1-\nome^{n_1+n_2}}\left[\frac{\nome^{n_1}}{1-\nome^{n_1}}+1\right]\nonumber\\
&=T(x_1,x_2;\nom)+\sum_{n_1,n_2>0}\frac{e^{2\pi i (n_1x_1+ n_2 x_2)}\nome^{n_1+n_2}}{1-\nome^{n_1+n_2}}\,.\label{RelationLength2}
\end{align} 
The last term can be rewritten as
\begin{align}
&\sum_{n_1,n_2>0}\frac{e^{2\pi i (n_1x_1+ n_2 x_2)}\nome^{n_1+n_2}}{1-\nome^{n_1+n_2}}=\sum_{n>0}e^{2\pi i x_1 n}\sum_{N=n+1}^\infty\frac{e^{2\pi i x_2 (N-n)}\nome^N}{1-\nome^N}\nonumber\\
&\hspace{1cm}=-\frac{1}{1-e^{-2\pi i(x_1-x_2)}}T(x_2;\nom)-\sum_{n=1}^\infty e^{2\pi i n(x_1-x_2)}\sum_{N=1}^n\frac{e^{2\pi i N x_2}\nome^N}{1-\nome^N}\,.
\end{align}
Here the last term is in fact part of the generating function of brackets of length 1. We can furthermore write it in the following fashion
\begin{align}
&\sum_{n=1}^\infty e^{2\pi i n(x_1-x_2)}\sum_{N=1}^n\frac{e^{2\pi i N x_2}\nome^N}{1-\nome^N}=\sum_{n=1}^\infty e^{2\pi i n(x_1-x_2)}\sum_{k=1}^\infty\sum_{N=1}^n e^{2\pi i N x_2}\,Q_\rho^{Nk}\nonumber\\
&=\sum_{n=1}^\infty e^{2\pi i n(x_1-x_2)}\sum_{k=1}^\infty e^{2\pi i x_2}Q_\rho^k\,\frac{1-(e^{2\pi i x_2}Q_\rho^k)^n}{1- e^{2\pi i x_2}Q_\rho^k}=\sum_{n=1}^\infty e^{2\pi i n(x_1-x_2)}\sum_{k=1}^\infty e^{2\pi i x_2}Q_\rho^k\,[n]_{e^{2\pi i x_2}Q_\rho^k}\,.
\end{align}
Here we have defined the $q$-analog of a non-negative integer $n$ \cite{Bradley}
\begin{align}
[n]_q:=\sum_{k=0}^{n-1}\sum_{k=0}^{n-1}q^k=\frac{1-q^n}{1-q}\,,&&\forall n\in\mathbb{N}&&\text{and} &&|q|\in [0,1]\,.
\end{align}
Summarising we can therefore write
\begin{align}
T(x_1-\nom,x_2;\nom)&=T(x_1,x_2;\nom)-\frac{1}{1-e^{-2\pi i(x_1-x_2)}}T(x_2;\nom)-\sum_{n=1}^\infty e^{2\pi i n(x_1-x_2)}\sum_{k=1}^\infty e^{2\pi i x_2}Q_\rho^k\,[n]_{e^{2\pi i x_2}Q_\rho^k}\,.\label{RelationLength2f}
\end{align}
Notice, that the relation (\ref{EllipticTrafoAllArgs}) as well as (\ref{RelationLength2f}) allow to treat all cases with $N=3$ in section~\ref{Sect:CaseN3}.
%%%%%%%%%%%%%%%%%%%%%%%%%%%%%%%%%%%
%%%%%%%%%%%%%%%%%%%%%%%%%%%%%%%%%%%
\section{Coefficients}
In this appendix we collect several explicit coefficients in the decomposition of the free energy that were too lengthy to be displayed in the main body of the paper.
%%%%%%%%%%%%
\subsection{Contributions $H_{(s_1,s_2)}^{(0,0,r)}(\rho,S)$}\label{App:OrbifoldSector}
In the following we list the coefficients $H_{(s_1,s_2)}^{(0,0,r)}(\rho,S)$ that make up what is called the reduced free energy for $N=2$ in \cite{Companion1}.
{\allowdisplaybreaks
\begin{align}
H_{(0,0)}^{(0,0,1)}(\rho,S)&=-\frac{1}{12}\phi_{-2,1}\left[\phi_{0,1}+2E_2\,\phi_{-2,1}\right]\,,\nonumber\\
H_{(2,0)}^{(0,0,1)}(\rho,S)&=\frac{1}{12\cdot 24} \left(E_4-2 E_2^2\right) \phi _{-2,1}^2+\frac{\phi _{0,1}^2}{2\cdot 24^2}\,,\nonumber\\
H_{(4,0)}^{(0,0,1)}(\rho,S)&=\frac{\left(-10 E_2^3-3 E_4 E_2+4 E_6\right) \phi _{-2,1}^2}{5\cdot 24^3}+\frac{\left(5
   E_2^2-7 E_4\right) \phi _{0,1} \phi _{-2,1}}{5\cdot 24^3}+\frac{E_2\, \phi _{0,1}^2}{4\cdot 24^3}\,,\nonumber\\
H_{(6,0)}^{(0,0,1)}(\rho,S)&=\frac{\left(-70 E_2^4-168 E_4 E_2^2-8 E_6 E_2+123 E_4^2\right)
   \phi _{-2,1}^2}{105\cdot 24^4}\nonumber\\
   &\hspace{1cm}+\frac{\left(35 E_2^3-21 E_4 E_2-29 E_6\right) \phi _{0,1}
   \phi _{-2,1}}{1260\cdot 24^3}+\frac{E_4 \phi _{0,1}^2}{160\cdot 24^3}\,,\nonumber\\
H_{(1,1)}^{(0,0,1)}(\rho,S)&=\frac{1}{6\cdot 24} \left(2 E_2^2-E_4\right) \phi _{-2,1}^2+\frac{1}{144} E_2 \phi _{0,1} \phi
   _{-2,1}+\frac{\phi _{0,1}^2}{24^2}\,,\nonumber\\
H_{(3,1)}^{(0,0,1)}(\rho,S)&=\frac{\left(5 E_2^3-3 E_4 E_2-2 E_6\right) }{15\cdot 24^2}\,\phi _{-2,1}^2\,,\nonumber\\
H_{(5,1)}^{(0,0,1)}(\rho,S)&=\frac{\left(35 E_2^4+21 E_4 E_2^2-26 E_6 E_2-30 E_4^2\right) \phi
   _{-2,1}^2}{210\cdot 24^3}+\frac{\left(-5 E_2^3+3 E_4 E_2+2 E_6\right) \phi _{0,1} \phi
   _{-2,1}}{60\cdot 24^3}\,,\nonumber\\
H_{(2,2)}^{(0,0,1)}(\rho,S)&=\frac{\left(-10 E_2^3+E_4 E_2+4 E_6\right) \phi
   _{-2,1}^2}{20\cdot 24^2}-\frac{\left(E_2^2+E_4\right) \phi _{0,1} \phi _{-2,1}}{12\cdot 24^2}-\frac{E_2\, \phi
   _{0,1}^2}{2\cdot 24^3}\,,\nonumber\\
H_{(4,2)}^{(0,0,1)}(\rho,S)&=\frac{\left(-350 E_2^4-168 E_4 E_2^2+120 E_6 E_2+279 E_4^2\right)
   \phi _{-2,1}^2}{35\cdot 24^4}\nonumber\\
&\hspace{1cm}+\frac{\left(5 E_2^3-3 E_4 E_2-7 E_6\right) \phi _{0,1}
   \phi _{-2,1}}{60\cdot 24^3}-\frac{E_4 \phi _{0,1}^2}{160\cdot 24^3}\,,\nonumber\\
H_{(3,3)}^{(0,0,1)}(\rho,S)&=\frac{\left(35 E_2^4+21 E_4 E_2^2-26 E_6 E_2-30 E_4^2\right) \phi
   _{-2,1}^2}{63\cdot 24^3}+\frac{\left(-5 E_2^3+3 E_4 E_2+2 E_6\right) \phi _{0,1} \phi
   _{-2,1}}{90\cdot 24^3}\,,\nonumber\\
\label{OrbifoldSectorContributions}
\end{align}}
Furthermore, we also have 
\begin{align}
&H_{(0,0)}^{(0,0,2)}(\rho,S)=\frac{\phi _{-2,1}}{60\cdot 24^2} \bigg[8 (E_2(\rho) (33 E_4(\rho)-48 E_4(2\rho))+2 E_2(2\rho) (64 E_4(2\rho)-59
   E_4(\rho))) \phi _{-2,1}^3\nonumber\\
   &\hspace{0.2cm}+4 \left(10 E_2(\rho)^2-40 E_2(2\rho)^2-9 E_4(\rho)+24 E_4(2\rho)\right) \phi
   _{0,1} \phi _{-2,1}^2-10 (E_2(\rho)+2 E_2(2\rho)) \phi _{0,1}^2 \phi _{-2,1}-5 \phi _{0,1}^3\bigg]\,.\label{OrbifoldSectorContributionR2}
\end{align}
as well as 
{\allowdisplaybreaks
\begin{align}
&H_{(0,0)}^{(0,0,3)}(\rho,S)=-\frac{\phi_{0,1}^5 \phi_{-2,1}}{3981312}-\frac{\phi_{0,1}^4 \phi_{-2,1}^2 (2 E_2(\rho)+3 E_2(3\rho))}{1990656}\nonumber\\
&\hspace{0.5cm}-\frac{\phi_{0,1}^3 \phi_{-2,1}^3
   \left(-15 E_2(\rho)^2+135 E_2(3\rho)^2+38 E_4(\rho)-108 E_4(3\rho)\right)}{4976640}\nonumber\\
&\hspace{0.5cm}-\frac{\phi_{0,1}^2
   \phi_{-2,1}^4 (3 E_2(3\rho) (977 E_4(\rho)-8577 E_4(3\rho))+E_2(\rho) (8091 E_4(3\rho)-311
   E_4(\rho))+15120 E_6(3\rho))}{4976640}\nonumber\\
&\hspace{0.5cm}-\frac{\phi_{0,1} \phi_{-2,1}^5}{60963840} \big[-1323 E_4(\rho)
   E_2(\rho)^2+4 (3727 E_6(\rho)+275454 E_6(3\rho)) E_2(\rho)+2231145 E_4(3\rho)^2\nonumber\\
   &\hspace{1.3cm}+3 E_2(3\rho) (3969
   E_2(3\rho) E_4(\rho)-68 (554 E_6(\rho)+15903 E_6(3\rho)))\big]\nonumber\\
&\hspace{0.5cm}-\frac{\phi_{-2,1}^6}{17418240} \big[210
   (E_4(\rho)-9 E_4(3\rho)) E_2(\rho)^3-360 E_6(\rho) E_2(\rho)^2-5 \big(701 E_4(\rho)^2\nonumber\\
&\hspace{1.3cm}+631971
   E_4(3\rho)^2\big) E_2(\rho)+15 E_2(3\rho) \big(-378 (E_4(\rho)-9 E_4(3\rho)) E_2(3\rho)^2+3071
   E_4(\rho)^2\nonumber\\
&\hspace{1.3cm}+633429 E_4(3\rho)^2\big)-24 \left(3645 E_2(3\rho)^2+15076 E_4(\rho)+249084 E_4(3\rho)\right)
   E_6(3\rho)\big])\,.\label{OrbifoldSectorContributionR3}
\end{align}}
%%%%%%%%%%%%%%%%%%%%%%%%%%%%%%%%%
\subsection{Coefficients $g^{i,(n,r=2)}_{(2,0)}(\rho)$}\label{App:CoefsgR2N2}
For completeness in the following we list the coefficients $g^{i,(n,2)}_{(2,0)}$, which constitute the next-to-leading order in $\epsilon_1$ for the free energy of $N=2$ to order $Q_R^2$.
{\allowdisplaybreaks
\begin{align}
g^{1,(n,2)}_{(2,0)}&= -\frac{n^7}{120 (1-Q_\rho^n)}+\frac{E_2 n^5}{288 (1-Q_\rho^n)}-\frac{53 E_4 n^3}{1440 (1-Q_\rho^n)}+\frac{n(E_2 E_4+2E_6)
   n}{144 (1-Q_\rho^n)}\nonumber\\
   &\hspace{0.5cm}-\left\{\begin{array}{lcl}0 & \text{if} & \text{gcd}(n,2)=1\,,\\[4pt]-\frac{n^3 \psi _2^2}{432 (1-Q_\rho^n)}+\frac{n(2E_6+\psi_2^2(2E_2+3\psi_2))}{3\cdot 24^2
   (1-Q_\rho^n)}\, & \text{if} & \text{gcd}(n,2)>1\,.\end{array}\right.\nonumber\\[10pt]
g^{2,(n,2)}_{(2,0)}&=-\frac{26 n^5-8E_2n^3+9 E_4 n}{2\cdot 24^2 (1-Q_\rho^n)}-\left\{\begin{array}{lcl} 0 & \text{if} & \text{gcd}(n,2)=1\,,\\[4pt]\frac{n^3 \psi _2}{432 (1-Q_\rho^n)}-\frac{n (6\psi _2^2+4E_2\psi_2-3E_4)}{6\cdot 24^2 (1-Q_\rho^n)}\, & \text{if} & \text{gcd}(n,2)>1\,,\end{array}\right.\nonumber\\[10pt]
g^{3,(n,2)}_{(2,0)}&= -\frac{n^3}{128 (1-Q_\rho^n)}+\frac{E_2 n}{1152 (1-Q_\rho^n)} -\left\{\begin{array}{lcl} 0 & \text{if} & \text{gcd}(n,2)=1\,,\\[4pt] -\frac{n^3}{1728 (1-Q_\rho^n)}+\frac{n (2E_2+3\psi _2)}{12\cdot 24^2 (1-Q_\rho^n)} & \text{if} & \text{gcd}(n,2)>1\,,\end{array}\right.\nonumber\\[10pt]
g^{4,(n,2)}_{(2,0)}&=-\frac{n}{8\cdot 24^2 (1-Q_\rho^n)}+\left\{\begin{array}{lcl} 0 & \text{if} & \text{gcd}(n,2)=1\,,\\[4pt] \frac{n}{24^3 (1-Q_\rho^n)}& \text{if} & \text{gcd}(n,2)>1\,,\end{array}\right.\nonumber\\[10pt]
g^{5,(n,2)}_{(2,0)}&=0\,.
\end{align}}
As we can see, these coefficients display similar characteristics as their counterparts $g^{i,(n,2)}_{(0,0)}$. To keep the presentation short, we shall not present the resummed free energy to this order.
%%%%%%%%%%%%%%%%%%%%%%%%%%%%%%%%%
\subsection{Coefficients $H_{(s_1,0)}^{(0,1)}(\rho,S)$}\label{App:CoefsHN1}
In order to verify self-similarity in the form of (\ref{SelfSimilarityGeneral}) we need the free energy for $N=1$. Indeed,the coefficients $H_{(s_1,0)}^{(0,1)}(\rho,S)$ for $s_1\in\{0,2,4,6\}$ are given by \cite{Hohenegger:2015btj}
\begin{align}
H_{(0,0)}^{(0,1)}(\rho,S)&=-\phi_{-2,1}(\rho,S)\,,\nonumber\\
H_{(2,0)}^{(0,1)}(\rho,S)&=\frac{1}{4\cdot 24}\left[\phi_{0,1}(\rho,S)-2\,E_2(\rho)\,\phi_{-2,1}(\rho,S)\right]\,,\nonumber\\
H_{(4,0)}^{(0,1)}(\rho,S)&=\frac{1}{40\cdot 24^2}\left[5\,E_2(\rho)\,\phi_{0,1}(\rho,S)-(13\,E_4(\rho)+5E_2^2(\rho))\,\phi_{-2,1}(\rho,S)\right]\,,\nonumber\\
H_{(6,0)}^{(0,1)}(\rho,S)&=\frac{1}{140\cdot 24^4}\big[21 \left(5 E_2(\rho)^2+7 E_4(\rho)\right) \phi _{0,1}(\rho,S)\nonumber\\
&\hspace{1cm}-2 \left(35 E_2(\rho)^3+273 E_4(\rho) E_2(\rho)+184
   E_6(\rho)\right)\, \phi _{-2,1}(\rho,S)\big]\,.\label{CoeffsH1NS}
\end{align}
\begin{align}
H_{(0,0)}^{(0,2)}(\rho,S)&=-\frac{1}{24}\phi_{-2,1}(\rho,S)\phi_{0,1}(\rho,S)+\frac{1}{12}\phi_{-2,1}^2(\rho,S)\,(E_2(\rho)-2E_2(2\rho))\,.\label{CoeffsH2NS}
\end{align}
\begin{align}
H_{(0,0)}^{(0,3)}(\rho,S)&=-\frac{\phi_{-2,1}\phi^2_{0,1}}{576}+\frac{\phi_{-2,1}^2\phi_{0,1}}{288}\,(E_2(\rho)-3E_2(3\rho))+\phi_{-2,1}^3\left[-\frac{37}{1440}E_4(\rho)+\frac{3}{160}E_4(3\rho)\right]\,.\label{CoeffsH3NS}
\end{align}
%%%%%%%%%%%%%%%%%%%%%%%%%%%%%%%%%%%
%%%%%%%%%%%%%%%%%%%%%%%%%%%%%%%%%%%
%%%%%%%%%%%%%%%%%%%%%%%%%%%%%%%%%%%
\subsection{Functions $G_{(0,0)}^{(i_1,i_2)}(R,S)$}\label{Sect:FuncGN2}
The $G_{(0,0)}^{(i_1,i_2)}(R,S)$ (introduced in (\ref{DefinitionG})) for $N=2$ for low values of $(i_1,i_2)$ as functions of $R$ take the form
\begin{align}
G_{(0,0)}^{(1,0)}(R,S)&=-\phi_{-2,1}\,,\nonumber\\
G_{(0,0)}^{(2,0)}(R,S)&=\frac{1}{6} (E_2(R)-E_2(2R)) \phi _{-2,1}^2\,,\nonumber\\
G_{(0,0)}^{(2,1)}(R,S)&=\frac{1}{576} \phi _{-2,1} \big[\left(8 E_2^2(R)-12 E_4(R)\right) \phi _{-2,1}^2-4 E_2(R) \phi _{0,1} \phi_{-2,1}-\phi _{0,1}^2\big]\,,\nonumber\\
G_{(0,0)}^{(3,0)}(R,S)&=-\frac{\phi _{-2,1}^2}{1440} \big[\left(20 E_2(R)^2+7 E_4(R)-27 E_4(3R)\right) \phi _{-2,1}+15
   (E_2(3R)-E_2(R)) \phi _{0,1}\big]\,,\nonumber\\
G_{(0,0)}^{(3,1)}(R,S)&=-\frac{\phi _{-2,1}^3}{1296} \big[4 \left(E_2(R)^3-E_6(R)\right) \phi _{-2,1}+3
   \left(E_4(R)-E_2(R)^2\right) \phi _{0,1}\big]\,,\nonumber\\
G_{(0,0)}^{(3,2)}(R,S)&=-\frac{\phi _{-2,1}}{3\cdot 24^4} \big[16 \left(4 E_2(R)^4-36 E_4(R) E_2(R)^2-64 E_6(R) E_2(R)+99
   E_4(R)^2\right) \phi _{-2,1}^4\nonumber\\
   &\hspace{1cm}-32 \left(2 E_2(R)^3-33 E_4(R) E_2(R)+28 E_6(R)\right)
   \phi _{0,1} \phi _{-2,1}^3\nonumber\\
   &\hspace{1cm}-24 \left(4 E_2(R)^2-7 E_4(R)\right) \phi _{0,1}^2 \phi _{-2,1}^2+24 E_2(R)
   \phi _{0,1}^3 \phi _{-2,1}+3 \phi _{0,1}^4\big]\,.
\end{align}
Here the arguments are $\phi_{-2,1}(R,S)$ and $\phi _{0,1}(R,S)$. With the results available, also $G_{(0,0)}^{(4,0)}(R,S)$ can be fixed up to a single constant, which, however, we shall not explicitly display. 
%%%%%%%%%%%%%%%%%%%%%%%%%%%%%%%%%%%
%%%%%%%%%%%%%%%%%%%%%%%%%%%%%%%%%%%
%%%%%%%%%%%%%%%%%%%%%%%%%%%%%%%%%%%

\end{document}